\documentclass[12pt]{article}

\usepackage{tikz}
\usepackage[utf8]{inputenc}
\usepackage{calculator}

\usepackage{graphicx}

\usepackage[title]{appendix}

\usepackage{graphicx}
\usepackage{amsfonts}
\usepackage{amsmath}
\usepackage{amssymb}
\usepackage{amsthm}
\usepackage{hyperref}
\usetikzlibrary{decorations.markings}

\numberwithin{equation}{section}
\newcommand{\nn}{\nonumber}
\newcommand{\Tr}{{\rm Tr}}
\newcommand{\Id}{{\rm Id}}
\newcommand{\ie}{i.e. }
\newcommand{\be}{\begin{equation} }
\newcommand{\ee}{\end{equation}}
\newcommand{\bea}{\begin{eqnarray}}
\newcommand{\eea}{\end{eqnarray}}

\newcommand{\cM}{\mathcal{M}}

\newcommand{\cN}{\mathcal{N}}

\newcommand{\OmS}{\Omega_{\rm S}}

\newlength{\abstractwidth}
\ifx\notheorems\undefined
\theoremstyle{plain}
\newtheorem{theorem}{Theorem}[section]
\newtheorem{proposition}[theorem]{Proposition}
\newtheorem{lemma}[theorem]{Lemma}
\newtheorem{assumption}[theorem]{Assumption}

\theoremstyle{definition}
\newtheorem{remark}[theorem]{Remark}
\newtheorem{definition}[theorem]{Definition}

\theoremstyle{remark}

\fi

\begin{document}


\thispagestyle{empty}

\begin{flushright}
arXiv:2110.06652v3
\end{flushright}
\vskip 0.3in

\begin{center}
{\Large \bf On the existence of scaling multi-centered black holes}
\vskip 0.2in

{\large Pierre Descombes and Boris Pioline}

\vskip 0.3in

{\it Laboratoire de Physique Th\'eorique et Hautes
Energies (LPTHE), UMR 7589 CNRS-Sorbonne Universit\'e,
Campus Pierre et Marie Curie,
4 place Jussieu, \\ F-75005 Paris, France} \\

\vskip 0.15in

{\tt \small descombes,pioline@lpthe.jussieu.fr}

\vskip 0.4in

\begin{abstract}
\vskip 0.1in
For suitable charges of the constituents, the phase space of multi-centered BPS black holes in $\cN=2$ four-dimensional supergravity famously exhibits scaling regions where the distances between the 
centers can be made arbitrarily small, so that the bound state becomes indistinguishable from a single-centered black hole.  In this note 
we establish necessary conditions on the Dirac product of charges for the existence of such regions for any number of centers, generalizing the standard triangular inequalities in the three-center case. Furthermore, we show the same conditions are necessary for the existence of multi-centered solutions at the attractor point. We prove
that similar conditions are also necessary for the existence of self-stable Abelian representations of the corresponding quiver, as suggested
by the duality between the Coulomb and Higgs branches of supersymmetric quantum mechanics.
\end{abstract}
\end{center}

\tableofcontents

\section{Introduction}

In string theory vacua with $\cN=2$ supersymmetry in 3+1 dimensions, determining
the exact number of single-particle BPS states with total electromagnetic charge $\gamma$
and arbitrary values of the moduli $z$ is still a daunting problem. Much progress has
been made on computing the BPS index $\Omega(\gamma,z)$, which counts the same states 
weighted by a sign $(-1)^{2J_3}$, where $J_3$ is the angular momentum. As a result, 
$\Omega(\gamma,z)$ becomes independent of the hypermultiplet moduli (in particular, of the string
coupling in type II string theories compactified on a Calabi-Yau threefold), but it may still
jump across certain codimension-one walls in vector moduli space -- a phenomenon 
known as wall-crossing \cite{Denef:2007vg}, first identified in the context of supersymmetric gauge
theories \cite{Seiberg:1994rs} and studied independently in the mathematical literature on Donaldson-Thomas
invariants \cite{ks,Joyce:2008pc}. 

At strong coupling, the jumps of $\Omega(\gamma,z)$ reflect the appearance or disappearance of multi-centered black hole solutions, where the total charge
$\gamma=\sum_{i=1}^n \gamma_i$ is distributed over $n$ centers with charge $\gamma_i$ \cite{Denef:2000nb,Denef:2007vg}.
The number of configurational bound states of such constituents can be determined by quantizing 
the phase space of such solutions \cite{deBoer:2008zn}, providing a transparent derivation of the wall-crossing formula \cite{Manschot:2010qz}. More generally, 
the knowledge of the number of configurational bound states allows 
to express the total index  $\Omega(\gamma,z)$ in terms of more elementary indices 
$\OmS(\gamma_i)$ which count single-centered  black holes \cite{Manschot:2011xc}, or in the context of quiver quantum mechanics, pure-Higgs states \cite{Bena:2012hf,Lee:2012sc,Manschot:2012rx}.
However, this procedure is complicated by the fact that for certain configurations of 
charges $\{\gamma_i\}$,  the space of solutions is non-compact, due to the existence of `scaling regions' where some of the centers become arbitrarily close to each other, and hard to distinguish from single-centered black holes. The modest goal of this note is to establish some necessary conditions on
the charges $\gamma_i$ for such a scaling region to arise. 

The distance between the centers of a multi-centered black hole solution are constrained by Denef's 
equations \cite{Denef:2000nb}
\be
\label{eqdenef}
\forall i=1\dots n, \qquad \sum_{j\neq i}\frac{\kappa_{ij}}{r_{ij}} = \zeta_i
\ee
where $r_{ij}=|\vec r_i - \vec r_j|$ and $\kappa_{ij}=\langle \gamma_i, \gamma_j \rangle$ is the Dirac-Schwinger-Zwanziger product of electromagnetic charges, and $\zeta_i$ are real parameters determined by the vector multiplet moduli. In order to analyze solutions to \eqref{eqdenef}, it is useful to introduce the quiver $Q$, obtained by associating one node $i$ to each center $\vec r_i$, and $\kappa_{ij}$ arrows from node $i$ to node $j$ whenever $\kappa_{ij}>0$. Indeed, the same equations \eqref{eqdenef} describe supersymmetric solutions on the Coulomb branch of the supersymmetric quantum mechanics specified by the quiver $Q$ \cite{Denef:2002ru}.
We assume that the quiver $Q$ is connected, since otherwise the $n$-body problem reduces to the same
problem for each connected component.\medskip

When non-empty, the space of solutions to \eqref{eqdenef} modulo overall translations, which we denote by $S_Q^{\zeta}$, is a real symplectic space  of dimension $2n-2$ \cite{deBoer:2008zn}. 
For generic non-zero values of $\zeta_i$ away from walls of marginal stability, it is easy to see that the maximal
distance $r_+={\rm sup}\{ r_{ij}, i<j \}$ is bounded from above, as appropriate for a classical bound state. In contrast, the minimal distance $r_-={\rm inf}\{  r_{ij}, i<j\}$ may vanish, due to some regions
in $S_Q^\zeta$ where a subset of the 
centers become arbitrarily close to each other \cite{Denef:2002ru,Bena:2006kb,Denef:2007vg}. The existence of such  `scaling regions' requires certain conditions on the signs of the Dirac products $\kappa_{ij}$, as well as on their magnitude. For example, if the $i$-th center satisfies $\kappa_{ij}\geq 0$ for all $j\neq i$, or 
$\kappa_{ij}\leq 0$ for all $j\neq i$, then 
it is clear from \eqref{eqdenef} that the distances $r_{ij}$ cannot be arbitrarily small. Thus, in order for all centers to coalesce at one point, the quiver $Q$ must have no source nor sink.
In particular, scaling solutions never exist for $n=2$ centers. For $n=3$ centers, it is well-known \cite{Bena:2006kb,Denef:2007vg} that scaling solutions exist whenever $\kappa_{12},\kappa_{23}, \kappa_{31}$ have the same sign (say positive) and moreover satisfy the triangular inequalities,
\be
\kappa_{12}\leq \kappa_{23} +\kappa_{31}, \quad 
\kappa_{23}\leq \kappa_{31} +\kappa_{12},
\quad \kappa_{31}\leq \kappa_{12} +\kappa_{23}
\ee
In that case, the equations \eqref{eqdenef} are solved by setting $r_{ij}=\lambda \kappa_{ij}+\dots$
with $\lambda\to 0$, with the dots corresponding to $\zeta$-dependent corrections which become irrelevant as $\lambda\to 0$.

In general, the existence of scaling regions where all centers coalesce is independent of the parameters $\zeta_i$, and can be analyzed by setting $\zeta_i=0$ in \eqref{eqdenef}, obtaining the
`conformal Denef equations'
\be
\label{eqdenefconf}
\forall i=1\dots n, \qquad \sum_{j\neq i} \frac{\kappa_{ij}}{r_{ij}} = 0
\ee
We refer to solutions of \eqref{eqdenefconf} as `scaling solutions'.
The case where only a subset of the $n$ centers coalesce can be 
analyzed by restricting \eqref{eqdenefconf} to those centers. 
We shall also be interested in solutions of \eqref{eqdenef}  at the special point $\zeta_i=-\sum_{j\neq i} \kappa_{ij}$,
which, as argued in \cite{MPSunpublished,Alexandrov:2018iao}, corresponds to the attractor point 
$z_\gamma$ for a black hole with total charge $\gamma=\sum_i \gamma_i$ \cite{Ferrara:1995ih}. At that point, one obtains the `attractor Denef equations'
\be
\label{eqdenefatt}
\forall i=1\dots n, \qquad \qquad \sum_{j\neq i}\kappa_{ij}
\left(1 
+\frac{1}{r_{ij}} \right)  = 0
\ee
We denote by $S_Q^0$ and $S_Q^\star$ the set of solutions to 
\eqref{eqdenefconf} and \eqref{eqdenefatt} modulo overall translations.

We shall now state a set of necessary conditions on $\kappa_{ij}$ 
for the existence of solutions to \eqref{eqdenefconf}. These conditions follow from geometric constraints on the centers, similarly to the three-center case, and are slightly more general than the conditions that were conjectured (and proven in some special cases) in \cite[\S 2.3]{Beaujard:2021fsk}. 
Quite remarkably,  we shall see that the same conditions follow from the existence of solutions to \eqref{eqdenefatt} at the attractor point.
 This is in agreement with the expectation that the only multi-centered solutions allowed at the attractor point are scaling solutions \cite{Alexandrov:2018iao}. We anticipate that these conditions will prove useful in further studies of black hole micro-states in Calabi-Yau compactifications (see \cite{Alexandrov:2018iao,Alexandrov:2018lgp,Alexandrov:2019rth,Chattopadhyaya:2021rdi} for recent progress in this direction). 

The first condition constrains the signs of the Dirac product $\kappa_{ij}$, and is easily stated in terms of the quiver $Q$: the quiver should be {\it strongly connected}, which means that for any pair of vertices $i,j$, there should be a path going from $i$ to $j$ and and a path going from $j$ to $i$. In particular, it implies that the $n$ centers cannot be separated into two disjoint groups $S$ and $\bar S$ such that all arrows go from $S$ to $\bar S$ \cite{MPSunpublished} (in particular $Q$ should have no source nor sink). 

The second condition constrains the magnitude of the Dirac products $\kappa_{ij}$, but requires some additional terminology and notation. Given a quiver $Q$, let $Q_0$ be the set of vertices, $Q_1$ the set of arrows and $Q_2$ the set of simple oriented cycles. A cut (respectively, a weak cut) is a subset $I \subset Q_1$ of the set of arrows 
such that each simple oriented cycle $w\in Q_2$ contains exactly (respectively, at most) one arrow. The second condition is that, for any weak cut $I$, the 'generalized triangular inequalities'
\be
\label{condcut}
|I|\leq |Q_1-I|
\ee
are satisfied, 
where $|I|$ denotes the number of arrows $i\to j$ in the weak cut $I$ (counted with multiplicity $\kappa_{ij}$), and similarly for 
the complement $Q_1-I$.
In the case where one of these inequalities is saturated, there are no attractor solutions, and if furthermore $Q$ is biconnected (also known as non-separable, see \S\ref{sec_conn} for the relevant definitions),  the only scaling solutions are collinear.\medskip

For example, in the special case of a cyclic quiver $v_1\to v_2 \to \dots \to v_n \to v_{n+1}=v_1$, each set of arrows $i\to i+1$ provides one possible cut, and  the condition
\eqref{condcut} reduces to the well-known conditions 
\be
\kappa_{n1} \leq  \kappa_{12} + \kappa_{23} + \dots + \kappa_{n-1,n}
\ee
and cyclic permutations thereof \cite{Manschot:2012rx}. More generally, if the quiver admits a cut as well as a cycle passing through all the nodes (which can be taken to be $v_1\to v_2 \to \dots \to v_n \to v_1$ at the expense of relabelling the nodes), then the condition 
\eqref{condcut} reduces to the conditions $\sum_{i<j} \kappa_{ij} \geq  0$ (and cyclic permutations thereof) conjectured in \cite[\S 2.3]{Beaujard:2021fsk}. We also study a natural generalization of these inequalities in the non-Abelian case, but we show that they cannot hold in full generality.
\medskip

We further  observe that the conditions stated above have a simple interpretation in terms of the Higgs branch of the supersymmetric quantum mechanics associated to the quiver $Q$ with generic potential $W$.
In that description, BPS states correspond to cohomology classes
of the moduli space of stable representations $Q$ satisfying the potential equations $\partial_a W=0$ for each arrow $a\in Q_1$, where
the stability condition is determined by the parameters $\zeta_i$ \cite{Denef:2002ru}. In the attractor (or self-stability) chamber 
$\zeta^i=-\sum_{j\neq i} \kappa_{ij}$, it is immediate to show
that stable representations cannot exist unless $Q$ is strongly connected. Moreover, for a quiver $(Q,W)$ with a cut and a generic potential, every cycle vanishes in any Abelian representation, hence each cycle contains a vanishing arrow. Let $I$ be the set of arrows such that the corresponding chiral fields vanish, and suppose $I$ is a cut \footnote{In the appendix, we adapt the proof of the final inequalities when this condition is not satisfied, \ie when there are more vanishing arrows than necessary: in this case, there are less than $|I|$ independent potential constraints}. The dimension of the moduli space of stable Abelian representations is then given by
\be
\dim_{\mathbb{C}}\cM^{\zeta,s}_{Q,W} = |Q_1-I| - |I| - (|Q_0| - 1)
\label{expectdima}
\ee
where the first term corresponds to the non-vanishing chiral fields, the second to the potential constraints induced by the vanishing chiral fields (which can be proven to be independent), and the last term to the quotient by the complexified gauge group $(\mathbb{C}^\ast)^{Q_0}$ modulo the diagonal $\mathbb{C}^\ast$ subgroup (which acts trivially). Requiring that the moduli space is not empty for some cut $I$, hence has positive dimension, we arrive at the inequality
\be\label{condcutstrong}
|I| \leq |Q_1-I| - |Q_0| +1 
\ee
In \S\ref{sec_Higgs} we show that the only cut which is compatible with the stability parameters $\zeta^i=-\sum_{j\neq i} \kappa_{ij}$ (corresponding to the attractor or self-stability condition) is the one which minimizes the expected dimension \eqref{expectdima}.\footnote{This minimization property was first observed in examples of quivers associated to non-compact Calabi-Yau threefolds in \cite{Beaujard:2020sgs}.} In 
that case,  the condition \eqref{condcutstrong} holds for any cut,
and produces a stronger\footnote{The fact that the inequalities
for the existence of self-stable representations are stronger than the one for  the existence of scaling solutions is consistent with the fact that the contribution of scaling solutions may cancel against the contribution of regular collinear solutions when computing the 
equivariant Dirac index of the phase space of multi-centered configurations using
localisation \cite{Manschot:2013sya}. \label{foocancel}} version of the condition \eqref{condcut} on the Coulomb branch. Moreover, 
we prove Proposition \ref{prophiggsweak} which says that the same inequality continues to holds when $I$ is a weak cut, as on the Coulomb branch side. We also examine a natural non-Abelian generalization of these inequalities, but conclude that it cannot hold in full generality.\medskip

The remainder of this note is organized as follows. In Section \ref{sec_Coulomb}, we  derive general constraints for the existence of scaling or attractor solutions on the Coulomb branch of Abelian quivers, and conjecture similar constraints in the 
non-Abelian case. 
In Section \ref{sec_Higgs} we discuss similar conditions for the existence of self-stable representations, which arise on the Higgs branch of the quiver quantum mechanics at the attractor point. Mathematical proofs of some
technical results are relegated to the Appendix.

\section{Existence of scaling and attractor solutions\label{sec_Coulomb}}

In this section, we derive general conditions for the existence of scaling or attractor solutions. The main idea is to reinterpret Denef's equations as current conservation, decompose each current into a sum of positive currents running around the simple cycles, and enforce generalized triangular inequalities on each cycle. In order to implement this idea, we need the  notions of 
biconnectedness, strong connectedness, cuts and R-charge, which we
introduce along the way.

\subsection{Denef's equations as current conservation}

Consider a quiver $Q=(Q_0,Q_1)$. Here $Q_0$ denotes the set of nodes of the quiver, and $Q_1$ denotes the set of arrows $a:i\to j$ of sources $s(a)=i\in Q_0$ and target $t(a)=j\in Q_0$. Let $Q_2$ be the set of simple oriented cycles of $Q$, \ie oriented cycles passing at most once through each node of $Q$. Consider the sequence:
\begin{align}\label{graphcoho2}
  \mathbb{R}^{Q_2}\overset{\partial_2}{\to}\mathbb{R}^{Q_1}\overset{\partial_1}{\to}\mathbb{R}^{Q_0}\overset{\partial_0}\to\mathbb{R}
\end{align}
with $\partial_2$, $\partial_1$ and $\partial_0$ defined by 
\be
\partial_2(w)=\sum_{a\in w}a\ ,\qquad
    \partial_1(a)=t(a)-s(a)\ ,\qquad\partial_0(i)=1
\ee
It is immediate to check that $\partial_1\circ\partial_2=0$ and $\partial_0\circ\partial_1=0$, so \eqref{graphcoho2} is a complex\footnote{This complex is related to the cellular homology complex $0{\to}\mathbb{R}^{Q_1}\overset{\partial_1}{\to}\mathbb{R}^{Q_0}\to 0$ of the graph considered as a cellular space. The cellular homology group $H^0$ gives the number of connected components of the graph, see \cite[Sec 2.2]{Hatch}.}.
We refer to elements of $\mathbb{R}^{Q_1}$ as `currents', and elements of $\ker(\partial_1)$ as `conserved currents'. 
The complex \eqref{graphcoho2} is exact at $\mathbb{R}^{Q_0}$ if and only if $Q$ is connected: we shall assume that $Q$ is connected unless otherwise specified.\medskip

The Denef equations \eqref{eqdenef} for general stability parameters $\zeta_i$ can be recast as current conservation by rewriting them as follows:
\begin{align}\label{FIstabC}
    \forall\;i\in Q_0\quad \sum_{(a:i\to j)\in Q_1}\frac{1}{r_{ij}}-\sum_{(a:j\to i)\in Q_1}\frac{1}{r_{ij}}=\zeta_i \nn\\
    \iff \partial_1\left( (\frac{1}{r_{ij}})_{(a:i\to j)\in Q_1} \right)=(\zeta_i)_{i\in Q_0}
\end{align}
Assuming (without loss of generality) that the quiver $Q$ is connected, and  summing over $i\in Q_0$, it is clear that solutions only exist if $\partial_0((\zeta_i)_{i\in Q_0})=\sum_{i\in Q_0}\zeta_i=0$. We can therefore choose $(\chi_a)_{a\in Q_1}$ such that $\partial_1((\chi_a)_{a\in Q_1})=(-\zeta_i)_{i\in Q_0}$. We define the current $\lambda^\chi_a$ running along the arrow $a:i\to j$ as
\begin{align}
    \lambda^\chi_a:=\chi_a+\frac{1}{r_{ij}}
\end{align}
The equations \eqref{FIstabC} then amount to conservation $\partial_1(\lambda^\chi)=0$ of the current $\lambda^\chi$.
For given $\zeta_i$, the $\chi_a$ are in general not unique, but for the conformal (resp. attractor) Denef equations there is a canonical choice, namely
\begin{align}
\label{denefcurrents}
    \lambda^0_a:=\frac{1}{r_{ij}}\ ,\qquad 
    \lambda^\star_a:= 1+\frac{1}{r_{ij}} 
\end{align}
such that the conserved current $\lambda^0$ and $\lambda^\star$ are strictly positive. This positivity will be crucial for deriving constraints on the existence of scaling and attractor solutions.
This property is preserved for a small perturbation of the 
attractor stability parameters,  obtained by replacing 
$\lambda^\star_a$ by $1+\delta_a+\frac{1}{r_{ij}}$ with $|\delta_a|\ll 1$. \medskip

\subsection{Biconnectedness and strong connectedness\label{sec_conn}}

In order to study the existence of conformal or
attractor solutions, it will be convenient to decompose the 
connected quiver $Q$ in two steps: first into biconnected components, and then into strongly connected components.\footnote{The notions of  biconnected graph (also known as 'non-separable'), biconnected components of a graph (also known as `blocks'), strongly connected directed graph and strongly connected components are standard in graph theory, see for example \cite[Sec 3,6]{Har}.}\medskip

For the first step, a quiver is said to be biconnected if it cannot be disconnected by removing one node. In general, a quiver can be decomposed into maximal biconnected subquivers called biconnected components joined by a shared node (see Figure \ref{figfc}). Consider the unoriented graph $K$ with one node for each biconnected component and one node for each node of the quiver shared between different biconnected components, and an edge between the node $i$ and the biconnected component $b$ if $i\in b$. In lemma \ref{lembicongraph} 
we show that $K$ is a connected tree, \ie has no cycle. Given a solution of Denef's equations at any stability for each biconnected component, one can join these solutions at shared nodes by identifying the corresponding centers: the fact that $K$ has no cycle ensures that this can be done consistently. According to lemma \ref{lembiconcons}, a conserved current running on a connected quiver is also conserved on each of its biconnected components, and the converse is also true: a solution of Denef's equations on a connected quiver is obtained by gluing together solutions for each of the biconnected components. The gluing at a common point in $\mathbb{R}^3$ freezes the relative translational degrees of freedom. Denoting by $B$ the set of biconnected components, one has:
\begin{align}
    S^\zeta_Q =\prod_{b\in B} S^\zeta_{Q^b}
\end{align}

\medskip

One can also decompose $Q$ into strongly connected components, defined as follows. A quiver is strongly connected if for each pair of nodes $i,j\in Q_0$ there is an oriented path from $i$ to $j$ and from $j$ to $i$. We define an equivalence relation between nodes by writing $i\sim j$ if there is a path from $i$ to $j$ and from $j$ to $i$: this relation is automatically symmetric, reflexive by considering the trivial path, and transitive by concatenation of paths. The  equivalence classes under this relation are called strongly connected components. Let us now draw an arrow from a strongly connected component to another if there is an arrow in $Q$ from one node in the first equivalence class to a node of the second equivalence class. The resulting graph $\Gamma$ is connected and has no cycle because there cannot exists paths going both ways between two strongly connected components. Thus $\Gamma$ defines a poset. In Figure
\ref{figsc}, we show an example of a quiver with its graph $\Gamma$ of strongly connected components.

From lemma \ref{lemgraphcoho}, a strictly positive conserved current can exist on a quiver $Q$ only if $Q$ is strongly connected, therefore:

\begin{proposition}\label{propstrongcon}
    A quiver which admits scaling or attractor solutions must be strongly connected.
\end{proposition}

\begin{figure}
\centering
\begin{tikzpicture}[x=0.75pt,y=0.75pt,yscale=-1,xscale=1]

\draw    (126,152.5) -- (183.38,110.68) ;
\draw [shift={(185,109.5)}, rotate = 503.91] [color={rgb, 255:red, 0; green, 0; blue, 0 }  ][line width=0.75]    (10.93,-3.29) .. controls (6.95,-1.4) and (3.31,-0.3) .. (0,0) .. controls (3.31,0.3) and (6.95,1.4) .. (10.93,3.29)   ;
\draw    (126,152.5) -- (182.17,177.68) ;
\draw [shift={(184,178.5)}, rotate = 204.15] [color={rgb, 255:red, 0; green, 0; blue, 0 }  ][line width=0.75]    (10.93,-3.29) .. controls (6.95,-1.4) and (3.31,-0.3) .. (0,0) .. controls (3.31,0.3) and (6.95,1.4) .. (10.93,3.29)   ;
\draw    (185,109.5) -- (184.03,176.5) ;
\draw [shift={(184,178.5)}, rotate = 270.83] [color={rgb, 255:red, 0; green, 0; blue, 0 }  ][line width=0.75]    (10.93,-3.29) .. controls (6.95,-1.4) and (3.31,-0.3) .. (0,0) .. controls (3.31,0.3) and (6.95,1.4) .. (10.93,3.29)   ;
\draw    (233,165.5) -- (186.3,111.02) ;
\draw [shift={(185,109.5)}, rotate = 409.4] [color={rgb, 255:red, 0; green, 0; blue, 0 }  ][line width=0.75]    (10.93,-3.29) .. controls (6.95,-1.4) and (3.31,-0.3) .. (0,0) .. controls (3.31,0.3) and (6.95,1.4) .. (10.93,3.29)   ;
\draw    (185,109.5) -- (254.01,102.7) ;
\draw [shift={(256,102.5)}, rotate = 534.37] [color={rgb, 255:red, 0; green, 0; blue, 0 }  ][line width=0.75]    (10.93,-3.29) .. controls (6.95,-1.4) and (3.31,-0.3) .. (0,0) .. controls (3.31,0.3) and (6.95,1.4) .. (10.93,3.29)   ;
\draw    (256,102.5) -- (292.69,144.99) ;
\draw [shift={(294,146.5)}, rotate = 229.18] [color={rgb, 255:red, 0; green, 0; blue, 0 }  ][line width=0.75]    (10.93,-3.29) .. controls (6.95,-1.4) and (3.31,-0.3) .. (0,0) .. controls (3.31,0.3) and (6.95,1.4) .. (10.93,3.29)   ;
\draw    (294,146.5) -- (234.91,164.91) ;
\draw [shift={(233,165.5)}, rotate = 342.7] [color={rgb, 255:red, 0; green, 0; blue, 0 }  ][line width=0.75]    (10.93,-3.29) .. controls (6.95,-1.4) and (3.31,-0.3) .. (0,0) .. controls (3.31,0.3) and (6.95,1.4) .. (10.93,3.29)   ;
\draw    (256,102.5) -- (233.69,163.62) ;
\draw [shift={(233,165.5)}, rotate = 290.06] [color={rgb, 255:red, 0; green, 0; blue, 0 }  ][line width=0.75]    (10.93,-3.29) .. controls (6.95,-1.4) and (3.31,-0.3) .. (0,0) .. controls (3.31,0.3) and (6.95,1.4) .. (10.93,3.29)   ;
\draw    (185,109.5) -- (121.55,57.76) ;
\draw [shift={(120,56.5)}, rotate = 399.19] [color={rgb, 255:red, 0; green, 0; blue, 0 }  ][line width=0.75]    (10.93,-3.29) .. controls (6.95,-1.4) and (3.31,-0.3) .. (0,0) .. controls (3.31,0.3) and (6.95,1.4) .. (10.93,3.29)   ;
\draw    (120,56.5) -- (174.11,38.14) ;
\draw [shift={(176,37.5)}, rotate = 521.26] [color={rgb, 255:red, 0; green, 0; blue, 0 }  ][line width=0.75]    (10.93,-3.29) .. controls (6.95,-1.4) and (3.31,-0.3) .. (0,0) .. controls (3.31,0.3) and (6.95,1.4) .. (10.93,3.29)   ;
\draw    (176,37.5) -- (184.75,107.52) ;
\draw [shift={(185,109.5)}, rotate = 262.87] [color={rgb, 255:red, 0; green, 0; blue, 0 }  ][line width=0.75]    (10.93,-3.29) .. controls (6.95,-1.4) and (3.31,-0.3) .. (0,0) .. controls (3.31,0.3) and (6.95,1.4) .. (10.93,3.29)   ;
\draw    (256,102.5) -- (316.02,94.76) ;
\draw [shift={(318,94.5)}, rotate = 532.65] [color={rgb, 255:red, 0; green, 0; blue, 0 }  ][line width=0.75]    (10.93,-3.29) .. controls (6.95,-1.4) and (3.31,-0.3) .. (0,0) .. controls (3.31,0.3) and (6.95,1.4) .. (10.93,3.29)   ;
\draw    (318,94.5) -- (306.41,38.46) ;
\draw [shift={(306,36.5)}, rotate = 438.31] [color={rgb, 255:red, 0; green, 0; blue, 0 }  ][line width=0.75]    (10.93,-3.29) .. controls (6.95,-1.4) and (3.31,-0.3) .. (0,0) .. controls (3.31,0.3) and (6.95,1.4) .. (10.93,3.29)   ;
\draw    (306,36.5) -- (244.98,44.25) ;
\draw [shift={(243,44.5)}, rotate = 352.76] [color={rgb, 255:red, 0; green, 0; blue, 0 }  ][line width=0.75]    (10.93,-3.29) .. controls (6.95,-1.4) and (3.31,-0.3) .. (0,0) .. controls (3.31,0.3) and (6.95,1.4) .. (10.93,3.29)   ;
\draw    (256,102.5) -- (243.44,46.45) ;
\draw [shift={(243,44.5)}, rotate = 437.37] [color={rgb, 255:red, 0; green, 0; blue, 0 }  ][line width=0.75]    (10.93,-3.29) .. controls (6.95,-1.4) and (3.31,-0.3) .. (0,0) .. controls (3.31,0.3) and (6.95,1.4) .. (10.93,3.29)   ;
\draw  [color={rgb, 255:red, 208; green, 2; blue, 27 }  ,draw opacity=1 ] (136.93,186.16) .. controls (116.96,173.64) and (111.54,146.33) .. (124.81,125.16) .. controls (138.09,103.99) and (165.04,96.98) .. (185,109.5) .. controls (204.96,122.02) and (210.39,149.33) .. (197.11,170.5) .. controls (183.84,191.67) and (156.89,198.68) .. (136.93,186.16) -- cycle ;
\draw  [color={rgb, 255:red, 208; green, 2; blue, 27 }  ,draw opacity=1 ] (183.58,112.63) .. controls (192.53,92.93) and (225.88,88.81) .. (258.06,103.43) .. controls (290.23,118.06) and (309.06,145.89) .. (300.11,165.59) .. controls (291.15,185.29) and (257.81,189.41) .. (225.63,174.79) .. controls (193.45,160.16) and (174.62,132.33) .. (183.58,112.63) -- cycle ;
\draw  [color={rgb, 255:red, 208; green, 2; blue, 27 }  ,draw opacity=1 ] (246,33.98) .. controls (262.16,12.3) and (292.84,7.83) .. (314.52,23.99) .. controls (336.2,40.15) and (340.67,70.82) .. (324.52,92.5) .. controls (308.36,114.18) and (277.68,118.66) .. (256,102.5) .. controls (234.32,86.34) and (229.84,55.66) .. (246,33.98) -- cycle ;
\draw  [color={rgb, 255:red, 208; green, 2; blue, 27 }  ,draw opacity=1 ] (139.57,23.85) .. controls (161.11,12.43) and (188.74,22.34) .. (201.29,45.99) .. controls (213.83,69.64) and (206.54,98.08) .. (185,109.5) .. controls (163.46,120.92) and (135.83,111.01) .. (123.29,87.36) .. controls (110.74,63.71) and (118.03,35.27) .. (139.57,23.85) -- cycle ;
\draw  [color={rgb, 255:red, 208; green, 2; blue, 27 }  ,draw opacity=1 ] (411.93,185.16) .. controls (391.96,172.64) and (386.54,145.33) .. (399.81,124.16) .. controls (413.09,102.99) and (440.04,95.98) .. (460,108.5) .. controls (479.96,121.02) and (485.39,148.33) .. (472.11,169.5) .. controls (458.84,190.67) and (431.89,197.68) .. (411.93,185.16) -- cycle ;
\draw  [color={rgb, 255:red, 208; green, 2; blue, 27 }  ,draw opacity=1 ] (458.58,111.63) .. controls (467.53,91.93) and (500.88,87.81) .. (533.06,102.43) .. controls (565.23,117.06) and (584.06,144.89) .. (575.11,164.59) .. controls (566.15,184.29) and (532.81,188.41) .. (500.63,173.79) .. controls (468.45,159.16) and (449.62,131.33) .. (458.58,111.63) -- cycle ;
\draw  [color={rgb, 255:red, 208; green, 2; blue, 27 }  ,draw opacity=1 ] (521,32.98) .. controls (537.16,11.3) and (567.84,6.83) .. (589.52,22.99) .. controls (611.2,39.15) and (615.67,69.82) .. (599.52,91.5) .. controls (583.36,113.18) and (552.68,117.66) .. (531,101.5) .. controls (509.32,85.34) and (504.84,54.66) .. (521,32.98) -- cycle ;
\draw  [color={rgb, 255:red, 208; green, 2; blue, 27 }  ,draw opacity=1 ] (414.57,22.85) .. controls (436.11,11.43) and (463.74,21.34) .. (476.29,44.99) .. controls (488.83,68.64) and (481.54,97.08) .. (460,108.5) .. controls (438.46,119.92) and (410.83,110.01) .. (398.29,86.36) .. controls (385.74,62.71) and (393.03,34.27) .. (414.57,22.85) -- cycle ;
\draw [color={rgb, 255:red, 0; green, 0; blue, 0 }  ,draw opacity=1 ]   (560.26,62.24) -- (533.06,102.43) ;
\draw [color={rgb, 255:red, 0; green, 0; blue, 0 }  ,draw opacity=1 ]   (533.06,102.43) -- (516.84,138.11) ;
\draw [color={rgb, 255:red, 0; green, 0; blue, 0 }  ,draw opacity=1 ]   (460,108.5) -- (516.84,138.11) ;
\draw [color={rgb, 255:red, 0; green, 0; blue, 0 }  ,draw opacity=1 ]   (437.29,65.67) -- (460,108.5) ;
\draw [color={rgb, 255:red, 0; green, 0; blue, 0 }  ,draw opacity=1 ]   (458.58,111.63) -- (435.96,146.83) ;

\end{tikzpicture}
\caption{Constructing the unoriented graph of biconnected components.\label{figfc}}
\end{figure}
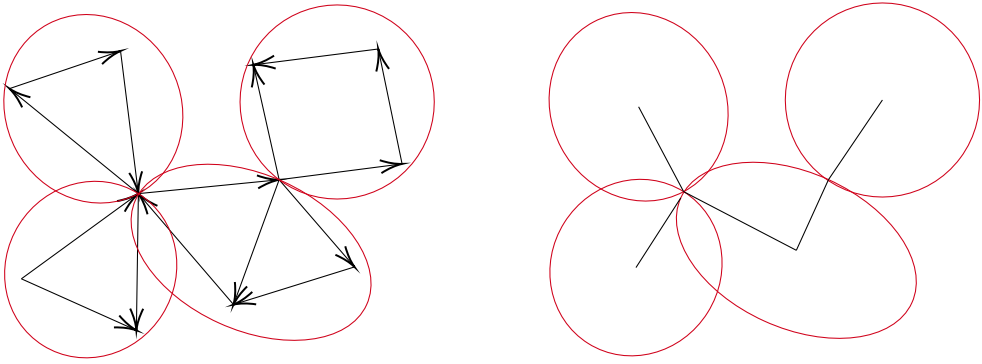

\begin{figure}
\begin{center}
\begin{tikzpicture}[x=0.75pt,y=0.75pt,yscale=-1,xscale=1]
\draw    (86,153.5) -- (113.79,117.09) ;
\draw [shift={(115,115.5)}, rotate = 487.35] [color={rgb, 255:red, 0; green, 0; blue, 0 }  ][line width=0.75]    (10.93,-3.29) .. controls (6.95,-1.4) and (3.31,-0.3) .. (0,0) .. controls (3.31,0.3) and (6.95,1.4) .. (10.93,3.29)   ;
\draw    (115,115.5) -- (132.11,149.71) ;
\draw [shift={(133,151.5)}, rotate = 243.43] [color={rgb, 255:red, 0; green, 0; blue, 0 }  ][line width=0.75]    (10.93,-3.29) .. controls (6.95,-1.4) and (3.31,-0.3) .. (0,0) .. controls (3.31,0.3) and (6.95,1.4) .. (10.93,3.29)   ;
\draw    (133,151.5) -- (88,153.41) ;
\draw [shift={(86,153.5)}, rotate = 357.56] [color={rgb, 255:red, 0; green, 0; blue, 0 }  ][line width=0.75]    (10.93,-3.29) .. controls (6.95,-1.4) and (3.31,-0.3) .. (0,0) .. controls (3.31,0.3) and (6.95,1.4) .. (10.93,3.29)   ;
\draw    (115,115.5) -- (160.16,96.28) ;
\draw [shift={(162,95.5)}, rotate = 516.95] [color={rgb, 255:red, 0; green, 0; blue, 0 }  ][line width=0.75]    (10.93,-3.29) .. controls (6.95,-1.4) and (3.31,-0.3) .. (0,0) .. controls (3.31,0.3) and (6.95,1.4) .. (10.93,3.29)   ;
\draw    (133,151.5) -- (161.08,97.28) ;
\draw [shift={(162,95.5)}, rotate = 477.38] [color={rgb, 255:red, 0; green, 0; blue, 0 }  ][line width=0.75]    (10.93,-3.29) .. controls (6.95,-1.4) and (3.31,-0.3) .. (0,0) .. controls (3.31,0.3) and (6.95,1.4) .. (10.93,3.29)   ;
\draw    (162,95.5) -- (188.64,66.96) ;
\draw [shift={(190,65.5)}, rotate = 493.03] [color={rgb, 255:red, 0; green, 0; blue, 0 }  ][line width=0.75]    (10.93,-3.29) .. controls (6.95,-1.4) and (3.31,-0.3) .. (0,0) .. controls (3.31,0.3) and (6.95,1.4) .. (10.93,3.29)   ;
\draw    (190,65.5) -- (211.86,96.86) ;
\draw [shift={(213,98.5)}, rotate = 235.12] [color={rgb, 255:red, 0; green, 0; blue, 0 }  ][line width=0.75]    (10.93,-3.29) .. controls (6.95,-1.4) and (3.31,-0.3) .. (0,0) .. controls (3.31,0.3) and (6.95,1.4) .. (10.93,3.29)   ;
\draw    (213,98.5) -- (184.49,124.16) ;
\draw [shift={(183,125.5)}, rotate = 318.01] [color={rgb, 255:red, 0; green, 0; blue, 0 }  ][line width=0.75]    (10.93,-3.29) .. controls (6.95,-1.4) and (3.31,-0.3) .. (0,0) .. controls (3.31,0.3) and (6.95,1.4) .. (10.93,3.29)   ;
\draw    (183,125.5) -- (163.15,97.14) ;
\draw [shift={(162,95.5)}, rotate = 415.01] [color={rgb, 255:red, 0; green, 0; blue, 0 }  ][line width=0.75]    (10.93,-3.29) .. controls (6.95,-1.4) and (3.31,-0.3) .. (0,0) .. controls (3.31,0.3) and (6.95,1.4) .. (10.93,3.29)   ;
\draw    (183,125.5) -- (189.77,67.49) ;
\draw [shift={(190,65.5)}, rotate = 456.65] [color={rgb, 255:red, 0; green, 0; blue, 0 }  ][line width=0.75]    (10.93,-3.29) .. controls (6.95,-1.4) and (3.31,-0.3) .. (0,0) .. controls (3.31,0.3) and (6.95,1.4) .. (10.93,3.29)   ;
\draw    (190,65.5) -- (224.48,35.81) ;
\draw [shift={(226,34.5)}, rotate = 499.27] [color={rgb, 255:red, 0; green, 0; blue, 0 }  ][line width=0.75]    (10.93,-3.29) .. controls (6.95,-1.4) and (3.31,-0.3) .. (0,0) .. controls (3.31,0.3) and (6.95,1.4) .. (10.93,3.29)   ;
\draw    (133,151.5) -- (174.45,185.24) ;
\draw [shift={(176,186.5)}, rotate = 219.14] [color={rgb, 255:red, 0; green, 0; blue, 0 }  ][line width=0.75]    (10.93,-3.29) .. controls (6.95,-1.4) and (3.31,-0.3) .. (0,0) .. controls (3.31,0.3) and (6.95,1.4) .. (10.93,3.29)   ;
\draw    (176,186.5) -- (182.77,127.49) ;
\draw [shift={(183,125.5)}, rotate = 456.55] [color={rgb, 255:red, 0; green, 0; blue, 0 }  ][line width=0.75]    (10.93,-3.29) .. controls (6.95,-1.4) and (3.31,-0.3) .. (0,0) .. controls (3.31,0.3) and (6.95,1.4) .. (10.93,3.29)   ;
\draw    (176,186.5) -- (212.22,100.34) ;
\draw [shift={(213,98.5)}, rotate = 472.8] [color={rgb, 255:red, 0; green, 0; blue, 0 }  ][line width=0.75]    (10.93,-3.29) .. controls (6.95,-1.4) and (3.31,-0.3) .. (0,0) .. controls (3.31,0.3) and (6.95,1.4) .. (10.93,3.29)   ;
\draw  [color={rgb, 255:red, 241; green, 8; blue, 57 }  ,draw opacity=1 ] (74,139.5) .. controls (74,120.45) and (89.45,105) .. (108.5,105) .. controls (127.55,105) and (143,120.45) .. (143,139.5) .. controls (143,158.55) and (127.55,174) .. (108.5,174) .. controls (89.45,174) and (74,158.55) .. (74,139.5) -- cycle ;
\draw  [color={rgb, 255:red, 241; green, 8; blue, 57 }  ,draw opacity=1 ] (146,95.5) .. controls (146,73.13) and (164.13,55) .. (186.5,55) .. controls (208.87,55) and (227,73.13) .. (227,95.5) .. controls (227,117.87) and (208.87,136) .. (186.5,136) .. controls (164.13,136) and (146,117.87) .. (146,95.5) -- cycle ;
\draw  [color={rgb, 255:red, 241; green, 8; blue, 57 }  ,draw opacity=1 ] (146.75,186.5) .. controls (146.75,170.35) and (159.85,157.25) .. (176,157.25) .. controls (192.15,157.25) and (205.25,170.35) .. (205.25,186.5) .. controls (205.25,202.65) and (192.15,215.75) .. (176,215.75) .. controls (159.85,215.75) and (146.75,202.65) .. (146.75,186.5) -- cycle ;
\draw  [color={rgb, 255:red, 241; green, 8; blue, 57 }  ,draw opacity=1 ] (204,34.5) .. controls (204,22.35) and (213.85,12.5) .. (226,12.5) .. controls (238.15,12.5) and (248,22.35) .. (248,34.5) .. controls (248,46.65) and (238.15,56.5) .. (226,56.5) .. controls (213.85,56.5) and (204,46.65) .. (204,34.5) -- cycle ;
\draw  [color={rgb, 255:red, 241; green, 8; blue, 57 }  ,draw opacity=1 ] (299,139.5) .. controls (299,120.45) and (314.45,105) .. (333.5,105) .. controls (352.55,105) and (368,120.45) .. (368,139.5) .. controls (368,158.55) and (352.55,174) .. (333.5,174) .. controls (314.45,174) and (299,158.55) .. (299,139.5) -- cycle ;
\draw  [color={rgb, 255:red, 241; green, 8; blue, 57 }  ,draw opacity=1 ] (371,95.5) .. controls (371,73.13) and (389.13,55) .. (411.5,55) .. controls (433.87,55) and (452,73.13) .. (452,95.5) .. controls (452,117.87) and (433.87,136) .. (411.5,136) .. controls (389.13,136) and (371,117.87) .. (371,95.5) -- cycle ;
\draw  [color={rgb, 255:red, 241; green, 8; blue, 57 }  ,draw opacity=1 ] (371.75,186.5) .. controls (371.75,170.35) and (384.85,157.25) .. (401,157.25) .. controls (417.15,157.25) and (430.25,170.35) .. (430.25,186.5) .. controls (430.25,202.65) and (417.15,215.75) .. (401,215.75) .. controls (384.85,215.75) and (371.75,202.65) .. (371.75,186.5) -- cycle ;
\draw  [color={rgb, 255:red, 241; green, 8; blue, 57 }  ,draw opacity=1 ] (429,34.5) .. controls (429,22.35) and (438.85,12.5) .. (451,12.5) .. controls (463.15,12.5) and (473,22.35) .. (473,34.5) .. controls (473,46.65) and (463.15,56.5) .. (451,56.5) .. controls (438.85,56.5) and (429,46.65) .. (429,34.5) -- cycle ;
\draw [color={rgb, 255:red, 252; green, 8; blue, 8 }  ,draw opacity=1 ]   (333.5,139.5) -- (409.76,96.48) ;
\draw [shift={(411.5,95.5)}, rotate = 510.57] [color={rgb, 255:red, 252; green, 8; blue, 8 }  ,draw opacity=1 ][line width=0.75]    (10.93,-3.29) .. controls (6.95,-1.4) and (3.31,-0.3) .. (0,0) .. controls (3.31,0.3) and (6.95,1.4) .. (10.93,3.29)   ;
\draw [color={rgb, 255:red, 252; green, 8; blue, 8 }  ,draw opacity=1 ]   (333.5,139.5) -- (399.36,185.36) ;
\draw [shift={(401,186.5)}, rotate = 214.85] [color={rgb, 255:red, 252; green, 8; blue, 8 }  ,draw opacity=1 ][line width=0.75]    (10.93,-3.29) .. controls (6.95,-1.4) and (3.31,-0.3) .. (0,0) .. controls (3.31,0.3) and (6.95,1.4) .. (10.93,3.29)   ;
\draw [color={rgb, 255:red, 252; green, 8; blue, 8 }  ,draw opacity=1 ]   (401,186.5) -- (411.27,97.49) ;
\draw [shift={(411.5,95.5)}, rotate = 456.58] [color={rgb, 255:red, 252; green, 8; blue, 8 }  ,draw opacity=1 ][line width=0.75]    (10.93,-3.29) .. controls (6.95,-1.4) and (3.31,-0.3) .. (0,0) .. controls (3.31,0.3) and (6.95,1.4) .. (10.93,3.29)   ;
\draw [color={rgb, 255:red, 252; green, 8; blue, 8 }  ,draw opacity=1 ]   (411.5,95.5) -- (449.91,36.18) ;
\draw [shift={(451,34.5)}, rotate = 482.92] [color={rgb, 255:red, 252; green, 8; blue, 8 }  ,draw opacity=1 ][line width=0.75]    (10.93,-3.29) .. controls (6.95,-1.4) and (3.31,-0.3) .. (0,0) .. controls (3.31,0.3) and (6.95,1.4) .. (10.93,3.29)   ;

\draw (255,119.4) node [anchor=north west][inner sep=0.75pt]    {$\Rightarrow $};
\end{tikzpicture}
\end{center}
\caption{Constructing the graph of strongly connected components.
\label{figsc}}
\end{figure}
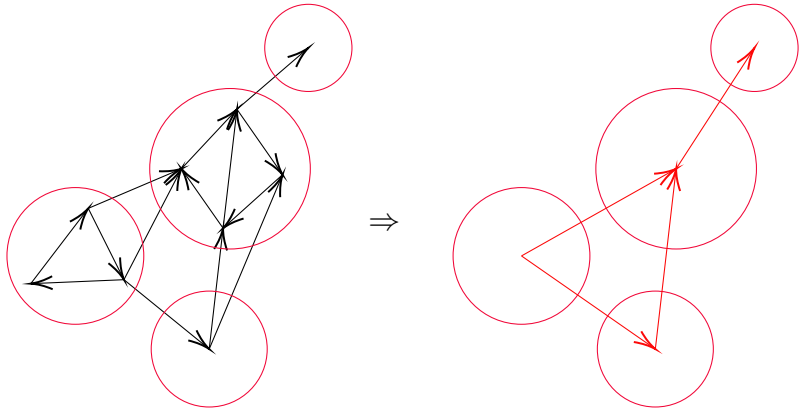

Notice that by $i)$ of lemma \ref{lembiconstrong}, a quiver is strongly connected if and only if its biconnected components are strongly connected. The biconnected components of a strongly connected quiver $Q$ have an alternative description as follows. Consider the equivalence relation $\sim$ on the arrows of $Q$ generated by $a\sim b$ if they are in a simple oriented cycle $w\in Q_2$.  From $ii)$ of lemma \ref{lembiconstrong}, the equivalence classes of $\sim$ are identical to the biconnected components of $Q$, therefore $Q$ is biconnected if and only if $\sim$ has a single equivalence class.

\subsection{Cuts, weak cuts and R-charge \label{sec_cuts}}

Consider a map $R:\mathbb{R}^Q_1\to \mathbb{R}$ assigning a charge $R(a)$ to each arrow $a\in Q_1$. For any simple cycle $w\in Q_2$,  one consider the charge of the cycle
$R\circ\partial_2(w)=\sum_{a\in w}R(a)$. We say that $R$ is an R-charge if all simple cycles have charge $R\circ\partial_2(w)=2$. For any subset $I\subset Q_1$ of arrows of $Q$, we define the charge $R_I$ by $R_I(a)=2$ if $a\in I$, and $R_I(a)=0$ if $a\in Q_1-I$.\medskip 

We define a cut (resp. a weak cut, a strong cut) as a subset $I$ of arrows of the quiver such that each simple oriented cycle $w\in Q_2$ contains exactly (resp. at most, at least) one arrow of $I$, or equivalently $R_I\circ\partial_2(w)=2$ (resp. $R_I\circ\partial_2(w)\leq2$, $R_I\circ\partial_2(w)\geq2$). We shall consider maximal weak cuts, \ie weak cuts which are maximal for the inclusion. There are two possibilities, as proven in lemma \ref{cutRcharge}:

\begin{itemize}
    \item either $Q$ admits a cut, and then all maximal weak cuts are cuts, and  $Q$ admits an $R$-charge, 
    \item or $Q$ admits no $R$-charge and in particular no cut.
\end{itemize}

\begin{figure}
\begin{center}
\begin{tikzpicture}
\begin{scope}[shift={(0,-0.5)}]  
    \node[draw,circle,inner sep=2pt,fill] (A) at (-2,0) {};
    \node[draw,circle,inner sep=2pt,fill] (B) at (0,0) {};
    \node[draw,circle,inner sep=2pt,fill] (C) at (-1,1) {};
    \draw[->,>=latex](A) -- (B) {};
    \draw[->,>=latex](B) -- (C) {};
    \draw[->,>=latex,red,dashed](C) -- (A) {};
\end{scope}
\begin{scope}[shift={(0,0)}] 
    \node[draw,circle,inner sep=2pt,fill] (D) at (3,0) {};
    \node[draw,circle,inner sep=2pt,fill] (E) at (5,0) {};
    \node[draw,circle,inner sep=2pt,fill] (F) at (4,1) {};
    \node[draw,circle,inner sep=2pt,fill] (G) at (4,-1) {};
    \draw[->,>=latex](F) -- (D) {};
    \draw[->,>=latex](F) -- (E) {};
    \draw[->,>=latex](E) -- (G) {};
    \draw[->,>=latex](D) -- (G) {};
    \draw[->,>=latex,dashed,red](G) -- (F) {};
    
    \node[draw,circle,inner sep=2pt,fill] (H) at (6,0) {};
    \node[draw,circle,inner sep=2pt,fill] (I) at (8,0) {};
    \node[draw,circle,inner sep=2pt,fill] (J) at (7,1) {};
    \node[draw,circle,inner sep=2pt,fill] (K) at (7,-1) {};https://www.overleaf.com/project/6138c32853baff4c6a09d9b6
    \draw[->,>=latex,dashed,red](J) -- (H) {};
    \draw[->,>=latex](J) -- (I) {};
    \draw[->,>=latex](H) -- (K) {};
    \draw[->,>=latex,dashed,red](I) -- (K) {};
    \draw[->,>=latex](K) -- (J) {};
    
    \node[draw,circle,inner sep=2pt,fill] (D1) at (9,0) {};
    \node[draw,circle,inner sep=2pt,fill] (E1) at (11,0) {};
    \node[draw,circle,inner sep=2pt,fill] (F1) at (10,1) {};
    \node[draw,circle,inner sep=2pt,fill] (G1) at (10,-1) {};
    \draw[->,>=latex,dashed,red](F1) -- (D1) {};
    \draw[->,>=latex,dashed,red](F1) -- (E1) {};
    \draw[->,>=latex](E1) -- (G1) {};
    \draw[->,>=latex](D1) -- (G1) {};
    \draw[->,>=latex](G1) -- (F1) {};
    \end{scope}
\end{tikzpicture}
\end{center}
\caption{Two examples of quivers with cuts. Arrows in the cut are drawn in dashed red.\label{Fig1}}
\end{figure}
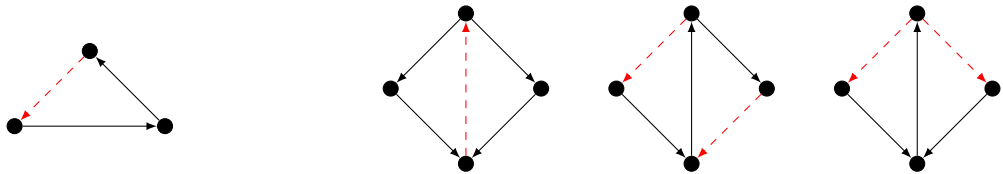

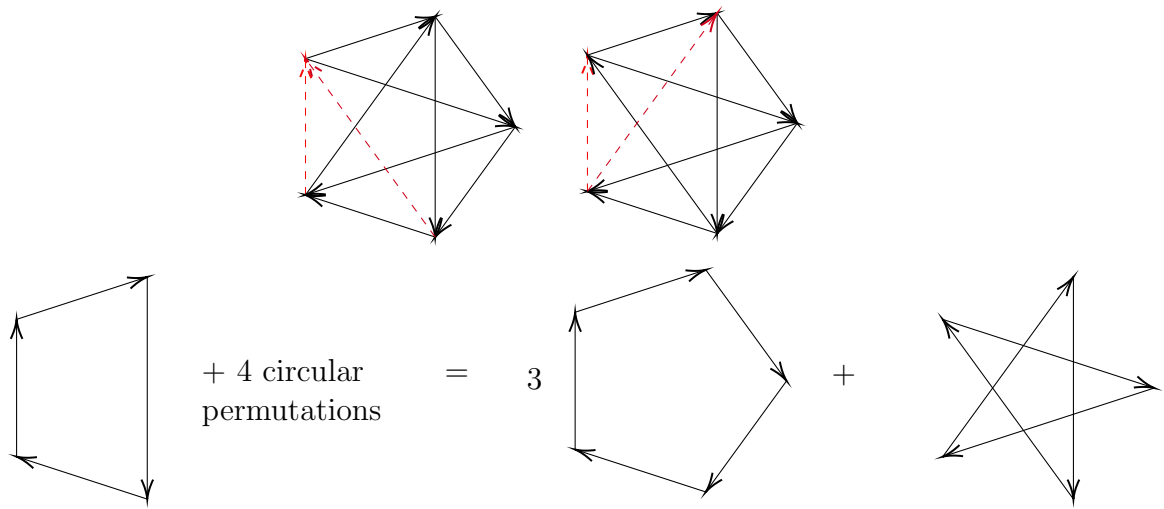
\begin{figure}
\begin{center}
\begin{tikzpicture}[x=0.75pt,y=0.75pt,yscale=-0.9,xscale=0.9]
\begin{scope}[shift={(0,0)}] 
\draw    (18.51,63) -- (89.84,39.82) ;
\draw [shift={(91.74,39.21)}, rotate = 522] [color={rgb, 255:red, 0; green, 0; blue, 0 }  ][line width=0.75]    (10.93,-3.29) .. controls (6.95,-1.4) and (3.31,-0.3) .. (0,0) .. controls (3.31,0.3) and (6.95,1.4) .. (10.93,3.29)   ;
\draw    (91.74,163.79) -- (20.41,140.62) ;
\draw [shift={(18.51,140)}, rotate = 378] [color={rgb, 255:red, 0; green, 0; blue, 0 }  ][line width=0.75]    (10.93,-3.29) .. controls (6.95,-1.4) and (3.31,-0.3) .. (0,0) .. controls (3.31,0.3) and (6.95,1.4) .. (10.93,3.29)   ;
\draw    (18.51,140) -- (18.51,65) ;
\draw [shift={(18.51,63)}, rotate = 450] [color={rgb, 255:red, 0; green, 0; blue, 0 }  ][line width=0.75]    (10.93,-3.29) .. controls (6.95,-1.4) and (3.31,-0.3) .. (0,0) .. controls (3.31,0.3) and (6.95,1.4) .. (10.93,3.29)   ;
\draw    (91.74,39.21) -- (91.74,161.79) ;
\draw [shift={(91.74,163.79)}, rotate = 270] [color={rgb, 255:red, 0; green, 0; blue, 0 }  ][line width=0.75]    (10.93,-3.29) .. controls (6.95,-1.4) and (3.31,-0.3) .. (0,0) .. controls (3.31,0.3) and (6.95,1.4) .. (10.93,3.29)   ;
\draw    (331.51,59) -- (402.84,35.82) ;
\draw [shift={(404.74,35.21)}, rotate = 522] [color={rgb, 255:red, 0; green, 0; blue, 0 }  ][line width=0.75]    (10.93,-3.29) .. controls (6.95,-1.4) and (3.31,-0.3) .. (0,0) .. controls (3.31,0.3) and (6.95,1.4) .. (10.93,3.29)   ;
\draw    (404.74,35.21) -- (448.82,95.88) ;
\draw [shift={(450,97.5)}, rotate = 234] [color={rgb, 255:red, 0; green, 0; blue, 0 }  ][line width=0.75]    (10.93,-3.29) .. controls (6.95,-1.4) and (3.31,-0.3) .. (0,0) .. controls (3.31,0.3) and (6.95,1.4) .. (10.93,3.29)   ;
\draw    (404.74,159.79) -- (333.41,136.62) ;
\draw [shift={(331.51,136)}, rotate = 378] [color={rgb, 255:red, 0; green, 0; blue, 0 }  ][line width=0.75]    (10.93,-3.29) .. controls (6.95,-1.4) and (3.31,-0.3) .. (0,0) .. controls (3.31,0.3) and (6.95,1.4) .. (10.93,3.29)   ;
\draw    (331.51,136) -- (331.51,61) ;
\draw [shift={(331.51,59)}, rotate = 450] [color={rgb, 255:red, 0; green, 0; blue, 0 }  ][line width=0.75]    (10.93,-3.29) .. controls (6.95,-1.4) and (3.31,-0.3) .. (0,0) .. controls (3.31,0.3) and (6.95,1.4) .. (10.93,3.29)   ;
\draw    (450,97.5) -- (405.92,158.18) ;
\draw [shift={(404.74,159.79)}, rotate = 306] [color={rgb, 255:red, 0; green, 0; blue, 0 }  ][line width=0.75]    (10.93,-3.29) .. controls (6.95,-1.4) and (3.31,-0.3) .. (0,0) .. controls (3.31,0.3) and (6.95,1.4) .. (10.93,3.29)   ;
\draw    (610.74,39.21) -- (610.74,161.79) ;
\draw [shift={(610.74,163.79)}, rotate = 270] [color={rgb, 255:red, 0; green, 0; blue, 0 }  ][line width=0.75]    (10.93,-3.29) .. controls (6.95,-1.4) and (3.31,-0.3) .. (0,0) .. controls (3.31,0.3) and (6.95,1.4) .. (10.93,3.29)   ;
\draw    (656,101.5) -- (539.41,139.38) ;
\draw [shift={(537.51,140)}, rotate = 342] [color={rgb, 255:red, 0; green, 0; blue, 0 }  ][line width=0.75]    (10.93,-3.29) .. controls (6.95,-1.4) and (3.31,-0.3) .. (0,0) .. controls (3.31,0.3) and (6.95,1.4) .. (10.93,3.29)   ;
\draw    (610.74,163.79) -- (538.68,64.62) ;
\draw [shift={(537.51,63)}, rotate = 414] [color={rgb, 255:red, 0; green, 0; blue, 0 }  ][line width=0.75]    (10.93,-3.29) .. controls (6.95,-1.4) and (3.31,-0.3) .. (0,0) .. controls (3.31,0.3) and (6.95,1.4) .. (10.93,3.29)   ;
\draw    (537.51,140) -- (609.57,40.82) ;
\draw [shift={(610.74,39.21)}, rotate = 486] [color={rgb, 255:red, 0; green, 0; blue, 0 }  ][line width=0.75]    (10.93,-3.29) .. controls (6.95,-1.4) and (3.31,-0.3) .. (0,0) .. controls (3.31,0.3) and (6.95,1.4) .. (10.93,3.29)   ;
\draw    (537.51,63) -- (654.1,100.88) ;
\draw [shift={(656,101.5)}, rotate = 198] [color={rgb, 255:red, 0; green, 0; blue, 0 }  ][line width=0.75]    (10.93,-3.29) .. controls (6.95,-1.4) and (3.31,-0.3) .. (0,0) .. controls (3.31,0.3) and (6.95,1.4) .. (10.93,3.29)   ;

\draw (121,84) node [anchor=north west][inner sep=0.75pt]   [align=left] {+ 4 circular \\permutations};
\draw (257,90) node [anchor=north west][inner sep=0.75pt]   [align=left] {=};
\draw (303,89) node [anchor=north west][inner sep=0.75pt]   [align=left] {3};
\draw (474,87) node [anchor=north west][inner sep=0.75pt]   [align=left] {+};
\end{scope}
\begin{scope}[shift={(0,-150)}] 

\draw    (180.41,66.79) -- (251.18,43.8) ;
\draw [shift={(253.09,43.18)}, rotate = 522] [color={rgb, 255:red, 0; green, 0; blue, 0 }  ][line width=0.75]    (10.93,-3.29) .. controls (6.95,-1.4) and (3.31,-0.3) .. (0,0) .. controls (3.31,0.3) and (6.95,1.4) .. (10.93,3.29)   ;
\draw    (253.09,43.18) -- (296.82,103.38) ;
\draw [shift={(298,105)}, rotate = 234] [color={rgb, 255:red, 0; green, 0; blue, 0 }  ][line width=0.75]    (10.93,-3.29) .. controls (6.95,-1.4) and (3.31,-0.3) .. (0,0) .. controls (3.31,0.3) and (6.95,1.4) .. (10.93,3.29)   ;
\draw    (298,105) -- (254.26,165.2) ;
\draw [shift={(253.09,166.82)}, rotate = 306] [color={rgb, 255:red, 0; green, 0; blue, 0 }  ][line width=0.75]    (10.93,-3.29) .. controls (6.95,-1.4) and (3.31,-0.3) .. (0,0) .. controls (3.31,0.3) and (6.95,1.4) .. (10.93,3.29)   ;
\draw    (253.09,166.82) -- (182.32,143.82) ;
\draw [shift={(180.41,143.21)}, rotate = 378] [color={rgb, 255:red, 0; green, 0; blue, 0 }  ][line width=0.75]    (10.93,-3.29) .. controls (6.95,-1.4) and (3.31,-0.3) .. (0,0) .. controls (3.31,0.3) and (6.95,1.4) .. (10.93,3.29)   ;
\draw [dashed, color={rgb, 255:red, 244; green, 12; blue, 12 }  ,draw opacity=1 ]   (180.41,143.21) -- (180.41,68.79) ;
\draw [dashed, shift={(180.41,66.79)}, rotate = 450] [color={rgb, 255:red, 244; green, 12; blue, 12 }  ,draw opacity=1 ][line width=0.75]    (10.93,-3.29) .. controls (6.95,-1.4) and (3.31,-0.3) .. (0,0) .. controls (3.31,0.3) and (6.95,1.4) .. (10.93,3.29)   ;
\draw    (180.41,66.79) -- (296.1,104.38) ;
\draw [shift={(298,105)}, rotate = 198] [color={rgb, 255:red, 0; green, 0; blue, 0 }  ][line width=0.75]    (10.93,-3.29) .. controls (6.95,-1.4) and (3.31,-0.3) .. (0,0) .. controls (3.31,0.3) and (6.95,1.4) .. (10.93,3.29)   ;
\draw    (253.09,43.18) -- (253.09,164.82) ;
\draw [shift={(253.09,166.82)}, rotate = 270] [color={rgb, 255:red, 0; green, 0; blue, 0 }  ][line width=0.75]    (10.93,-3.29) .. controls (6.95,-1.4) and (3.31,-0.3) .. (0,0) .. controls (3.31,0.3) and (6.95,1.4) .. (10.93,3.29)   ;
\draw    (298,105) -- (182.32,142.59) ;
\draw [shift={(180.41,143.21)}, rotate = 342] [color={rgb, 255:red, 0; green, 0; blue, 0 }  ][line width=0.75]    (10.93,-3.29) .. controls (6.95,-1.4) and (3.31,-0.3) .. (0,0) .. controls (3.31,0.3) and (6.95,1.4) .. (10.93,3.29)   ;
\draw [dashed, color={rgb, 255:red, 208; green, 2; blue, 27 }  ,draw opacity=1 ]   (253.09,166.82) -- (181.59,68.41) ;
\draw [dashed, shift={(180.41,66.79)}, rotate = 414] [color={rgb, 255:red, 208; green, 2; blue, 27 }  ,draw opacity=1 ][line width=0.75]    (10.93,-3.29) .. controls (6.95,-1.4) and (3.31,-0.3) .. (0,0) .. controls (3.31,0.3) and (6.95,1.4) .. (10.93,3.29)   ;
\draw    (180.41,143.21) -- (251.91,44.8) ;
\draw [shift={(253.09,43.18)}, rotate = 486] [color={rgb, 255:red, 0; green, 0; blue, 0 }  ][line width=0.75]    (10.93,-3.29) .. controls (6.95,-1.4) and (3.31,-0.3) .. (0,0) .. controls (3.31,0.3) and (6.95,1.4) .. (10.93,3.29)   ;
\draw    (338.41,64.79) -- (409.18,41.8) ;
\draw [shift={(411.09,41.18)}, rotate = 522] [color={rgb, 255:red, 0; green, 0; blue, 0 }  ][line width=0.75]    (10.93,-3.29) .. controls (6.95,-1.4) and (3.31,-0.3) .. (0,0) .. controls (3.31,0.3) and (6.95,1.4) .. (10.93,3.29)   ;
\draw    (411.09,41.18) -- (454.82,101.38) ;
\draw [shift={(456,103)}, rotate = 234] [color={rgb, 255:red, 0; green, 0; blue, 0 }  ][line width=0.75]    (10.93,-3.29) .. controls (6.95,-1.4) and (3.31,-0.3) .. (0,0) .. controls (3.31,0.3) and (6.95,1.4) .. (10.93,3.29)   ;
\draw    (456,103) -- (412.26,163.2) ;
\draw [shift={(411.09,164.82)}, rotate = 306] [color={rgb, 255:red, 0; green, 0; blue, 0 }  ][line width=0.75]    (10.93,-3.29) .. controls (6.95,-1.4) and (3.31,-0.3) .. (0,0) .. controls (3.31,0.3) and (6.95,1.4) .. (10.93,3.29)   ;
\draw    (411.09,164.82) -- (340.32,141.82) ;
\draw [shift={(338.41,141.21)}, rotate = 378] [color={rgb, 255:red, 0; green, 0; blue, 0 }  ][line width=0.75]    (10.93,-3.29) .. controls (6.95,-1.4) and (3.31,-0.3) .. (0,0) .. controls (3.31,0.3) and (6.95,1.4) .. (10.93,3.29)   ;
\draw [dashed, color={rgb, 255:red, 244; green, 12; blue, 12 }  ,draw opacity=1 ]   (338.41,141.21) -- (338.41,66.79) ;
\draw [dashed, shift={(338.41,64.79)}, rotate = 450] [color={rgb, 255:red, 244; green, 12; blue, 12 }  ,draw opacity=1 ][line width=0.75]    (10.93,-3.29) .. controls (6.95,-1.4) and (3.31,-0.3) .. (0,0) .. controls (3.31,0.3) and (6.95,1.4) .. (10.93,3.29)   ;
\draw    (338.41,64.79) -- (454.1,102.38) ;
\draw [shift={(456,103)}, rotate = 198] [color={rgb, 255:red, 0; green, 0; blue, 0 }  ][line width=0.75]    (10.93,-3.29) .. controls (6.95,-1.4) and (3.31,-0.3) .. (0,0) .. controls (3.31,0.3) and (6.95,1.4) .. (10.93,3.29)   ;
\draw    (411.09,41.18) -- (411.09,162.82) ;
\draw [shift={(411.09,164.82)}, rotate = 270] [color={rgb, 255:red, 0; green, 0; blue, 0 }  ][line width=0.75]    (10.93,-3.29) .. controls (6.95,-1.4) and (3.31,-0.3) .. (0,0) .. controls (3.31,0.3) and (6.95,1.4) .. (10.93,3.29)   ;
\draw    (456,103) -- (340.32,140.59) ;
\draw [shift={(338.41,141.21)}, rotate = 342] [color={rgb, 255:red, 0; green, 0; blue, 0 }  ][line width=0.75]    (10.93,-3.29) .. controls (6.95,-1.4) and (3.31,-0.3) .. (0,0) .. controls (3.31,0.3) and (6.95,1.4) .. (10.93,3.29)   ;
\draw [color={rgb, 255:red, 0; green, 0; blue, 0 }  ,draw opacity=1 ]   (411.09,164.82) -- (339.59,66.41) ;
\draw [shift={(338.41,64.79)}, rotate = 414] [color={rgb, 255:red, 0; green, 0; blue, 0 }  ,draw opacity=1 ][line width=0.75]    (10.93,-3.29) .. controls (6.95,-1.4) and (3.31,-0.3) .. (0,0) .. controls (3.31,0.3) and (6.95,1.4) .. (10.93,3.29)   ;
\draw [dashed, color={rgb, 255:red, 208; green, 2; blue, 27 }  ,draw opacity=1 ]   (338.41,141.21) -- (409.91,42.8) ;
\draw [shift={(411.09,41.18)}, rotate = 486] [color={rgb, 255:red, 208; green, 2; blue, 27 }  ,draw opacity=1 ][line width=0.75]    (10.93,-3.29) .. controls (6.95,-1.4) and (3.31,-0.3) .. (0,0) .. controls (3.31,0.3) and (6.95,1.4) .. (10.93,3.29)   ;
\end{scope}
\end{tikzpicture}
\end{center}
\caption{The pentagram quiver does not admit any cut, due to a relation between simple cycles. The maximal weak cuts are shown in red, up to cyclic permutation of the nodes. \label{figpenta}}
\end{figure}

\begin{figure}
\begin{center}
\begin{tikzpicture}[x=0.75pt,y=0.75pt,yscale=-0.8,xscale=0.8]
\begin{scope}
\draw    (48.75,151.76) -- (19,100.23) ;
\draw [shift={(18,98.5)}, rotate = 420] [color={rgb, 255:red, 0; green, 0; blue, 0 }  ][line width=0.75]    (10.93,-3.29) .. controls (6.95,-1.4) and (3.31,-0.3) .. (0,0) .. controls (3.31,0.3) and (6.95,1.4) .. (10.93,3.29)   ;
\draw    (18,98.5) -- (47.75,46.97) ;
\draw [shift={(48.75,45.24)}, rotate = 480] [color={rgb, 255:red, 0; green, 0; blue, 0 }  ][line width=0.75]    (10.93,-3.29) .. controls (6.95,-1.4) and (3.31,-0.3) .. (0,0) .. controls (3.31,0.3) and (6.95,1.4) .. (10.93,3.29)   ;
\draw    (48.75,45.24) -- (48.75,149.76) ;
\draw [shift={(48.75,151.76)}, rotate = 270] [color={rgb, 255:red, 0; green, 0; blue, 0 }  ][line width=0.75]    (10.93,-3.29) .. controls (6.95,-1.4) and (3.31,-0.3) .. (0,0) .. controls (3.31,0.3) and (6.95,1.4) .. (10.93,3.29)   ;
\draw    (101.75,40.24) -- (161.25,40.24) ;
\draw [shift={(163.25,40.24)}, rotate = 540] [color={rgb, 255:red, 0; green, 0; blue, 0 }  ][line width=0.75]    (10.93,-3.29) .. controls (6.95,-1.4) and (3.31,-0.3) .. (0,0) .. controls (3.31,0.3) and (6.95,1.4) .. (10.93,3.29)   ;
\draw    (163.25,40.24) -- (193,91.77) ;
\draw [shift={(194,93.5)}, rotate = 240] [color={rgb, 255:red, 0; green, 0; blue, 0 }  ][line width=0.75]    (10.93,-3.29) .. controls (6.95,-1.4) and (3.31,-0.3) .. (0,0) .. controls (3.31,0.3) and (6.95,1.4) .. (10.93,3.29)   ;
\draw    (194,93.5) -- (103.48,41.24) ;
\draw [shift={(101.75,40.24)}, rotate = 390] [color={rgb, 255:red, 0; green, 0; blue, 0 }  ][line width=0.75]    (10.93,-3.29) .. controls (6.95,-1.4) and (3.31,-0.3) .. (0,0) .. controls (3.31,0.3) and (6.95,1.4) .. (10.93,3.29)   ;
\draw    (313,104.5) -- (283.25,156.03) ;
\draw [shift={(282.25,157.76)}, rotate = 300] [color={rgb, 255:red, 0; green, 0; blue, 0 }  ][line width=0.75]    (10.93,-3.29) .. controls (6.95,-1.4) and (3.31,-0.3) .. (0,0) .. controls (3.31,0.3) and (6.95,1.4) .. (10.93,3.29)   ;
\draw    (282.25,157.76) -- (222.75,157.76) ;
\draw [shift={(220.75,157.76)}, rotate = 360] [color={rgb, 255:red, 0; green, 0; blue, 0 }  ][line width=0.75]    (10.93,-3.29) .. controls (6.95,-1.4) and (3.31,-0.3) .. (0,0) .. controls (3.31,0.3) and (6.95,1.4) .. (10.93,3.29)   ;
\draw    (220.75,157.76) -- (311.27,105.5) ;
\draw [shift={(313,104.5)}, rotate = 510] [color={rgb, 255:red, 0; green, 0; blue, 0 }  ][line width=0.75]    (10.93,-3.29) .. controls (6.95,-1.4) and (3.31,-0.3) .. (0,0) .. controls (3.31,0.3) and (6.95,1.4) .. (10.93,3.29)   ;
\draw    (434.75,42.24) -- (494.25,42.24) ;
\draw [shift={(496.25,42.24)}, rotate = 540] [color={rgb, 255:red, 0; green, 0; blue, 0 }  ][line width=0.75]    (10.93,-3.29) .. controls (6.95,-1.4) and (3.31,-0.3) .. (0,0) .. controls (3.31,0.3) and (6.95,1.4) .. (10.93,3.29)   ;
\draw    (496.25,42.24) -- (526,93.77) ;
\draw [shift={(527,95.5)}, rotate = 240] [color={rgb, 255:red, 0; green, 0; blue, 0 }  ][line width=0.75]    (10.93,-3.29) .. controls (6.95,-1.4) and (3.31,-0.3) .. (0,0) .. controls (3.31,0.3) and (6.95,1.4) .. (10.93,3.29)   ;
\draw    (527,95.5) -- (497.25,147.03) ;
\draw [shift={(496.25,148.76)}, rotate = 300] [color={rgb, 255:red, 0; green, 0; blue, 0 }  ][line width=0.75]    (10.93,-3.29) .. controls (6.95,-1.4) and (3.31,-0.3) .. (0,0) .. controls (3.31,0.3) and (6.95,1.4) .. (10.93,3.29)   ;
\draw    (496.25,148.76) -- (436.75,148.76) ;
\draw [shift={(434.75,148.76)}, rotate = 360] [color={rgb, 255:red, 0; green, 0; blue, 0 }  ][line width=0.75]    (10.93,-3.29) .. controls (6.95,-1.4) and (3.31,-0.3) .. (0,0) .. controls (3.31,0.3) and (6.95,1.4) .. (10.93,3.29)   ;
\draw    (434.75,148.76) -- (405,97.23) ;
\draw [shift={(404,95.5)}, rotate = 420] [color={rgb, 255:red, 0; green, 0; blue, 0 }  ][line width=0.75]    (10.93,-3.29) .. controls (6.95,-1.4) and (3.31,-0.3) .. (0,0) .. controls (3.31,0.3) and (6.95,1.4) .. (10.93,3.29)   ;
\draw    (404,95.5) -- (433.75,43.97) ;
\draw [shift={(434.75,42.24)}, rotate = 480] [color={rgb, 255:red, 0; green, 0; blue, 0 }  ][line width=0.75]    (10.93,-3.29) .. controls (6.95,-1.4) and (3.31,-0.3) .. (0,0) .. controls (3.31,0.3) and (6.95,1.4) .. (10.93,3.29)   ;
\draw    (684,90.5) -- (593.48,38.24) ;
\draw [shift={(591.75,37.24)}, rotate = 390] [color={rgb, 255:red, 0; green, 0; blue, 0 }  ][line width=0.75]    (10.93,-3.29) .. controls (6.95,-1.4) and (3.31,-0.3) .. (0,0) .. controls (3.31,0.3) and (6.95,1.4) .. (10.93,3.29)   ;
\draw    (591.75,37.24) -- (591.75,141.76) ;
\draw [shift={(591.75,143.76)}, rotate = 270] [color={rgb, 255:red, 0; green, 0; blue, 0 }  ][line width=0.75]    (10.93,-3.29) .. controls (6.95,-1.4) and (3.31,-0.3) .. (0,0) .. controls (3.31,0.3) and (6.95,1.4) .. (10.93,3.29)   ;
\draw    (591.75,143.76) -- (682.27,91.5) ;
\draw [shift={(684,90.5)}, rotate = 510] [color={rgb, 255:red, 0; green, 0; blue, 0 }  ][line width=0.75]    (10.93,-3.29) .. controls (6.95,-1.4) and (3.31,-0.3) .. (0,0) .. controls (3.31,0.3) and (6.95,1.4) .. (10.93,3.29)   ;
\draw (218,89) node [anchor=north west][inner sep=0.75pt]   [align=left] {+};
\draw (558,85) node [anchor=north west][inner sep=0.75pt]   [align=left] {+};
\draw (82,88) node [anchor=north west][inner sep=0.75pt]   [align=left] {+};
\draw (360,86) node [anchor=north west][inner sep=0.75pt]   [align=left] {=};
\end{scope}
\begin{scope}[shift={(0,-150)}] 
\draw    (374.75,47.24) -- (434.25,47.24) ;
\draw [shift={(436.25,47.24)}, rotate = 540] [color={rgb, 255:red, 0; green, 0; blue, 0 }  ][line width=0.75]    (10.93,-3.29) .. controls (6.95,-1.4) and (3.31,-0.3) .. (0,0) .. controls (3.31,0.3) and (6.95,1.4) .. (10.93,3.29)   ;
\draw    (436.25,47.24) -- (466,98.77) ;
\draw [shift={(467,100.5)}, rotate = 240] [color={rgb, 255:red, 0; green, 0; blue, 0 }  ][line width=0.75]    (10.93,-3.29) .. controls (6.95,-1.4) and (3.31,-0.3) .. (0,0) .. controls (3.31,0.3) and (6.95,1.4) .. (10.93,3.29)   ;
\draw    (467,100.5) -- (437.25,152.03) ;
\draw [shift={(436.25,153.76)}, rotate = 300] [color={rgb, 255:red, 0; green, 0; blue, 0 }  ][line width=0.75]    (10.93,-3.29) .. controls (6.95,-1.4) and (3.31,-0.3) .. (0,0) .. controls (3.31,0.3) and (6.95,1.4) .. (10.93,3.29)   ;
\draw    (436.25,153.76) -- (376.75,153.76) ;
\draw [shift={(374.75,153.76)}, rotate = 360] [color={rgb, 255:red, 0; green, 0; blue, 0 }  ][line width=0.75]    (10.93,-3.29) .. controls (6.95,-1.4) and (3.31,-0.3) .. (0,0) .. controls (3.31,0.3) and (6.95,1.4) .. (10.93,3.29)   ;
\draw    (374.75,153.76) -- (345,102.23) ;
\draw [shift={(344,100.5)}, rotate = 420] [color={rgb, 255:red, 0; green, 0; blue, 0 }  ][line width=0.75]    (10.93,-3.29) .. controls (6.95,-1.4) and (3.31,-0.3) .. (0,0) .. controls (3.31,0.3) and (6.95,1.4) .. (10.93,3.29)   ;
\draw [dashed, color={rgb, 255:red, 241; green, 10; blue, 10 }  ,draw opacity=1 ]   (344,100.5) -- (373.75,48.97) ;
\draw [shift={(374.75,47.24)}, rotate = 480] [color={rgb, 255:red, 241; green, 10; blue, 10 }  ,draw opacity=1 ][line width=0.75]    (10.93,-3.29) .. controls (6.95,-1.4) and (3.31,-0.3) .. (0,0) .. controls (3.31,0.3) and (6.95,1.4) .. (10.93,3.29)   ;
\draw    (467,100.5) -- (376.48,48.24) ;
\draw [shift={(374.75,47.24)}, rotate = 390] [color={rgb, 255:red, 0; green, 0; blue, 0 }  ][line width=0.75]    (10.93,-3.29) .. controls (6.95,-1.4) and (3.31,-0.3) .. (0,0) .. controls (3.31,0.3) and (6.95,1.4) .. (10.93,3.29)   ;
\draw    (374.75,47.24) -- (374.75,151.76) ;
\draw [shift={(374.75,153.76)}, rotate = 270] [color={rgb, 255:red, 0; green, 0; blue, 0 }  ][line width=0.75]    (10.93,-3.29) .. controls (6.95,-1.4) and (3.31,-0.3) .. (0,0) .. controls (3.31,0.3) and (6.95,1.4) .. (10.93,3.29)   ;
\draw [dashed, color={rgb, 255:red, 247; green, 11; blue, 11 }  ,draw opacity=1 ]   (374.75,153.76) -- (465.27,101.5) ;
\draw [shift={(467,100.5)}, rotate = 510] [color={rgb, 255:red, 247; green, 11; blue, 11 }  ,draw opacity=1 ][line width=0.75]    (10.93,-3.29) .. controls (6.95,-1.4) and (3.31,-0.3) .. (0,0) .. controls (3.31,0.3) and (6.95,1.4) .. (10.93,3.29)   ;
\draw    (526.75,47.24) -- (586.25,47.24) ;
\draw [shift={(588.25,47.24)}, rotate = 540] [color={rgb, 255:red, 0; green, 0; blue, 0 }  ][line width=0.75]    (10.93,-3.29) .. controls (6.95,-1.4) and (3.31,-0.3) .. (0,0) .. controls (3.31,0.3) and (6.95,1.4) .. (10.93,3.29)   ;
\draw    (588.25,47.24) -- (618,98.77) ;
\draw [shift={(619,100.5)}, rotate = 240] [color={rgb, 255:red, 0; green, 0; blue, 0 }  ][line width=0.75]    (10.93,-3.29) .. controls (6.95,-1.4) and (3.31,-0.3) .. (0,0) .. controls (3.31,0.3) and (6.95,1.4) .. (10.93,3.29)   ;
\draw    (619,100.5) -- (589.25,152.03) ;
\draw [shift={(588.25,153.76)}, rotate = 300] [color={rgb, 255:red, 0; green, 0; blue, 0 }  ][line width=0.75]    (10.93,-3.29) .. controls (6.95,-1.4) and (3.31,-0.3) .. (0,0) .. controls (3.31,0.3) and (6.95,1.4) .. (10.93,3.29)   ;
\draw    (588.25,153.76) -- (528.75,153.76) ;
\draw [shift={(526.75,153.76)}, rotate = 360] [color={rgb, 255:red, 0; green, 0; blue, 0 }  ][line width=0.75]    (10.93,-3.29) .. controls (6.95,-1.4) and (3.31,-0.3) .. (0,0) .. controls (3.31,0.3) and (6.95,1.4) .. (10.93,3.29)   ;
\draw    (526.75,153.76) -- (497,102.23) ;
\draw [shift={(496,100.5)}, rotate = 420] [color={rgb, 255:red, 0; green, 0; blue, 0 }  ][line width=0.75]    (10.93,-3.29) .. controls (6.95,-1.4) and (3.31,-0.3) .. (0,0) .. controls (3.31,0.3) and (6.95,1.4) .. (10.93,3.29)   ;
\draw [dashed, color={rgb, 255:red, 241; green, 10; blue, 10 }  ,draw opacity=1 ]   (496,100.5) -- (525.75,48.97) ;
\draw [shift={(526.75,47.24)}, rotate = 480] [color={rgb, 255:red, 241; green, 10; blue, 10 }  ,draw opacity=1 ][line width=0.75]    (10.93,-3.29) .. controls (6.95,-1.4) and (3.31,-0.3) .. (0,0) .. controls (3.31,0.3) and (6.95,1.4) .. (10.93,3.29)   ;
\draw [dashed, color={rgb, 255:red, 208; green, 2; blue, 27 }  ,draw opacity=1 ]   (619,100.5) -- (528.48,48.24) ;
\draw [shift={(526.75,47.24)}, rotate = 390] [color={rgb, 255:red, 208; green, 2; blue, 27 }  ,draw opacity=1 ][line width=0.75]    (10.93,-3.29) .. controls (6.95,-1.4) and (3.31,-0.3) .. (0,0) .. controls (3.31,0.3) and (6.95,1.4) .. (10.93,3.29)   ;
\draw    (526.75,47.24) -- (526.75,151.76) ;
\draw [shift={(526.75,153.76)}, rotate = 270] [color={rgb, 255:red, 0; green, 0; blue, 0 }  ][line width=0.75]    (10.93,-3.29) .. controls (6.95,-1.4) and (3.31,-0.3) .. (0,0) .. controls (3.31,0.3) and (6.95,1.4) .. (10.93,3.29)   ;
\draw [color={rgb, 255:red, 0; green, 0; blue, 0 }  ,draw opacity=1 ]   (526.75,153.76) -- (617.27,101.5) ;
\draw [shift={(619,100.5)}, rotate = 510] [color={rgb, 255:red, 0; green, 0; blue, 0 }  ,draw opacity=1 ][line width=0.75]    (10.93,-3.29) .. controls (6.95,-1.4) and (3.31,-0.3) .. (0,0) .. controls (3.31,0.3) and (6.95,1.4) .. (10.93,3.29)   ;
\draw    (73.75,51.24) -- (133.25,51.24) ;
\draw [shift={(135.25,51.24)}, rotate = 540] [color={rgb, 255:red, 0; green, 0; blue, 0 }  ][line width=0.75]    (10.93,-3.29) .. controls (6.95,-1.4) and (3.31,-0.3) .. (0,0) .. controls (3.31,0.3) and (6.95,1.4) .. (10.93,3.29)   ;
\draw    (135.25,51.24) -- (165,102.77) ;
\draw [shift={(166,104.5)}, rotate = 240] [color={rgb, 255:red, 0; green, 0; blue, 0 }  ][line width=0.75]    (10.93,-3.29) .. controls (6.95,-1.4) and (3.31,-0.3) .. (0,0) .. controls (3.31,0.3) and (6.95,1.4) .. (10.93,3.29)   ;
\draw    (166,104.5) -- (136.25,156.03) ;
\draw [shift={(135.25,157.76)}, rotate = 300] [color={rgb, 255:red, 0; green, 0; blue, 0 }  ][line width=0.75]    (10.93,-3.29) .. controls (6.95,-1.4) and (3.31,-0.3) .. (0,0) .. controls (3.31,0.3) and (6.95,1.4) .. (10.93,3.29)   ;
\draw    (135.25,157.76) -- (75.75,157.76) ;
\draw [shift={(73.75,157.76)}, rotate = 360] [color={rgb, 255:red, 0; green, 0; blue, 0 }  ][line width=0.75]    (10.93,-3.29) .. controls (6.95,-1.4) and (3.31,-0.3) .. (0,0) .. controls (3.31,0.3) and (6.95,1.4) .. (10.93,3.29)   ;
\draw [dashed, color={rgb, 255:red, 208; green, 2; blue, 27 }  ,draw opacity=1 ]   (73.75,157.76) -- (44,106.23) ;
\draw [shift={(43,104.5)}, rotate = 420] [color={rgb, 255:red, 208; green, 2; blue, 27 }  ,draw opacity=1 ][line width=0.75]    (10.93,-3.29) .. controls (6.95,-1.4) and (3.31,-0.3) .. (0,0) .. controls (3.31,0.3) and (6.95,1.4) .. (10.93,3.29)   ;
\draw [color={rgb, 255:red, 0; green, 0; blue, 0 }  ,draw opacity=1 ]   (43,104.5) -- (72.75,52.97) ;
\draw [shift={(73.75,51.24)}, rotate = 480] [color={rgb, 255:red, 0; green, 0; blue, 0 }  ,draw opacity=1 ][line width=0.75]    (10.93,-3.29) .. controls (6.95,-1.4) and (3.31,-0.3) .. (0,0) .. controls (3.31,0.3) and (6.95,1.4) .. (10.93,3.29)   ;
\draw    (166,104.5) -- (75.48,52.24) ;
\draw [shift={(73.75,51.24)}, rotate = 390] [color={rgb, 255:red, 0; green, 0; blue, 0 }  ][line width=0.75]    (10.93,-3.29) .. controls (6.95,-1.4) and (3.31,-0.3) .. (0,0) .. controls (3.31,0.3) and (6.95,1.4) .. (10.93,3.29)   ;
\draw    (73.75,51.24) -- (73.75,155.76) ;
\draw [shift={(73.75,157.76)}, rotate = 270] [color={rgb, 255:red, 0; green, 0; blue, 0 }  ][line width=0.75]    (10.93,-3.29) .. controls (6.95,-1.4) and (3.31,-0.3) .. (0,0) .. controls (3.31,0.3) and (6.95,1.4) .. (10.93,3.29)   ;
\draw [dashed, color={rgb, 255:red, 247; green, 11; blue, 11 }  ,draw opacity=1 ]   (73.75,157.76) -- (164.27,105.5) ;
\draw [shift={(166,104.5)}, rotate = 510] [color={rgb, 255:red, 247; green, 11; blue, 11 }  ,draw opacity=1 ][line width=0.75]    (10.93,-3.29) .. controls (6.95,-1.4) and (3.31,-0.3) .. (0,0) .. controls (3.31,0.3) and (6.95,1.4) .. (10.93,3.29)   ;
\draw    (221.75,48.24) -- (281.25,48.24) ;
\draw [shift={(283.25,48.24)}, rotate = 540] [color={rgb, 255:red, 0; green, 0; blue, 0 }  ][line width=0.75]    (10.93,-3.29) .. controls (6.95,-1.4) and (3.31,-0.3) .. (0,0) .. controls (3.31,0.3) and (6.95,1.4) .. (10.93,3.29)   ;
\draw    (283.25,48.24) -- (313,99.77) ;
\draw [shift={(314,101.5)}, rotate = 240] [color={rgb, 255:red, 0; green, 0; blue, 0 }  ][line width=0.75]    (10.93,-3.29) .. controls (6.95,-1.4) and (3.31,-0.3) .. (0,0) .. controls (3.31,0.3) and (6.95,1.4) .. (10.93,3.29)   ;
\draw    (314,101.5) -- (284.25,153.03) ;
\draw [shift={(283.25,154.76)}, rotate = 300] [color={rgb, 255:red, 0; green, 0; blue, 0 }  ][line width=0.75]    (10.93,-3.29) .. controls (6.95,-1.4) and (3.31,-0.3) .. (0,0) .. controls (3.31,0.3) and (6.95,1.4) .. (10.93,3.29)   ;
\draw    (283.25,154.76) -- (223.75,154.76) ;
\draw [shift={(221.75,154.76)}, rotate = 360] [color={rgb, 255:red, 0; green, 0; blue, 0 }  ][line width=0.75]    (10.93,-3.29) .. controls (6.95,-1.4) and (3.31,-0.3) .. (0,0) .. controls (3.31,0.3) and (6.95,1.4) .. (10.93,3.29)   ;
\draw [dashed, color={rgb, 255:red, 208; green, 2; blue, 27 }  ,draw opacity=1 ]   (221.75,154.76) -- (192,103.23) ;
\draw [shift={(191,101.5)}, rotate = 420] [color={rgb, 255:red, 208; green, 2; blue, 27 }  ,draw opacity=1 ][line width=0.75]    (10.93,-3.29) .. controls (6.95,-1.4) and (3.31,-0.3) .. (0,0) .. controls (3.31,0.3) and (6.95,1.4) .. (10.93,3.29)   ;
\draw [color={rgb, 255:red, 0; green, 0; blue, 0 }  ,draw opacity=1 ]   (191,101.5) -- (220.75,49.97) ;
\draw [shift={(221.75,48.24)}, rotate = 480] [color={rgb, 255:red, 0; green, 0; blue, 0 }  ,draw opacity=1 ][line width=0.75]    (10.93,-3.29) .. controls (6.95,-1.4) and (3.31,-0.3) .. (0,0) .. controls (3.31,0.3) and (6.95,1.4) .. (10.93,3.29)   ;
\draw [dashed, color={rgb, 255:red, 208; green, 2; blue, 27 }  ,draw opacity=1 ]   (314,101.5) -- (223.48,49.24) ;
\draw [shift={(221.75,48.24)}, rotate = 390] [color={rgb, 255:red, 208; green, 2; blue, 27 }  ,draw opacity=1 ][line width=0.75]    (10.93,-3.29) .. controls (6.95,-1.4) and (3.31,-0.3) .. (0,0) .. controls (3.31,0.3) and (6.95,1.4) .. (10.93,3.29)   ;
\draw    (221.75,48.24) -- (221.75,152.76) ;
\draw [shift={(221.75,154.76)}, rotate = 270] [color={rgb, 255:red, 0; green, 0; blue, 0 }  ][line width=0.75]    (10.93,-3.29) .. controls (6.95,-1.4) and (3.31,-0.3) .. (0,0) .. controls (3.31,0.3) and (6.95,1.4) .. (10.93,3.29)   ;
\draw [color={rgb, 255:red, 0; green, 0; blue, 0 }  ,draw opacity=1 ]   (221.75,154.76) -- (312.27,102.5) ;
\draw [shift={(314,101.5)}, rotate = 510] [color={rgb, 255:red, 0; green, 0; blue, 0 }  ,draw opacity=1 ][line width=0.75]    (10.93,-3.29) .. controls (6.95,-1.4) and (3.31,-0.3) .. (0,0) .. controls (3.31,0.3) and (6.95,1.4) .. (10.93,3.29)   ;
\end{scope}
\end{tikzpicture}
\end{center}
\caption{The hexagram quiver does not admit any cut, due to a relation between simple cycles. The maximal weak cuts are shown in red, up to cyclic permutation of the nodes. \label{fighexa}}
\end{figure}
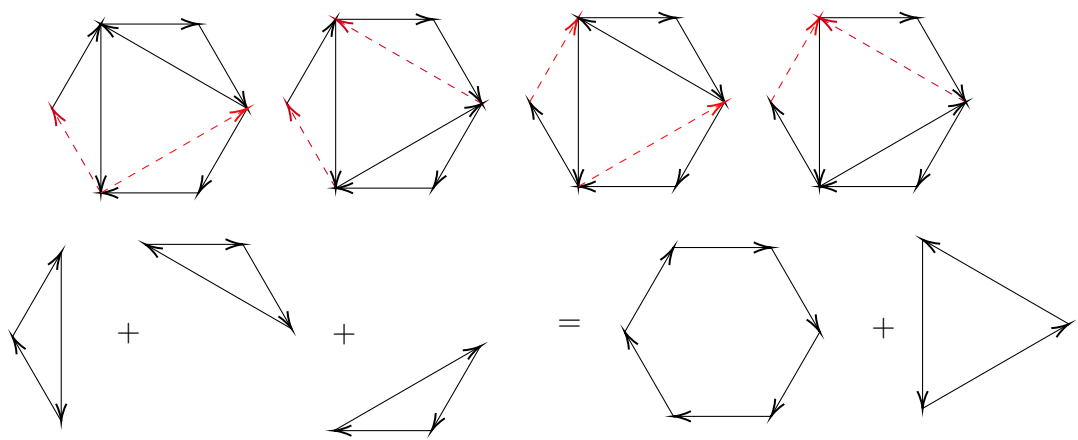

While quivers with few arrows typically have cuts (see Fig. \ref{Fig1}), it is easy to 
find examples of quivers which do not have any cuts. Our main examples are the 'pentagram' and 'hexagram' quivers, shown in Figure
\ref{figpenta} and \ref{fighexa}. In both cases, the 
relations between the cycles forbids the existence of an R-charge,
and therefore of a cut (see Figures \ref{figpenta} and \ref{fighexa}). Each of these quivers nonetheless admits maximal weak cuts, as shown in dashed red in the respective figure.

\begin{remark}
The notion of R-charge is standard in the physics literature on quiver gauge theories, but the details may vary. Here we do not impose any condition on the total R-charge of the arrows ending or starting at a given node. The notion of cut is also standard in the mathematical and physic literature about quivers with potential. The notions of 
weak and strong cut introduced here appear to be new.  R-charges and cuts are usually defined with respect to all cycles appearing in the potential, not only the simple cycles. Here we always assume that the potential is a generic sum of simple cycles, so this distinction is unessential. 
\end{remark}

\begin{figure}
\centering
\begin{tikzpicture}[x=0.75pt,y=0.75pt,yscale=-1,xscale=1]

\draw[fill] (62,54) circle (3pt);
\draw    (62,54) -- (158,54) ;
\draw [shift={(160,54)}, rotate = 180] [color={rgb, 255:red, 0; green, 0; blue, 0 }  ][line width=0.75]    (10.93,-3.29) .. controls (6.95,-1.4) and (3.31,-0.3) .. (0,0) .. controls (3.31,0.3) and (6.95,1.4) .. (10.93,3.29)   ;
\draw    (160,54) -- (160,150) ;
\draw [shift={(160,152)}, rotate = 270] [color={rgb, 255:red, 0; green, 0; blue, 0 }  ][line width=0.75]    (10.93,-3.29) .. controls (6.95,-1.4) and (3.31,-0.3) .. (0,0) .. controls (3.31,0.3) and (6.95,1.4) .. (10.93,3.29)   ;
\draw    (160,152) -- (64,152) ;
\draw [shift={(62,152)}, rotate = 360] [color={rgb, 255:red, 0; green, 0; blue, 0 }  ][line width=0.75]    (10.93,-3.29) .. controls (6.95,-1.4) and (3.31,-0.3) .. (0,0) .. controls (3.31,0.3) and (6.95,1.4) .. (10.93,3.29)   ;
\draw [dashed, color={rgb, 255:red, 208; green, 2; blue, 27 }  ,draw opacity=1 ]   (62,152) -- (62,56) ;
\draw [shift={(62,54)}, rotate = 450] [color={rgb, 255:red, 208; green, 2; blue, 27 }  ,draw opacity=1 ][line width=0.75]    (10.93,-3.29) .. controls (6.95,-1.4) and (3.31,-0.3) .. (0,0) .. controls (3.31,0.3) and (6.95,1.4) .. (10.93,3.29)   ;
\draw    (62,54) -- (158.59,150.59) ;
\draw [shift={(160,152)}, rotate = 225] [color={rgb, 255:red, 0; green, 0; blue, 0 }  ][line width=0.75]    (10.93,-3.29) .. controls (6.95,-1.4) and (3.31,-0.3) .. (0,0) .. controls (3.31,0.3) and (6.95,1.4) .. (10.93,3.29)   ;
\draw    (160,54) -- (63.41,150.59) ;
\draw [shift={(62,152)}, rotate = 315] [color={rgb, 255:red, 0; green, 0; blue, 0 }  ][line width=0.75]    (10.93,-3.29) .. controls (6.95,-1.4) and (3.31,-0.3) .. (0,0) .. controls (3.31,0.3) and (6.95,1.4) .. (10.93,3.29)   ;
\draw [dashed, color={rgb, 255:red, 208; green, 2; blue, 27 }  ,draw opacity=1 ]   (212,53) -- (308,53) ;
\draw[fill] (310,53) circle (3pt);
\draw [shift={(310,53)}, rotate = 180] [color={rgb, 255:red, 208; green, 2; blue, 27 }  ,draw opacity=1 ][line width=0.75]    (10.93,-3.29) .. controls (6.95,-1.4) and (3.31,-0.3) .. (0,0) .. controls (3.31,0.3) and (6.95,1.4) .. (10.93,3.29)   ;
\draw    (310,53) -- (310,149) ;
\draw [shift={(310,151)}, rotate = 270] [color={rgb, 255:red, 0; green, 0; blue, 0 }  ][line width=0.75]    (10.93,-3.29) .. controls (6.95,-1.4) and (3.31,-0.3) .. (0,0) .. controls (3.31,0.3) and (6.95,1.4) .. (10.93,3.29)   ;
\draw    (310,151) -- (214,151) ;
\draw [shift={(212,151)}, rotate = 360] [color={rgb, 255:red, 0; green, 0; blue, 0 }  ][line width=0.75]    (10.93,-3.29) .. controls (6.95,-1.4) and (3.31,-0.3) .. (0,0) .. controls (3.31,0.3) and (6.95,1.4) .. (10.93,3.29)   ;
\draw [color={rgb, 255:red, 0; green, 0; blue, 0 }  ,draw opacity=1 ]   (212,151) -- (212,55) ;
\draw [shift={(212,53)}, rotate = 450] [color={rgb, 255:red, 0; green, 0; blue, 0 }  ,draw opacity=1 ][line width=0.75]    (10.93,-3.29) .. controls (6.95,-1.4) and (3.31,-0.3) .. (0,0) .. controls (3.31,0.3) and (6.95,1.4) .. (10.93,3.29)   ;
\draw [dashed, color={rgb, 255:red, 208; green, 2; blue, 27 }  ,draw opacity=1 ]   (212,53) -- (308.59,149.59) ;
\draw [shift={(310,151)}, rotate = 225] [color={rgb, 255:red, 208; green, 2; blue, 27 }  ,draw opacity=1 ][line width=0.75]    (10.93,-3.29) .. controls (6.95,-1.4) and (3.31,-0.3) .. (0,0) .. controls (3.31,0.3) and (6.95,1.4) .. (10.93,3.29)   ;
\draw    (310,53) -- (213.41,149.59) ;
\draw [shift={(212,151)}, rotate = 315] [color={rgb, 255:red, 0; green, 0; blue, 0 }  ][line width=0.75]    (10.93,-3.29) .. controls (6.95,-1.4) and (3.31,-0.3) .. (0,0) .. controls (3.31,0.3) and (6.95,1.4) .. (10.93,3.29)   ;
\draw    (372,54) -- (468,54) ;
\draw[fill] (470,154) circle (3pt);
\draw [shift={(470,54)}, rotate = 180] [color={rgb, 255:red, 0; green, 0; blue, 0 }  ][line width=0.75]    (10.93,-3.29) .. controls (6.95,-1.4) and (3.31,-0.3) .. (0,0) .. controls (3.31,0.3) and (6.95,1.4) .. (10.93,3.29)   ;
\draw [dashed, color={rgb, 255:red, 208; green, 2; blue, 27 }  ,draw opacity=1 ]   (470,54) -- (470,150) ;
\draw [shift={(470,152)}, rotate = 270] [color={rgb, 255:red, 208; green, 2; blue, 27 }  ,draw opacity=1 ][line width=0.75]    (10.93,-3.29) .. controls (6.95,-1.4) and (3.31,-0.3) .. (0,0) .. controls (3.31,0.3) and (6.95,1.4) .. (10.93,3.29)   ;
\draw    (470,152) -- (374,152) ;
\draw [shift={(372,152)}, rotate = 360] [color={rgb, 255:red, 0; green, 0; blue, 0 }  ][line width=0.75]    (10.93,-3.29) .. controls (6.95,-1.4) and (3.31,-0.3) .. (0,0) .. controls (3.31,0.3) and (6.95,1.4) .. (10.93,3.29)   ;
\draw [color={rgb, 255:red, 0; green, 0; blue, 0 }  ,draw opacity=1 ]   (372,152) -- (372,56) ;
\draw [shift={(372,54)}, rotate = 450][color={rgb, 255:red, 0; green, 0; blue, 0 }  ] [line width=0.75]    (10.93,-3.29) .. controls (6.95,-1.4) and (3.31,-0.3) .. (0,0) .. controls (3.31,0.3) and (6.95,1.4) .. (10.93,3.29)   ;
\draw [dashed, color={rgb, 255:red, 208; green, 2; blue, 27 }  ,draw opacity=1 ]  (372,54) -- (468.59,150.59) ;
\draw [shift={(470,152)}, rotate = 225] [color={rgb, 255:red, 208; green, 2; blue, 27 }  ,draw opacity=1 ][line width=0.75]    (10.93,-3.29) .. controls (6.95,-1.4) and (3.31,-0.3) .. (0,0) .. controls (3.31,0.3) and (6.95,1.4) .. (10.93,3.29)   ;
\draw [dashed, color={rgb, 255:red, 208; green, 2; blue, 27 }  ,draw opacity=1 ]   (470,54) -- (373.41,150.59) ;
\draw [shift={(372,152)}, rotate = 315] [color={rgb, 255:red, 208; green, 2; blue, 27 }  ,draw opacity=1 ][line width=0.75]    (10.93,-3.29) .. controls (6.95,-1.4) and (3.31,-0.3) .. (0,0) .. controls (3.31,0.3) and (6.95,1.4) .. (10.93,3.29)   ;
\draw    (534,52) -- (630,52) ;
\draw[fill] (532,152) circle (3pt);
\draw [shift={(632,52)}, rotate = 180] [color={rgb, 255:red, 0; green, 0; blue, 0 }  ][line width=0.75]    (10.93,-3.29) .. controls (6.95,-1.4) and (3.31,-0.3) .. (0,0) .. controls (3.31,0.3) and (6.95,1.4) .. (10.93,3.29)   ;
\draw    (632,52) -- (632,148) ;
\draw [shift={(632,150)}, rotate = 270] [color={rgb, 255:red, 0; green, 0; blue, 0 }  ][line width=0.75]    (10.93,-3.29) .. controls (6.95,-1.4) and (3.31,-0.3) .. (0,0) .. controls (3.31,0.3) and (6.95,1.4) .. (10.93,3.29)   ;
\draw [dashed, color={rgb, 255:red, 208; green, 2; blue, 27 }  ,draw opacity=1 ]   (632,150) -- (536,150) ;
\draw [shift={(534,150)}, rotate = 360] [color={rgb, 255:red, 208; green, 2; blue, 27 }  ,draw opacity=1 ][line width=0.75]    (10.93,-3.29) .. controls (6.95,-1.4) and (3.31,-0.3) .. (0,0) .. controls (3.31,0.3) and (6.95,1.4) .. (10.93,3.29)   ;
\draw [color={rgb, 255:red, 0; green, 0; blue, 0 }  ,draw opacity=1 ]   (534,150) -- (534,54) ;
\draw [shift={(534,52)}, rotate = 450][line width=0.75]    (10.93,-3.29) .. controls (6.95,-1.4) and (3.31,-0.3) .. (0,0) .. controls (3.31,0.3) and (6.95,1.4) .. (10.93,3.29)   ;
\draw    (534,52) -- (630.59,148.59) ;
\draw [shift={(632,150)}, rotate = 225] [line width=0.75]    (10.93,-3.29) .. controls (6.95,-1.4) and (3.31,-0.3) .. (0,0) .. controls (3.31,0.3) and (6.95,1.4) .. (10.93,3.29)   ;
\draw [dashed, color={rgb, 255:red, 208; green, 2; blue, 27 }  ,draw opacity=1 ]   (632,52) -- (535.41,148.59) ;
\draw [shift={(534,150)}, rotate = 315] [color={rgb, 255:red, 208; green, 2; blue, 27 }  ,draw opacity=1 ][line width=0.75]    (10.93,-3.29) .. controls (6.95,-1.4) and (3.31,-0.3) .. (0,0) .. controls (3.31,0.3) and (6.95,1.4) .. (10.93,3.29)   ;

\draw (49,36) node [anchor=north west][inner sep=0.75pt]   [align=left] {1};
\draw (44,152) node [anchor=north west][inner sep=0.75pt]   [align=left] {4};
\draw (162,155) node [anchor=north west][inner sep=0.75pt]   [align=left] {3};
\draw (163,36) node [anchor=north west][inner sep=0.75pt]   [align=left] {2};
\draw (199,35) node [anchor=north west][inner sep=0.75pt]   [align=left] {1};
\draw (194,151) node [anchor=north west][inner sep=0.75pt]   [align=left] {4};
\draw (312,154) node [anchor=north west][inner sep=0.75pt]   [align=left] {3};
\draw (313,35) node [anchor=north west][inner sep=0.75pt]   [align=left] {2};
\draw (359,36) node [anchor=north west][inner sep=0.75pt]   [align=left] {1};
\draw (354,152) node [anchor=north west][inner sep=0.75pt]   [align=left] {4};
\draw (472,155) node [anchor=north west][inner sep=0.75pt]   [align=left] {3};
\draw (473,36) node [anchor=north west][inner sep=0.75pt]   [align=left] {2};
\draw (521,34) node [anchor=north west][inner sep=0.75pt]   [align=left] {1};
\draw (516,150) node [anchor=north west][inner sep=0.75pt]   [align=left] {4};
\draw (634,153) node [anchor=north west][inner sep=0.75pt]   [align=left] {3};
\draw (635,34) node [anchor=north west][inner sep=0.75pt]   [align=left] {2};
\end{tikzpicture}
\caption{Structure of the cuts for a quiver with a cycle $w_0$ passing through all the nodes. For each choice of origin along the cycles, (marked by a black node), arrows in the cut (which are dashed) are those which go backward along the cycle.\label{figthrough}}
\end{figure}
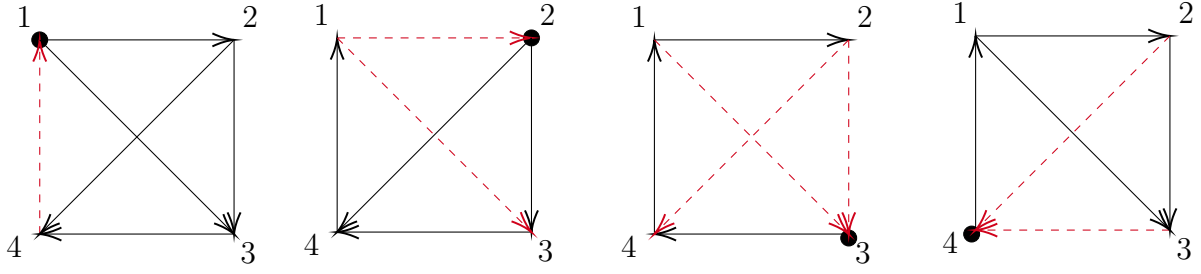

The cuts of a quiver are particularly easy to describe when $Q$ contains a cycle $w_0$ passing through all the nodes, and admits an R-charge. In particular $Q$ is strongly connected in this case, and also biconnected: if one removes a node $i$ of the quiver, two other nodes are still connected by the path on the cycle $w_0$ avoiding $i$. One can then label the nodes using $w_0$, such that this cycle passes by the nodes $1\to 2\to...\to n\to 1$. According to lemma \ref{cutcycle}, the cuts are then given by $I=\{a:i\to j|i>j\}$ and circular permutations thereof (\ie when we choose the order $r<r+1<...<n<1<...<r-1$ on $\{1,...,n\}$). Said differently, the cuts consist of the arrows 'going the wrong way' according to one of the $n$ choices of circular ordering on the nodes. 
An example is shown in Figure \ref{figthrough}. 

\subsection{Constraints on the existence of scaling and attractor solutions}


Having introduced the necessary notions, we are now ready to derive necessary conditions for the existence of scaling and attractor solutions. The key point is the positivity
of the currents $\lambda^\bullet$ in \eqref{denefcurrents}, whose conservation corresponds with the conformal ($\bullet=0$) or attractor ($\bullet=\star$)
Denef equation. We can then apply  part $ii)$ of lemma \ref{lemgraphcoho}, to conclude that $\lambda^\bullet$
is a  sum of positive currents circulating on simple cycles of the quiver, namely there exists $(\mu_w)_{w\in Q_2}$ such that 
\be
\label{lambdamu}
\lambda^\bullet_a=\sum_{w\ni a}\mu_w\ ,\qquad \mu_w>0 \quad\forall \ w\in Q_2
\ee
Now, for each simple cycle $w$, and every  arrow $(a:i\to j)\in w$, 
we have the (generalized) triangular inequality 
\begin{align}\label{confineq}
    r_{ij}\leq \sum_{(b:k\to l)\in w | b\neq a}r_{kl}
\end{align}
Since the function $f:x\mapsto (1+\frac{1}{x})^{-1}$ is increasing on $\mathbb{R}^\star_+$, we deduce
\begin{align}\label{attrineq}
    \left(1+\frac{1}{r_{ij}}\right)^{-1}&\leq 
    \left(1+\frac{1}{\sum_{(b:k\to l)\in w | b\neq a}r_{kl}}\right)^{-1}=\frac{\sum_{(b:k\to l)\in w | b\neq a}r_{kl}}{1+\sum_{(b:k\to l)\in w | b\neq a}r_{kl}}\nn\\&
    <\sum_{(b:k\to l)\in w | b\neq a}\frac{r_{kl}}{1+r_{kl}} =\sum_{(b:k\to l)\in w | b\neq a} \left(1+\frac{1}{r_{kl}}\right)^{-1}
\end{align}
Thus, we have 
\begin{align}
 \frac{1}{\lambda^\bullet_a}&\leq\sum_{b\in w | b\neq a}\frac{1}{\lambda^\bullet_b}
 \label{geneq}
\end{align}
both for $\bullet=0$ and $\bullet=\star$. The data of a (weak) cut $I$ allows one to choose consistently an arrow in each simple cycle (resp. in a set of cycles). One denotes for convenience:
\begin{align}
    \epsilon_a^{I}=\left\{
    \begin{array}{ll}
        -1 & \mbox{if}\; a\in I\\
        1  & \mbox{if}\; a\in Q_1-I
    \end{array}
\right.
\end{align}
It then follows from \eqref{geneq} that, for any weak cut, 
\begin{align}\label{ineqs}
    &\sum_{a\in w}\frac{\epsilon^I_a}{\lambda^\bullet_a}\geq 0 
\end{align}
the inequality being strict when $w$ contains no arrow of $I$. One can then take a linear combination of the inequalities \eqref{ineqs} with the same strictly positive coefficients $\mu_w$ appearing in \eqref{lambdamu}, obtaining
\begin{align}
   0 \leq &\sum_{w\in Q_2}\mu_w\sum_{a\in w} \frac{\epsilon_a^{I} }{\lambda^\bullet_a}
    =\sum_{a\in Q_1}\frac{\epsilon_a^{ I}}{\sum_{w\ni a}\mu_w} \left(\sum_{w\ni a}\mu_w\right)
    =\sum_{a\in Q_1}\epsilon_a^{ I}
    =|Q_1-I|-|I|
    \label{ineqwcut}
\end{align}
Hence, the existence of scaling or attractor solutions implies,
for every weak cut, the inequality 
\begin{align}
\label{ineqI}
  |I|\leq |Q_1-I|
\end{align}

A few remarks are in order: 
\begin{itemize}
 
\item The inequality \eqref{ineqI} can only be saturated for scaling solutions ($\bullet=0$), and if each simple cycle contains an arrow of $I$, \ie $I$ is a cut. In that case, \eqref{confineq} is an equality for each $w\in Q_2, a\in I\cap w$. In particular, the centers along each oriented cycle of the scaling solution must be collinear. One has in particular that \eqref{confineq} is a strict inequality for $a\in w-(I\cap w)$, and therefore $|J|<|Q_1-J|$
for all cuts $J$ distinct from $I$. Moreover, two arrows contained in the same cycle $w$ are necessarily collinear. If $Q$ is biconnected and strongly connected, from lemma \ref{lembiconstrong} one can connect any two arrows by a sequence of arrows such that two consecutive arrows are contained in a common cycle, and in that case the whole scaling solution is collinear.

   \item 
    For a small perturbation of the attractor stability condition, such that the current is perturbed to $\lambda_a = 1+\delta_a + 1/r_{ij}$ with $|\delta_a|\ll 1$, the inequality \eqref{attrineq} remains true up to small corrections of order $\delta_a (1+1/r_{kl})^{-2}$. 
After multiplying by $\mu^\star_w$, which is of order $1+1/r_{kl}$,
the resulting correction to \eqref{ineqI} remains small even for $r_{kl}\to 0$. Since the leading terms are integer, this correction 
does not affect the inequality \eqref{ineqI}, although it affects 
the analysis of the cases where the inequality is saturated. 

\end{itemize}

We thus obtain the main result of this note:

\begin{proposition}\label{cutscaling}
    If a quiver $Q$ admits attractor or scaling solutions, then for each weak cut $I$ one has:
\begin{align}
    |I|\leq|Q_1-I|
\end{align}
If scaling solutions exist, the inequality can be saturated for at most one cut. When this is the case and if $Q$ is biconnected, then the scaling solutions are collinear.
\end{proposition}

We shall now compare our result to the conjecture 
put forward  (and proven for some simple cases) in \cite{Beaujard:2021fsk}. In that reference, it
was assumed that  $Q$ admits an R-charge and contains a cycle $w_0\in Q_2$ passing through all the nodes, giving a cyclic ordering $Q_0 \simeq \mathbb{Z}/n\mathbb{Z}$. As show in lemma \ref{cutcycle}, all the cuts of $Q$ are then of the form $I=\{a:i\to j|j>i\}$ and cyclic permutations. The conjecture in \cite{Beaujard:2021fsk} is therefore
a consequence of Proposition \ref{cutscaling},

\begin{proposition}\label{propcutcoul}
    For a quiver with an R-charge and a cycle passing through all the nodes, scaling or attractor solutions can only exist if
    \begin{align}
        \sum_{i<j}\kappa_{ij}\geq 0 \qquad
        \mbox{and cyclic permutations}
    \end{align}
\end{proposition}

\subsection{Non-Abelian scaling and attractor solutions}\label{subnonab}

We now consider a non-Abelian quiver $Q$ with dimension vector $(d_i)_{i\in Q_0}$, supposing $d_i\geq 1$ for $i\in Q_0$. On the Coulomb branch, the gauge group $\prod_{i\in Q_0} U(d_i)$ is broken to a Cartan subgroup $\prod_{i\in Q_0} U(1)^{d_i}$, and the scalar fields in the vector multiplets associated to this subgroup satisfy similar equations \eqref{eqdenef} as in the Abelian case with $n=\sum_{i\in Q_0} d_i$ centers, with $d_i$ of them carrying the same charge $\gamma_i$ and stability parameter $\zeta_i$ for each $i\in Q_0$. Labelling by $(i,k)$ the $k$-th center with charge $\gamma_i$, for $1\leq k\leq d_i$, the distances between centers satisfy a non-Abelian version of Denef's equations:
\begin{align}
 \forall i=1\dots n, \quad \forall k=1\dots d_i, \quad    \sum_{j\neq i}
 \sum_{k'=1}^{d_j}\frac{\kappa_{ij}}{|\vec r_{(i,k)} - \vec r_{(j,k')}|}=\zeta_i
    \label{eqdenefna}
\end{align}
As before, solutions may only exist if $\sum_{i\in Q_0} \zeta_i d_i=0$. The same equations arise from the `totally Abelianized quiver' $Q^d$ with nodes $Q^d_0:=\{(i,k)|i\in Q_0,1\leq k\leq d_i\}$, and arrows $Q^d_1:=\{(a,k,k'): (i,k) \to (j,k'))|(a: i\to j) \in Q_1,1\leq k\leq d_i,1\leq k'\leq d_j \}$. On special loci where $m$ of the centers attached to the same node $i\in Q_0$ coincide, the gauge factor $U(1)^{d_i}$ is enhanced to $U(m) \times U(1)^{d_i-m}$,  leading to partially Abelianized quivers with one node carrying charge $m \gamma_i$ and $d_i-m$ nodes carrying charge $\gamma_i$ \cite{Manschot:2010qz}. We can focus on the Denef equations associated to the totally Abelianized quiver $Q^d$, since solutions relevant for partially Abelianized quivers arise as special cases.\medskip

It is clear from this construction that the Abelianized quiver $Q^d$ is connected (resp. strongly connected) if and only if $Q$ is. In particular, if $Q$ admits a non-Abelian $d$-dimensional scaling or attractor solution, then $Q^d$ is strongly connected, and therefore also $Q$:

\begin{proposition}\label{propstrongconnonab}
    A non-Abelian quiver which admits scaling or attractor  solutions must be strongly connected.
\end{proposition}

If $Q$ is biconnected, then $Q^d$ is biconnected too, but the converse is not true: if $i$ is a node shared between two biconnected components of $Q$, those two components merge into a single biconnected component in $Q^d$ when $d_i\geq 2$ (see the example of the butterfly quiver with dimension vector $(2,1,1,1,1)$, Figure \ref{figbutscal}). In particular, the decomposition into biconnected components of the solutions to Denef's equations does not hold in the non-Abelian case, because biconnected components of $Q$ are not in correspondence with those of $Q^d$.\medskip

We shall now attempt to derive constraints on the magnitude
of the Dirac products $\kappa_{ij}$ for the existence of 
non-Abelian scaling or attractor solutions. Let $\lambda^\bullet$ 
be the corresponding strictly positive 
conserved current on $Q^d$.
For $a$ an arrow on $Q^d$,  we denote by $p(a)$ its projection to $Q$;
similarly, for  a path $v$ on $Q^d$, we denote by $p(v)$ its projection to $Q$. For $I$ a weak cut of $Q$, each cycle of $p^{-1}(Q_2)$ (\ie projecting on a simple cycle of $Q$) contains at most one arrow of $p^{-1}(I)$. By the same argument as in the proof of proposition \ref{cutscaling}, the triangular inequality in the corresponding polygon in the non-Abelian scaling or attractor solutions gives:
\begin{align}\label{ineqspI}
    &\sum_{a\in w}\frac{\epsilon^{p^{-1}(I)}_a}{\lambda^\bullet_a}\geq 0 
\end{align}
Now, suppose that the current could be expressed as a sum 
$\lambda^\bullet_a=\sum_{w\in p^{-1}(Q_2)|w\in a}\mu_w$
of  positive currents $\mu_w$ circulating on simple cycles of $Q^d$
which project to simple cycles of $Q$. By taking
the linear combination of the inequalities \eqref{ineqspI} with coefficients $\mu_w$, one would
conclude that $|p^{-1}(I)|\leq|Q^d_1-p^{-1}(I)|$, or equivalently
\begin{align} \label{ineqsnonab}
    \sum_{(a:i\to j)\in I}d_id_j\leq\sum_{(a:i\to j)\in Q_1-I}d_id_j
\end{align}
In the special case where $p^{-1}(Q_2)=Q^d_2$, \ie when each cycle of $Q^d$ projects to a simple cycle of $Q$, the existence of the positive
currents $\mu_w$ follows directly from $ii)$ of lemma \ref{lemgraphcoho}, and the therefore the condition \eqref{ineqsnonab} must hold. This the case for example for the triangular quiver
where one of the entries in the dimension vector is equal to one, as in the cases considered in \cite[\S 6]{Manschot:2012rx}\cite[\S 3]{Lee:2013yka} \cite[\S 4]{Kim:2015fba} \cite{Messamah:2020ldn}.
More generally, when the quiver $Q$ is biconnected, lemma \ref{lemnonab} allows to express $\lambda^\bullet$ as a sum of (possibly negative) currents circulating on the cycles of $p^{-1}(Q_2)$. While we are not able to show that these currents can be taken to be positive, it seems plausible to propose that for a biconnected quiver, the existence of a $d$-dimensional scaling or attractor solution implies that for each weak cut $I$, the condition
\eqref{ineqsnonab} is satisfied.

\begin{figure}
\centering
\begin{tikzpicture}[x=0.75pt,y=0.75pt,yscale=-1,xscale=1]

\draw    (37.72,71.98) -- (128.03,125.58) ;
\draw [shift={(129.75,126.6)}, rotate = 210.69] [color={rgb, 255:red, 0; green, 0; blue, 0 }  ][line width=0.75]    (10.93,-3.29) .. controls (6.95,-1.4) and (3.31,-0.3) .. (0,0) .. controls (3.31,0.3) and (6.95,1.4) .. (10.93,3.29)   ;
\draw    (221.67,181.4) -- (131.47,127.63) ;
\draw [shift={(129.75,126.6)}, rotate = 390.8] [color={rgb, 255:red, 0; green, 0; blue, 0 }  ][line width=0.75]    (10.93,-3.29) .. controls (6.95,-1.4) and (3.31,-0.3) .. (0,0) .. controls (3.31,0.3) and (6.95,1.4) .. (10.93,3.29)   ;
\draw    (220.82,70.4) -- (221.66,179.4) ;
\draw [shift={(221.67,181.4)}, rotate = 269.56] [color={rgb, 255:red, 0; green, 0; blue, 0 }  ][line width=0.75]    (10.93,-3.29) .. controls (6.95,-1.4) and (3.31,-0.3) .. (0,0) .. controls (3.31,0.3) and (6.95,1.4) .. (10.93,3.29)   ;
\draw    (38.79,182.98) -- (37.74,73.98) ;
\draw [shift={(37.72,71.98)}, rotate = 449.45] [color={rgb, 255:red, 0; green, 0; blue, 0 }  ][line width=0.75]    (10.93,-3.29) .. controls (6.95,-1.4) and (3.31,-0.3) .. (0,0) .. controls (3.31,0.3) and (6.95,1.4) .. (10.93,3.29)   ;
\draw    (129.75,126.6) -- (40.49,181.92) ;
\draw [shift={(38.79,182.98)}, rotate = 328.21000000000004] [color={rgb, 255:red, 0; green, 0; blue, 0 }  ][line width=0.75]    (10.93,-3.29) .. controls (6.95,-1.4) and (3.31,-0.3) .. (0,0) .. controls (3.31,0.3) and (6.95,1.4) .. (10.93,3.29)   ;
\draw    (129.75,126.6) -- (219.12,71.45) ;
\draw [shift={(220.82,70.4)}, rotate = 508.32] [color={rgb, 255:red, 0; green, 0; blue, 0 }  ][line width=0.75]    (10.93,-3.29) .. controls (6.95,-1.4) and (3.31,-0.3) .. (0,0) .. controls (3.31,0.3) and (6.95,1.4) .. (10.93,3.29)   ;

\draw (18,115) node [anchor=north west][inner sep=0.75pt]   [align=left] {p};
\draw (232,113) node [anchor=north west][inner sep=0.75pt]   [align=left] {p};
\draw (122,136) node [anchor=north west][inner sep=0.75pt]   [align=left] {1};
\draw (224,53) node [anchor=north west][inner sep=0.75pt]   [align=left] {2};
\draw (223.67,184.4) node [anchor=north west][inner sep=0.75pt]   [align=left] {3};
\draw (27,184) node [anchor=north west][inner sep=0.75pt]   [align=left] {4};
\draw (25,56) node [anchor=north west][inner sep=0.75pt]   [align=left] {5};
\end{tikzpicture}
\caption{An example of non biconnected quiver: the butterfly 
quiver. \label{figbutter}}
\end{figure}
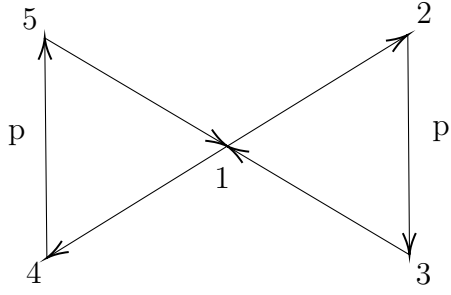

\begin{figure}
\begin{center}
\begin{tikzpicture}[x=0.75pt,y=0.75pt,yscale=-1,xscale=1]

\draw    (250,30.5) -- (397,30.5) ;
\draw [shift={(399,30.5)}, rotate = 180] [color={rgb, 255:red, 0; green, 0; blue, 0 }  ][line width=0.75]    (10.93,-3.29) .. controls (6.95,-1.4) and (3.31,-0.3) .. (0,0) .. controls (3.31,0.3) and (6.95,1.4) .. (10.93,3.29)   ;
\draw    (399,30.5) -- (399,157.5) ;
\draw [shift={(399,159.5)}, rotate = 270] [color={rgb, 255:red, 0; green, 0; blue, 0 }  ][line width=0.75]    (10.93,-3.29) .. controls (6.95,-1.4) and (3.31,-0.3) .. (0,0) .. controls (3.31,0.3) and (6.95,1.4) .. (10.93,3.29)   ;
\draw    (399,159.5) -- (253,158.51) ;
\draw [shift={(251,158.5)}, rotate = 360.39] [color={rgb, 255:red, 0; green, 0; blue, 0 }  ][line width=0.75]    (10.93,-3.29) .. controls (6.95,-1.4) and (3.31,-0.3) .. (0,0) .. controls (3.31,0.3) and (6.95,1.4) .. (10.93,3.29)   ;
\draw    (251,158.5) -- (103,159.49) ;
\draw [shift={(101,159.5)}, rotate = 359.62] [color={rgb, 255:red, 0; green, 0; blue, 0 }  ][line width=0.75]    (10.93,-3.29) .. controls (6.95,-1.4) and (3.31,-0.3) .. (0,0) .. controls (3.31,0.3) and (6.95,1.4) .. (10.93,3.29)   ;
\draw    (101,159.5) -- (101,32.5) ;
\draw [shift={(101,30.5)}, rotate = 450] [color={rgb, 255:red, 0; green, 0; blue, 0 }  ][line width=0.75]    (10.93,-3.29) .. controls (6.95,-1.4) and (3.31,-0.3) .. (0,0) .. controls (3.31,0.3) and (6.95,1.4) .. (10.93,3.29)   ;
\draw    (101,30.5) -- (248,30.5) ;
\draw [shift={(250,30.5)}, rotate = 180] [color={rgb, 255:red, 0; green, 0; blue, 0 }  ][line width=0.75]    (10.93,-3.29) .. controls (6.95,-1.4) and (3.31,-0.3) .. (0,0) .. controls (3.31,0.3) and (6.95,1.4) .. (10.93,3.29)   ;
\draw    (251,158.5) -- (397.49,31.81) ;
\draw [shift={(399,30.5)}, rotate = 499.14] [color={rgb, 255:red, 0; green, 0; blue, 0 }  ][line width=0.75]    (10.93,-3.29) .. controls (6.95,-1.4) and (3.31,-0.3) .. (0,0) .. controls (3.31,0.3) and (6.95,1.4) .. (10.93,3.29)   ;
\draw    (399,159.5) -- (251.51,31.81) ;
\draw [shift={(250,30.5)}, rotate = 400.89] [color={rgb, 255:red, 0; green, 0; blue, 0 }  ][line width=0.75]    (10.93,-3.29) .. controls (6.95,-1.4) and (3.31,-0.3) .. (0,0) .. controls (3.31,0.3) and (6.95,1.4) .. (10.93,3.29)   ;
\draw    (102,29.5) -- (249.49,157.19) ;
\draw [shift={(251,158.5)}, rotate = 220.89] [color={rgb, 255:red, 0; green, 0; blue, 0 }  ][line width=0.75]    (10.93,-3.29) .. controls (6.95,-1.4) and (3.31,-0.3) .. (0,0) .. controls (3.31,0.3) and (6.95,1.4) .. (10.93,3.29)   ;
\draw    (250,30.5) -- (102.51,158.19) ;
\draw [shift={(101,159.5)}, rotate = 319.11] [color={rgb, 255:red, 0; green, 0; blue, 0 }  ][line width=0.75]    (10.93,-3.29) .. controls (6.95,-1.4) and (3.31,-0.3) .. (0,0) .. controls (3.31,0.3) and (6.95,1.4) .. (10.93,3.29)   ;
\draw    (70,31) -- (71.97,155.5) ;
\draw [shift={(72,157.5)}, rotate = 269.09000000000003] [color={rgb, 255:red, 0; green, 0; blue, 0 }  ][line width=0.75]    (10.93,-3.29) .. controls (6.95,-1.4) and (3.31,-0.3) .. (0,0) .. controls (3.31,0.3) and (6.95,1.4) .. (10.93,3.29)   ;
\draw    (72,157.5) -- (70.03,33) ;
\draw [shift={(70,31)}, rotate = 449.09] [color={rgb, 255:red, 0; green, 0; blue, 0 }  ][line width=0.75]    (10.93,-3.29) .. controls (6.95,-1.4) and (3.31,-0.3) .. (0,0) .. controls (3.31,0.3) and (6.95,1.4) .. (10.93,3.29)   ;
\draw    (102,9) -- (242,9.49) ;
\draw [shift={(244,9.5)}, rotate = 180.2] [color={rgb, 255:red, 0; green, 0; blue, 0 }  ][line width=0.75]    (10.93,-3.29) .. controls (6.95,-1.4) and (3.31,-0.3) .. (0,0) .. controls (3.31,0.3) and (6.95,1.4) .. (10.93,3.29)   ;
\draw    (244,9.5) -- (104,9.01) ;
\draw [shift={(102,9)}, rotate = 360.2] [color={rgb, 255:red, 0; green, 0; blue, 0 }  ][line width=0.75]    (10.93,-3.29) .. controls (6.95,-1.4) and (3.31,-0.3) .. (0,0) .. controls (3.31,0.3) and (6.95,1.4) .. (10.93,3.29)   ;

\draw (56,87) node [anchor=north west][inner sep=0.75pt]   [align=left] {y};
\draw (161,-13) node [anchor=north west][inner sep=0.75pt]   [align=left] {$\displaystyle x$};
\draw (103,83) node [anchor=north west][inner sep=0.75pt]   [align=left] {$\displaystyle p$};
\draw (402,83) node [anchor=north west][inner sep=0.75pt]   [align=left] {$\displaystyle p$};
\draw (246,12.5) node [anchor=north west][inner sep=0.75pt]   [align=left] {1};
\draw (246,166) node [anchor=north west][inner sep=0.75pt]   [align=left] {$\displaystyle 1'$};
\draw (402,14) node [anchor=north west][inner sep=0.75pt]   [align=left] {2};
\draw (404,158) node [anchor=north west][inner sep=0.75pt]   [align=left] {3};
\draw (87,158) node [anchor=north west][inner sep=0.75pt]   [align=left] {4};
\draw (84,17) node [anchor=north west][inner sep=0.75pt]   [align=left] {5};

\end{tikzpicture}

\end{center}
\caption{Scaling solution for the butterfly quiver with $d=(2,1,1,1,1)$. \label{figbutscal}}
\end{figure}
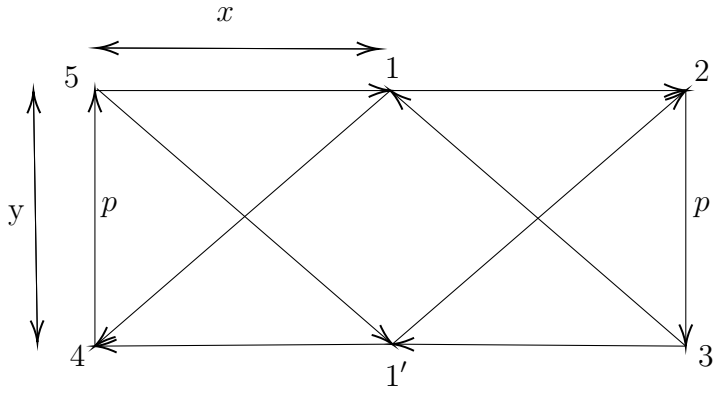

We can construct a  counterexample to the inequalities \eqref{ineqsnonab} for a non biconnected quiver, namely the 'butterfly' quiver shown in Figure \ref{figbutter}. We choose  multiplicities  $\kappa_{23}=\kappa_{45}=p\geq 1$, and multiplicity $1$ for other arrows, and dimension vector $(2,1,1,1,1)$. The corresponding Abelianized quiver $Q^d$ is obtained by splitting the central node $1$ into two nodes which we denote by $1$ and $1'$. In particular for the simple cycle $w:1\to 2\to 3\to 1'\to 4\to 5\to 1$, its projection on $Q$ $p(w):=1\to 2\to 3\to 1\to 4\to 5\to 1$ is not simple, and $w$ cannot be expressed as a linear combination of cycles projection to a simple cycle of $Q$. We consider solutions inscribed in a rectangle with height $y$ and length $2x$ as shown in Figure \ref{figbutscal}. The scaling (resp. attractor) Denef equations are trivially verified at the node $1$ and $1'$, and require 
\be
\frac{p}{y}=\frac{1}{x}+\frac{1}{\sqrt{x^2+y^2}}\ ,\qquad 
\mbox{resp.} \qquad 
p \left(1+\frac{1}{y}\right)=2+\frac{1}{x}+\frac{1}{\sqrt{x^2+y^2}}
\ee
at the other nodes. Solutions manifestly exist for any $p\geq 1$. In contrast, the inequalities \eqref{ineqsnonab}
for the cut $I=\{2\to 3, 4\to 5\}$ would require
\begin{align}
    \sum_{(a:i\to j)\in I}d_id_j=2p \leq \sum_{(a:i\to j)\in Q_1- I}d_id_j = 8
\end{align}
which is false. This shows that the assumption of biconnectedness is important for the inequalities \eqref{ineqsnonab} to hold. Whether the inequalities are verified in the biconnected case is left as an interesting problem for future work.

\medskip


\section{Existence of self-stable representations \label{sec_Higgs}}

In this section, we discuss constraints for the existence of vacuum solutions on the Higgs branch at the attractor point, which mathematically correspond to self-stable representations. Physically, we expect that such solutions only exist when
the Coulomb branch admits scaling solutions (except for simple 
representations associated to the nodes of the quiver, which 
exist for arbitrary stability condition). Indeed, we shall demonstrate that for quivers with generic potential, the existence of self-stable Abelian representations requires similar but slightly stronger conditions as for existence of scaling solutions
on the Coulomb branch.

\subsection{Conserved current on the Higgs branch}

We consider the quiver quantum mechanics associated to a connected quiver $Q$, dimension vector $(d_i)\in (\mathbb{N}^\star)^{Q_0}$, Fayet-Iliopoulos parameters $(\zeta_i)\in \mathbb{R}^{Q_0}$ and a generic potential $W$. 
On the Higgs branch, the expectation value of the 
chiral multiplets   associated to the arrows $a\in Q_1$ 
breaks the gauge group $\prod_{i\in Q_0} U(d_i)$
to the center $U(1)^{Q_0}$. After integrating out the vector multiplets
one is left with an effective quantum mechanics for the chiral multiplet scalars $(\phi_a)_{a\in Q_1}$. Supersymmetric vacua 
are given by solutions of the  D-term and F-term equations,
\begin{align}\label{FIstab}
    \sum_{(a:i\to j)\in Q_1}\phi_a^\dagger\phi_a-\sum_{(a:j\to i)\in Q_1}\phi_a\phi_a^\dagger=\zeta_i\, \Id_{d_i}\quad\forall\;i\in Q_0
\end{align}
\begin{align}
    \partial_{\phi_a} W=0
\end{align}
modulo the action of the gauge group. From \cite[Prop 6.5]{King}, such solutions are in one-to-one correspondence with polystable representations
of the quiver, \ie direct sums of stable representations, where the stability condition is determined by the
slope $\sum_{i\in Q_0} d_i \zeta_i/\sum_{i\in Q_0} d_i$. For
primitive dimension vector and generic stability parameters, polystable representations are automatically stable. 

Summing up the relations \eqref{FIstab} over $i\in Q_0$, it is
clear that solutions only exist when $\partial_0((d_i\zeta_i)_{i\in Q_0}=\sum_{i\in Q_0} d_i \zeta_i =0$. Thus, one can choose $(\chi_a)_{a\in Q_1}\in \mathbb{R}^{Q_1}$ such that $\partial_1((\chi_a)_{a\in Q_1}=(-d_i\zeta_i)_{i\in Q_0}$. The trace of the D-term equations \eqref{FIstab} then implies the conservation at each node of $Q$ of the current:
\begin{align}
\label{currentH}
    \lambda_a:=\chi_a+\Tr(\phi_a\phi_a^\dagger)
\end{align}
The choice of $\chi_a$ is non unique, unless $Q$ has no oriented cycle. For the attractor stability condition (also known as self-stability), one may choose $\chi_a=d_id_j$ such that the current $\lambda$ is strictly positive (arrows $a:i\to j$ such that $d_i d_j=0$ do not support any chiral multiplet and may be removed from the quiver $Q$). By applying $i)$ of lemma \ref{lemgraphcoho}, it then immediately follows that, similarly
to the Coulomb branch case,

\begin{proposition}\label{propstronghiggs}
A quiver which admits self-stable representation of dimension $(d_i)\in(\mathbb{N}^\star)^{Q_0}$ must be strongly connected.
\end{proposition}

We note that for Abelian representations, the D-terms equations are equivalent to the conservation of the current $\lambda$. As on the Coulomb branch, one has then, according to lemma \eqref{lembiconcons}, that an Abelian representation of $Q$ satisfies the D-term equations at the attractor stability condition if and only if the induced representations on the biconnected components of $Q$ satisfy the induced D-terms equations. Moreover, by definition, the F-term equations $\partial_aW=0$ constrain only the arrows which share a cycle with $a$, and therefore are in the same biconnected component as $a$. Thus, a representation satisfies the F-terms equations if and only if the induced representations satisfy the induced F-terms equations on each  biconnected component. To summarize, stable Abelian representations of the quiver with potential $(Q,W)$ at any stability condition are obtained by gluing such representations for the biconnected components of $Q$. 
As we shall see below, the example of the 'butterfly quiver' 
shows that this is no longer true in the  non-Abelian case.

\subsection{Stronger constraints in the Abelian case}

In order to obtain conditions on the number of chiral fields, we shall rely on lemma \ref{lemgenvan}, which applies to any quiver 
with potential $(Q,W)$  such that $W=\sum_{w\in C}\nu_w w$ is a linear combination of a subset of simple cycles $C\subset Q_2$
with generic coefficients, and such that there exists a subset of arrows $I\subset Q_2$ such that each cycle of $C$ contains exactly one arrow of $I$. Under these
assumptions, any Abelian representation of $(Q,W)$ is such that 
every each cycle in $C$ vanishes when evaluated on this representation. The proof relies on Bertini's theorem for linear systems on algebraic varieties, and generalizes the informal argument outlined in \cite[\S 5.2.3]{Denef:2007vg} in the case of a triangular cyclic quiver. 

Now, consider a quiver $Q$ with a cut $I$ with potential $W=\sum_{w\in Q_2}\nu_w w$ which is generic in the sense of definition \ref{defgenpot}, and $\phi$ an Abelian representation of $(Q,W)$. From lemma \ref{lemgenvan}, it follows that there is a strong cut $J$ such that the arrows of $J$ vanish in $\phi$ while no other arrows vanish: for simplicity, we consider here the case where $J$ is a cut, the general case being treated in the appendix. The representation $\phi$ is therefore a representation of the quiver with relations $(Q_J,\partial_J W)$, where $Q_J$ is the subquiver of $Q$ where one has removed the arrows of $J$, and the relations are $\partial_a W=0$ for $a\in J$.\medskip

Suppose that $\phi$ is $\zeta$-stable. The moduli space $M^{\zeta,s}_{Q_J}$ of $\zeta$-stable representations of $Q_J$ is smooth of dimension $|Q_1-J|-|Q_0|+1$. 
Since the moduli space $M^{\zeta,s}_{Q_J,\partial_J W_J}$ of $\zeta$-stable Abelian representations of $(Q_J,\partial_J W)$ is given by the zero locus of the $|J|$ relations $\partial_a W$ in $M^{\zeta,s}_{Q_J}$, one expects then that its complex dimension is given by
\begin{align}
    d_J=|Q_1-J|-|Q_0|+1-|J|\ ,
\end{align}
unless there are linear relations between the relations. We shall
now prove that there a no relation between the relations infinitesimally near $\phi$.\medskip

For a cycle $w$ of $W$, we denote its arrows by $w=b_1^w...b_{r_w}^w$. The linearization of $\partial_a W$ on the tangent space at $\phi$ gives:
\begin{align}
    \delta(\partial_a W):(\delta\phi_b)_{b\in Q_1}\mapsto\sum_{w,i|b^w_i=a}\nu_w\sum_{i+1\leq j\leq i-1}\phi_{b^w_{i+1}}\cdots\phi_{b^w_{j-1}}\, \delta\phi_{b^w_j}\,\phi_{b^w_{j+1}} \cdots \phi_{b^w_{i-1}}
\end{align}
Here we use the cyclic ordering of the cycle. Consider a linear relation $(\tilde{\phi}_a)_{a\in Q_1}$ between these relations such that:
\begin{align}
    \sum_{a\in Q_1}\tilde{\phi}_a \delta (\partial_a W)=0
\end{align}
One computes:
\begin{align}\label{cyclictrace}
    0=\sum_{a\in Q_1}\tilde{\phi}_a \delta(\partial_a W)
    &=\sum_{w\in Q_2}\nu_w\sum_{1\leq i\neq j\leq r_w}\tilde{\phi}_{b^w_i}\phi_{b^w_{i+1}}\cdots\phi_{b^w_{j-1}}\, \delta\phi_{b^w_j}\,\phi_{b^w_{j+1}} \cdots \phi_{b^w_{i-1}}\nn\\
    &=\sum_{w\in Q_2}\nu_w\sum_{1\leq i\neq j\leq r_w}\phi_{b^w_{j+1}} \cdots\phi_{b^w_{i-1}}\tilde{\phi}_{b^w_i}\phi_{b^w_{i+1}}\cdots\phi_{b^w_{j-1}}\,\delta\phi_{b^w_j}
\end{align}
Consider now the representation of $Q$ $\bar{\phi}=\phi+\epsilon\tilde{\phi}$ with $\epsilon^2=0$ (hence it is a representation over the ring $\mathbb{C}[\epsilon]/\epsilon^2$). Because $\phi$ satisfies the equation of the potential, one has, multiplying by $\epsilon$:
\begin{align}
    0&=\sum_{w\in Q_2}\nu_w\sum_{1\leq j\leq r_w}(\phi+\epsilon\tilde{\phi})_{b^w_{j+1}} \cdots\cdots(\phi+\epsilon\tilde{\phi})_{b^w_{j-1}}\, \delta\phi_{b^w_j}\nn\\
    &=\sum_{b\in Q_1}\partial_b W|_{\bar{\phi}}\delta\phi_b\label{relrel}
\end{align}
Now we consider a linear relation between the relations $\sum_{a\in J}\tilde{\phi}_a \delta (\partial_a W)=0$ where one restricts to deformations with vanishing arrows of $J$, hence $\delta\phi_b=0$ for $b\in J$. One obtains in this case $\bar{\phi}=(\epsilon\tilde{\phi}_a,\phi_b)_{a_\in J,b\in Q_1-J}$, and $\partial_b W|_{\bar{\phi}}$ for $b\in Q_1-J$. Moreover, for $a\in J$, $\partial_a W_J|_{\bar{\phi}}=\partial_a W|_{\phi}=0$, hence $\bar{\phi}$ is a representation of $(Q,W)$. Because $W$ is generic, according to Lemma \ref{lemgenvan}, the cycles of $Q_2$ vanish in $\bar{\phi}$. Each cycle in $Q_2$ contains exactly one arrow in $J$ and other arrows in $Q_1-J$: because the arrows of $Q_1-J$ do not vanish in $\phi$, it follows that the arrows of $J$ vanish in $\tilde{\phi}$. Hence there are no nontrivial linear relations between the $|J|$ differential forms, and the dimension of the tangent space at $\phi$ in $M^{\zeta,s}_{Q_J,\partial_I W_J}$ is:
\begin{align}
    0\leq d=|Q_1-J|-|Q_0|+1-|J|
\end{align}
Now $R_J\circ\partial_2=R_I\circ\partial_2$. Suppose that $\zeta$ is the attractor stability condition: the positive conserved current $\lambda=(1+|\phi_a|^2)_{a\in Q_1}$ can then be written as $\lambda=\sum_w \mu_w \partial_2(w)$ with $\mu_w>0$ from $ii)$ of lemma \ref{lemgraphcoho}, and then because the arrows of $J$  vanish on $\phi$:
\be
    2|I|\leq R_I(\lambda)=R_J(\lambda)=2|J|
\ee
hence
\be
    2|I|\leq|Q_1|-|Q_0|+1
\ee 
We show in the appendix how to remove the assumption that $J$ is a cut, and how to generalize the above argument when $I$ is only a weak cut, and obtain finally an equivalent of Proposition \ref{cutscaling} on the Higgs branch:

\begin{proposition}(Proposition \ref{prophiggsweak})
    If a quiver with generic potential $(Q,W)$ admits a self-stable Abelian representation, then for each weak cut $I$:
    \begin{align}
        |I|\leq|Q_1-I|-|Q_0|+1
    \end{align}
\end{proposition}

The case were $I$ is a cut but $J$ is not necessarily a cut is easy to deal with. Namely, because there is too much arrows vanishing, some relations of the potential $\delta_a W=0$ for $a\in J$ can become trivial, but this is counterpoised exactly by the fact that there is more relations of the form $\phi_a=0$ for $a\in J$. The case where $I$ is only a weak cut and not a cut case is more subtle, because for a quiver without cut and generic potential, the cycles do not necessarily vanish in an Abelian representation, hence the first part of the proof does not work. One must set some arrows to zero to apply a similar argument, but one must ensure that the quiver stays connected during this procedure (such that the action of the gauge group up to global scaling is still free), and that it is done in a consistent way (such that one does not lose too many relations of the quiver). We refer to the poof in the Appendix for all the details.

\subsection{Stronger constraints in the non-Abelian case}

One can try to obtain stronger constraints for the existence of non-Abelian self-stable representations, by assuming that non-Abelian analogue of Lemma \ref{lemgenvan}:

\begin{assumption}\label{assumvan}\medskip
In a stable representation $\phi$ of a quiver with potential $(Q,W)$ with $W$ generic, each cycle contains an arrow which vanishes on $\phi$.
\end{assumption}

If this assumptions was true, one could generalize the inequalities for cuts from proposition \ref{prophiggsweak} to the non-Abelian setting as follows. Let $\phi$ be a $d$-dimensional self-stable representation of a quiver $Q$ with generic potential $W$, and $I$ a cut of $Q$: under assumption \ref{assumvan}, the arrows of a strong cut $J$ of $Q$ vanish in $\phi$. We consider the simpler situation where $J$ is a cut, the general case being treated as in \ref{prophiggsweak}. $\phi$ is a $d$-dimensional stable representation of the quiver with relation $(Q_J,\partial_J W)$. The moduli space $M^{\zeta,s}_{Q_J,d}$ of $d$-dimensional representations of $Q_J$ is smooth of dimension $\sum_{(a:i\to j)\in Q_1-J}d_id_j-\sum_{i\in Q_0}d_i^2+1$, and the space $M^{\zeta,s}_{Q_J,\partial_J W,d}$ of stable representations of $(Q,\partial_J W)$ is the zero set of $\sum_{(a:i\to j)\in J}d_id_j$ equations inside $M^{\zeta,s}_{Q_J,d}$. Using the cyclicity of the trace, one can derive non-Abelian analogues of equations \ref{cyclictrace}, giving that a linear relation $(\tilde{\phi}_a)_{a\in J}$ between these equations at the tangent space of $\phi$ gives a representation $\bar{\phi}=(\phi_a,\epsilon\tilde{\phi}_b)_{a\in Q_1-J,b\in J}$ of $(Q_J,W)$ over $\mathbb{C}[\epsilon]/\epsilon^2$, which is still self-stable because some vanishing arrows in the self-stable representation $\phi$ have been set to a non-vanishing value. Then under the assumption \ref{assumvan}, the cycles of $Q_2$ contain an arrow vanishing in $\bar{\phi}$, hence $\tilde{\phi}_a=0$ for $a\in J$, hence there are no relations between the relations. The dimension of the tangent space at $\phi$ inside $M^{\zeta,s}_{Q_J,\partial_J W,d}$ is therefore:
\begin{align}
    0\leq d= \sum_{(a:i\to j)\in Q_1}d_id_j-\sum_{i\in Q_0}d_i^2+1-\sum_{(a:i\to j)\in J}d_id_j-\sum_{(a:i\to j)\in J}d_id_j
\end{align}
Applying the same arguments as in the proof of Proposition \ref{prophiggsweak} using the conserved current $(d_id_j+\Tr(\phi_a^\dagger\phi_a))_{(a:i\to j)\in Q_1}$, one  obtains \be
2\sum_{(a:i\to j)\in I}d_id_j\leq 2\sum_{(a:i\to j)\in J}d_id_j
\ee
hence for any cut $I$:
\begin{align}\label{ineqsnonabhiggs}
    \sum_{(a:i\to j)\in I}d_id_j\leq\sum_{(a:i\to j)\in Q_1-I}d_id_j-\sum_{i\in Q_0}d_i^2+1
\end{align}

\begin{figure}
\centering
\begin{tikzpicture}[x=0.75pt,y=0.75pt,yscale=-1,xscale=1]

\draw    (37.72,71.98) -- (131,71.51) ;
\draw [shift={(133,71.5)}, rotate = 539.71] [color={rgb, 255:red, 0; green, 0; blue, 0 }  ][line width=0.75]    (10.93,-3.29) .. controls (6.95,-1.4) and (3.31,-0.3) .. (0,0) .. controls (3.31,0.3) and (6.95,1.4) .. (10.93,3.29)   ;
\draw    (133,71.5) -- (218.72,71.97) ;
\draw [shift={(220.72,71.98)}, rotate = 180.31] [color={rgb, 255:red, 0; green, 0; blue, 0 }  ][line width=0.75]    (10.93,-3.29) .. controls (6.95,-1.4) and (3.31,-0.3) .. (0,0) .. controls (3.31,0.3) and (6.95,1.4) .. (10.93,3.29)   ;
\draw    (220.72,71.98) -- (220.72,182.98) ;
\draw [shift={(220.72,184.98)}, rotate = 270] [color={rgb, 255:red, 0; green, 0; blue, 0 }  ][line width=0.75]    (10.93,-3.29) .. controls (6.95,-1.4) and (3.31,-0.3) .. (0,0) .. controls (3.31,0.3) and (6.95,1.4) .. (10.93,3.29)   ;
\draw    (220.72,184.98) -- (135,185.49) ;
\draw [shift={(133,185.5)}, rotate = 359.65999999999997] [color={rgb, 255:red, 0; green, 0; blue, 0 }  ][line width=0.75]    (10.93,-3.29) .. controls (6.95,-1.4) and (3.31,-0.3) .. (0,0) .. controls (3.31,0.3) and (6.95,1.4) .. (10.93,3.29)   ;
\draw    (133,185.5) -- (39.72,184.99) ;
\draw [shift={(37.72,184.98)}, rotate = 360.31] [color={rgb, 255:red, 0; green, 0; blue, 0 }  ][line width=0.75]    (10.93,-3.29) .. controls (6.95,-1.4) and (3.31,-0.3) .. (0,0) .. controls (3.31,0.3) and (6.95,1.4) .. (10.93,3.29)   ;
\draw    (37.72,184.98) -- (37.72,73.98) ;
\draw [shift={(37.72,71.98)}, rotate = 450] [color={rgb, 255:red, 0; green, 0; blue, 0 }  ][line width=0.75]    (10.93,-3.29) .. controls (6.95,-1.4) and (3.31,-0.3) .. (0,0) .. controls (3.31,0.3) and (6.95,1.4) .. (10.93,3.29)   ;

\draw (123,50) node [anchor=north west][inner sep=0.75pt]   [align=left] {1};
\draw (224,53) node [anchor=north west][inner sep=0.75pt]   [align=left] {2};
\draw (223.67,184.4) node [anchor=north west][inner sep=0.75pt]   [align=left] {3};
\draw (27,184) node [anchor=north west][inner sep=0.75pt]   [align=left] {4};
\draw (25,56) node [anchor=north west][inner sep=0.75pt]   [align=left] {5};
\draw (126,188) node [anchor=north west][inner sep=0.75pt]   [align=left] {1'};
\draw (20,118) node [anchor=north west][inner sep=0.75pt]   [align=left] {$\displaystyle p$};
\draw (227,121) node [anchor=north west][inner sep=0.75pt]   [align=left] {$\displaystyle p$};

\end{tikzpicture}

\caption{\label{figbuthigg}
A stable representation of the butterfly quiver with dimension vector
$(2,1,1,1,1)$.}
\end{figure}
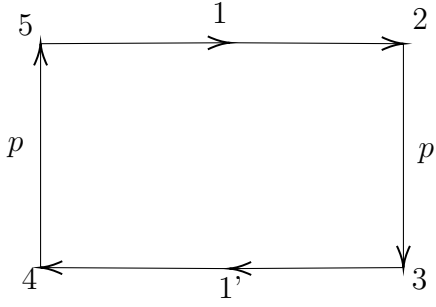

However, Assumption \ref{assumvan} fails for non-Abelian  quivers which are not biconnected. Consider as in subsection \ref{subnonab} the 'butterfly' quiver with $p$ arrows $2\to 3$ and $p\geq 2$ arrows $4\to 5$, and dimension vector $2$ on the central node $1$, and $1$ on the other nodes. For a generic potential $W$, the $p$ equations corresponding with the arrows $2\to 3$ (resp. $4\to 5$) impose that $\phi_{12}\phi_{31}=0$ (resp. $\phi_{14}\phi_{51}=0$). Denote $w^i$ (resp. $\bar{w}^i$) the cycle $a_{31}a^i_{23}a_{12}$ (resp. $a_{51}a^i_{45}a_{14}$), and $W=\sum_{i=1}^p\nu_i w_i+\sum_{i=1}^p\bar{\nu}_i\bar{w}_i$. Take $(\phi^i_{23})_{1\leq i\leq p}$ (resp. $(\phi^i_{45})_{1\leq i\leq p}$) as a generic vector satisfying $\sum_i\nu_i\phi^i_{23}=0$ (resp. $\sum_i\nu_i\phi^i_{45}=0$), in particular all these arrows are non-vanishing because $W$ is generic. We fix then:
\begin{align}
    \phi_{51}=\begin{pmatrix}1\\0\end{pmatrix}\;,\;\phi_{12}=(1,0)\;,\;\phi_{31}=\begin{pmatrix}0\\1\end{pmatrix}\;,\;\phi_{14}=(0,1)
\end{align}
The resulting representation $\phi$ satisfies the F-term relations for the potential $W$. As can be seen on the figure \ref{figbuthigg}, where the arrows correspond to nonzero matrix elements and the points $1,1'$ correspond to the two elements of the basis of the two dimensional vector space of the node $1$, each vector generates the whole representation, \ie the only subobjects of $\phi$ are the trivial ones, and then $\phi$ is stable for any stability condition. 
We have therefore constructed a self-stable representation of a quiver with generic potential $(Q,W)$ admitting a cut such that all the arrows are non-vanishing. Moreover, considering the cut $I$ containing the arrows $2\to 3$ and $4\to 5$: for $p\geq 1$, one has:
\begin{align}
    \sum_{(a:i\to j)\in I}d_id_j=2p>1=\sum_{(a:i\to j)\in Q_1- I}d_id_j-\left(\sum_{i\in Q_0}d_i^2-1\right)
\end{align}
\ie the stronger triangular inequalities are not necessarily true for non biconnected quivers. Notice the similarity of this construction 
with the counterexample discussed in \S\ref{subnonab}. \medskip

We leave it as an interesting problem to study if the assumption \ref{assumvan}, and then also the stronger inequalities \eqref{ineqsnonabhiggs} in the non-Abelian case, hold when one restricts to biconnected quivers.\medskip

\medskip

\noindent {\bf Acknowledgments}: We are grateful to Guillaume Beaujard, Jan Manschot, Swapnamay Mondal and Olivier Schiffmann for useful discussions, and to the anonymous referees for their careful reading and valuable suggestions. The research of BP is supported by Agence Nationale de la Recherche
under contract number ANR-21-CE31-0021.

\appendix

\section{Proofs}

In this Appendix, we provide mathematical proofs for some of the technical results used in the body of the paper.

\subsection{Conserved currents and graph homology}

A strictly positive conserved current is an element of $\ker(\partial_1)\cap(\mathbb{R}^\star_+)^{Q_1}$. A quiver is strongly connected if and only if for any two nodes $i,j\in Q_0$ , there is an oriented path from $i$ to $j$ and from $j$ to $i$. The key result which allows us to derive constraints on the existence of scaling or attractor solutions is the following:

\begin{lemma}\label{lemgraphcoho}
    $i)$ If $Q$ admits a strictly positive conserved current, then $Q$ is strongly connected.\medskip
    
    $ii)$ A strictly positive conserved current $\lambda$ can be expressed as a sum of strictly positive conserved currents circulating on all simple oriented cycles of the quiver, \ie $\lambda=\sum_{w\in Q_2}\mu_w\partial_2(w)$, with $\mu_w>0$.
\end{lemma}

Proof: 
\begin{description}
    \item[i)] Suppose that $Q$ is not strongly connected. The strongly connected components of $Q$ form a connected tree with at least two nodes: one has then a nontrivial partition $Q_0=Q_0'\sqcup Q_0''$ such that there is at least one arrow $Q_0'\to Q_0''$, and no arrow $Q_0''\to Q_0'$. Consider $\lambda\in \ker(\partial_1)\cap\mathbb{R}^{Q_1}$: by summing the conservation of $\lambda$ at each node of $Q_0'$, one has $\sum_{(a:Q_0'\to Q_0'')\in Q_1}\lambda_a=0$, and then $\lambda\not\in(\mathbb{R}^\star_+)^{Q_1}$.
    
    \item[ii)] Suppose now that $Q$ is strongly connected. We must show that $\ker(\partial_1)\cap(\mathbb{R}^\star_+)^{Q_1}\subset \partial_2((\mathbb{R}^\star_+)^{Q_2})$. We begin by showing $\ker(\partial_1)\cap(\mathbb{R}_+)^{Q_1}\subset\partial_2((\mathbb{R}_+)^{Q_2})$: for this we reason inductively on the number of arrows which carry a non-vanishing current. For $\lambda\in\ker(\partial_1)\cap(\mathbb{R}_+)^{Q_1}$, we consider $Q^\lambda=(Q_0,\{a\in Q_1|\lambda_a>0\})$, the (possibly non connected) quiver where we keep only the arrows with a non-vanishing current. If $Q^\lambda_1\neq\emptyset$, take $(a:i\to j)\in Q^\lambda_1$ such that $\lambda_a>0$ is minimal. One has $\lambda\in\ker(\partial_1)\cap(\mathbb{R}^\star_+)^{Q^\lambda_1}$, and then each connected component of $Q^\lambda$ is strongly connected from $i)$: there is then a path $v:j\to i$ in $Q^\lambda$, and then a simple oriented cycle $w:=av\in Q_2$ containing $a$, such that the arrows of $w$ are in $Q^\lambda$. One has $\lambda':=\lambda-\partial_2(\lambda_a w)\in\ker(\partial_1)$, and, because $\lambda_a$ is minimal, $\lambda'\in(\mathbb{R}_+)^{Q_1}$. Moreover, $\lambda'_a=0$, and then $Q^{\lambda'}_1\subsetneq Q^\lambda_1$: by induction on the number of arrows of $Q^\lambda$, one has $\lambda'\in\partial_2((\mathbb{R}_+)^{Q_2})$, and then $\lambda\in\partial_2((\mathbb{R}_+)^{Q_2})$. We have then $\ker(\partial_1)\cap(\mathbb{R}_+)^{Q_1}\subset\partial_2((\mathbb{R}_+)^{Q_2})$. Consider $\lambda\in \ker(\partial_1)\cap(\mathbb{R}^\star_+)^{Q_1}$: one has $0<\epsilon\ll1$ such that $\lambda-\partial_2(\epsilon\sum_{w\in Q_2}w)\in \ker(\partial_1)\cap(\mathbb{R}_+)^{Q_1}$, and then $\lambda=\partial_2\sum_{w\in Q_2}(\mu_w+\epsilon)w$ with $\mu_w\in\mathbb{R}_+$, \ie $\lambda\in\partial_2((\mathbb{R}^\star_+)^{Q_2})$.
\end{description} $\Box$

\subsection{Biconnected components of a quiver}

A quiver is biconnected if there is no node $i$ of the quiver such that removing $i$ (and then also the arrows with source or target $i$) disconnects the quiver. On a quiver $Q$, the biconnected components are defined as the maximal subquivers of $Q$ being biconnected. We prove the following fact about biconnected components:

\begin{lemma}\label{lembicongraph}
$i)$ The biconnected components give a partition of the arrows of the quiver $Q$, and two different biconnected components of a quiver can share at most one node.\medskip

$ii)$ Define $K$ as the unoriented graph with one node for each biconnected component and one node for each node of the quiver shared between different biconnected components, and an edge between the node $i$ and the biconnected component $b$ if $i\in b$. Then $K$ is a connected tree, \ie has no loops.
\end{lemma}

Proof:
This result can be deduced from \cite[Prop 3.5]{Har}, but we prove it here from clarity and completeness. Suppose that there is a sequence of distinct biconnected components $b_1,...,b_p$ and a sequence of distinct nodes $i_1,...,i_p$ such that $i_k\in b_k\cup b_{k+1}$, $i_p\in b_p\cup b_1$, with $p\geq 2$. Consider the subquiver $b_1\cup...\cup b_p$ of $Q$, and remove a node $i$: one can consider up to a circular permutation that $i\neq i_1$,..., $i\neq i_{p-1}$. Consider two nodes $j,j'$ such that $j\in b_k$, $j'\in b_{k'}$, and suppose up to exchanging $j$ and $j'$ that $k\leq k'$. Because the $b_{k''}$ are biconnected, there is an unoriented path between $j$ and $i_k$ in $b_k$ avoiding $i$, unoriented paths between $i_{k''}$ and $i_{k''+1}$ in $b_{k''}$ avoiding $i$ for $k\leq k''\leq k'$ and an unoriented path between $i_{k'}$ and $j'$ in $b_{k'}$ avoiding $i$. By concatenation, these give an unoriented path between $j$ and $j'$ in $b_1\cup...\cup b_p$ avoiding $i$, \ie $b_1\cup...\cup b_p$ is still connected when one has removed $i$. One obtains then that $b_1\cup...\cup b_p$ is biconnected, but it is strictly bigger that the $b_k$, which were assumed to be maximal biconnected subquivers of $Q$, giving a contradiction.\medskip
    
In particular, for $p=2$, one obtains that two different biconnected components can share at most one node: because an arrow is adjacent to two nodes, two different biconnected components cannot share an arrow, proving $i)$. The argument above for general $p$ gives exactly that $K$ has no cycle, \ie is a tree. Consider two biconnected components $b,b'$, and two nodes $i\in b,i'\in b'$: because $Q$ is connected, there is an unoriented path in $Q$ between $i$ and $i'$: by denoting $b,b_1,...,b_n,b'$ the biconnected components crossed by these paths, $b$ and $b'$ are then connected by a sequence of biconnected components sharing a node, \ie the graph $K$ is connected. This concludes the proof of $ii)$. $\Box$\medskip

Denote by $B$ the set of biconnected components of a quiver, and $Q^b$ the quiver associated to a biconnected component. Two biconnected components have distinct arrows, and then one has a decomposition $\mathbb{R}^{Q_1}=\bigoplus_{b\in B}\mathbb{R}^{Q^b_1}$. Consider a simple unoriented cycle of $Q$: it cannot pass through different biconnected components, since otherwise it would project to a cycle in $K$, which is forbiddent by the above lemma, \ie it is included in a single biconnected component. In particular this applies to simple oriented cycles, and then one has $\mathbb{R}^{Q_2}=\bigoplus_{b\in B}\mathbb{R}^{Q^b_2}$, giving a decomposition of the complex:
\begin{align}
  \bigoplus_{b\in B}\mathbb{R}^{Q^b_2}\overset{\sum_{b\in B}\partial^b_2}{\to}\bigoplus_{b\in B}\mathbb{R}^{Q^b_1}\overset{\sum_{b\in B}\partial^b_1}{\to}\mathbb{R}^{Q_0}\overset{\partial_0}\to\mathbb{R}\to 0
\end{align}

The cellular homology group $H^1$ of the graph gives the loops of the quiver (see \cite[Sec 2.2]{Hatch}), hence $\ker(\partial_1)$ is generated by the simple unoriented cycles of $Q$, each one lying in a single biconnected component. It gives then the decomposition $\ker(\partial_1)=\bigoplus_{b\in B}\ker(\partial^b_1)$:

\begin{lemma}\label{lembiconcons}
    A current on a quiver is conserved if and only if its restriction to each biconnected component is a conserved current.
\end{lemma}

\begin{lemma}\label{lembiconstrong}
$i)$ A quiver is strongly connected if and only if all its biconnected components are strongly connected.\medskip

$ii)$Suppose that $Q$ is strongly connected. Consider the equivalence class $\sim$ on the arrows of $Q$ generated by $a\sim b$ if $a,b\in w$ for $w\in Q_2$. The equivalence classes of $\sim$ correspond to the biconnected components of $Q$.
\end{lemma}

Proof:
\begin{enumerate}
    \item[$i)$] Suppose that $Q$ is strongly connected. Consider $i,i'$ in a biconnected component $b$. There is an oriented path $v:i\to i'$ in $Q$, giving a cycle $\bar{v}:b\to b$ on the graph $K$. If this cycle is trivial, then $v$ stays in the biconnected component $b$. If it is not trivial, since $K$ is a tree, the cycle $\bar{v}$ must be of the form $b\to j\to...\to j\to b$, \ie the path $v$ can be decomposed as $i\overset{v_1}{\to}j\overset{v_2}{\to}j\overset{v_3}{\to}i'$, with $v_3$ and $v_1$ oriented paths in $b$. One obtains then an oriented path $v_3v_1:i\to i'$ in $b$, \ie $b$ is strongly connected.\medskip
    
    Suppose that each biconnected component of $Q$ is strongly connected. Consider $i,i'$ two nodes of $Q$, with $i\in b,i'\in b'$. The graph $K$ is connected, consider a path $b\ni i_1\in b_1...i_{n-1}\in b_{n-1}\ni i_n\in b' $. Because $b,b_k,b'$ are strongly connected, consider oriented paths $v:i\to i_1,v_k:i_k\to i_{k+1},v':i_n\to i'$ respectively in $b,b_k,b'$. This gives an oriented path $v'v_{n-1}...v_1v:i\to i'$ in $Q$, \ie $Q$ is strongly connected.\medskip
    
    \item[$ii)$]Because a simple oriented cycle $w\in Q_2$ is contained in a single biconnected component of $Q$, an equivalence class of $\sim$ is contained in a single biconnected component of $Q$. One has then to show that a biconnected quiver contains only one single class of $\sim$. Two equivalence classes do not share any arrows by definition: one can construct an unoriented graph $\tilde{K}$ with a node corresponding to each equivalence class $s$ of $\sim$, and a node for each node $i\in Q_0$ shared between different equivalence classes, and an edge between $i$ and $s$ when $i\in s$.\medskip
    
    Considering two nodes $i,j$, there is a path between $i$ and $j$ in $Q$: the sequence of equivalence classes of $\sim$ crossed by this path gives a path in $\tilde{K}$ between $i$ and $j$, \ie $\tilde{K}$ is connected. Consider a cycle $i_1\in s_1\ni i_2...i_n\in s_n\ni i_1$ with no node or edge repeated in $\tilde{K}$. Because each equivalence class is strongly connected, one can choose an oriented path $v_k:i_k\to i_{k+1}$ with no node repeated in each equivalence class $s_k$, the concatenation giving an oriented cycle $v_n...v_1$ in $Q$. Because each $v_k$ contains no repeated node, there is a subcycle $w$ of $v_n...v_1$ which is simple and contains arrows of different equivalence classes of $\sim$, giving a contradiction. The graph $\tilde{K}$ is then a connected tree. If there were more than one equivalence classes of $\sim$, then there would be a node $i\in \tilde{K}$ such that removing $i$ disconnects $\tilde{K}$, and then disconnects the quiver $Q$, a contradiction because $Q$ was assumed to be biconnected. There is therefore a single equivalence class of $\sim$ in a biconnected quiver $Q$.
\end{enumerate}$\Box$

\subsection{Cuts and R-charge}

\begin{lemma}\label{cutRcharge}
The following assertions are equivalent:\\
$i)$ $Q$ admits a cut.\\
$ii)$ $Q$ admits an $R$-charge.\\
$iii)$ Each maximal weak cut of $Q$ is a cut.
\end{lemma}

Proof: $iii)\Rightarrow i)$: $\emptyset$ is a weak cut, and the poset of weak cuts is finite, then $Q$ admits a maximal weak cut, which is a cut by assumption.\medskip

$i)\Rightarrow ii)$ Consider a cut $I$ of $Q$. Recall that we have defined the homomorphism $R_I:\mathbb{R}^{Q_1}\mapsto\mathbb{R}$ by $R_I(a)=2$ if $a\in I$ and $R_I(a)=0$ otherwise. Each cycle $w\in Q_2$ contains exactly one arrow $a\in I$, and $R_I \circ \partial_2(w)=R_I(a)=2$, \ie $R_I$ is a R-charge.\medskip

\begin{figure}
\centering
\begin{tikzpicture}[x=0.75pt,y=0.75pt,yscale=-1,xscale=1]

\draw    (183.44,136.73) -- (207.53,61.68) ;
\draw [shift={(208.14,59.78)}, rotate = 467.79] [color={rgb, 255:red, 0; green, 0; blue, 0 }  ][line width=0.75]    (10.93,-3.29) .. controls (6.95,-1.4) and (3.31,-0.3) .. (0,0) .. controls (3.31,0.3) and (6.95,1.4) .. (10.93,3.29)   ;
\draw    (208.14,59.78) -- (286.96,59.49) ;
\draw [shift={(288.96,59.48)}, rotate = 539.79] [color={rgb, 255:red, 0; green, 0; blue, 0 }  ][line width=0.75]    (10.93,-3.29) .. controls (6.95,-1.4) and (3.31,-0.3) .. (0,0) .. controls (3.31,0.3) and (6.95,1.4) .. (10.93,3.29)   ;
\draw    (288.96,59.48) -- (313.59,134.36) ;
\draw [shift={(314.21,136.26)}, rotate = 251.79000000000002] [color={rgb, 255:red, 0; green, 0; blue, 0 }  ][line width=0.75]    (10.93,-3.29) .. controls (6.95,-1.4) and (3.31,-0.3) .. (0,0) .. controls (3.31,0.3) and (6.95,1.4) .. (10.93,3.29)   ;
\draw    (314.21,136.26) -- (250.62,182.82) ;
\draw [shift={(249,184)}, rotate = 323.78999999999996] [color={rgb, 255:red, 0; green, 0; blue, 0 }  ][line width=0.75]    (10.93,-3.29) .. controls (6.95,-1.4) and (3.31,-0.3) .. (0,0) .. controls (3.31,0.3) and (6.95,1.4) .. (10.93,3.29)   ;
\draw    (249,184) -- (185.07,137.9) ;
\draw [shift={(183.44,136.73)}, rotate = 395.78999999999996] [color={rgb, 255:red, 0; green, 0; blue, 0 }  ][line width=0.75]    (10.93,-3.29) .. controls (6.95,-1.4) and (3.31,-0.3) .. (0,0) .. controls (3.31,0.3) and (6.95,1.4) .. (10.93,3.29)   ;
\draw [color={rgb, 255:red, 208; green, 2; blue, 27 }  ,draw opacity=1 ]   (130,78) .. controls (129.01,83.42) and (121.24,94.17) .. (139.16,123.16) ;
\draw [shift={(140,124.5)}, rotate = 237.65] [color={rgb, 255:red, 208; green, 2; blue, 27 }  ,draw opacity=1 ][line width=0.75]    (10.93,-3.29) .. controls (6.95,-1.4) and (3.31,-0.3) .. (0,0) .. controls (3.31,0.3) and (6.95,1.4) .. (10.93,3.29)   ;
\draw    (208.14,59.78) .. controls (187.32,37.83) and (139.46,41.39) .. (130.39,76.38) ;
\draw [shift={(130,78)}, rotate = 282.36] [color={rgb, 255:red, 0; green, 0; blue, 0 }  ][line width=0.75]    (10.93,-3.29) .. controls (6.95,-1.4) and (3.31,-0.3) .. (0,0) .. controls (3.31,0.3) and (6.95,1.4) .. (10.93,3.29)   ;
\draw    (140,124.5) .. controls (145.91,128.44) and (146.97,139.17) .. (181.83,136.85) ;
\draw [shift={(183.44,136.73)}, rotate = 535.6600000000001] [color={rgb, 255:red, 0; green, 0; blue, 0 }  ][line width=0.75]    (10.93,-3.29) .. controls (6.95,-1.4) and (3.31,-0.3) .. (0,0) .. controls (3.31,0.3) and (6.95,1.4) .. (10.93,3.29)   ;
\draw [color={rgb, 255:red, 208; green, 2; blue, 27 }  ,draw opacity=1 ]   (248,-9.55) .. controls (242.54,-8.8) and (229.9,-12.85) .. (207.92,13.19) ;
\draw [shift={(206.91,14.41)}, rotate = 309.53999999999996] [color={rgb, 255:red, 208; green, 2; blue, 27 }  ,draw opacity=1 ][line width=0.75]    (10.93,-3.29) .. controls (6.95,-1.4) and (3.31,-0.3) .. (0,0) .. controls (3.31,0.3) and (6.95,1.4) .. (10.93,3.29)   ;
\draw    (289.6,59.05) .. controls (303.99,32.44) and (285.74,-11.94) .. (249.66,-9.69) ;
\draw [shift={(248,-9.55)}, rotate = 354.25] [color={rgb, 255:red, 0; green, 0; blue, 0 }  ][line width=0.75]    (10.93,-3.29) .. controls (6.95,-1.4) and (3.31,-0.3) .. (0,0) .. controls (3.31,0.3) and (6.95,1.4) .. (10.93,3.29)   ;
\draw    (206.91,14.41) .. controls (205,21.25) and (195.13,25.59) .. (208.17,58) ;
\draw [shift={(208.78,59.5)}, rotate = 247.55] [color={rgb, 255:red, 0; green, 0; blue, 0 }  ][line width=0.75]    (10.93,-3.29) .. controls (6.95,-1.4) and (3.31,-0.3) .. (0,0) .. controls (3.31,0.3) and (6.95,1.4) .. (10.93,3.29)   ;
\draw [color={rgb, 255:red, 208; green, 2; blue, 27 }  ,draw opacity=1 ]   (177.81,217.98) .. controls (182.72,220.47) and (190.81,230.99) .. (223.71,222.08) ;
\draw [shift={(225.23,221.66)}, rotate = 524.22] [color={rgb, 255:red, 208; green, 2; blue, 27 }  ,draw opacity=1 ][line width=0.75]    (10.93,-3.29) .. controls (6.95,-1.4) and (3.31,-0.3) .. (0,0) .. controls (3.31,0.3) and (6.95,1.4) .. (10.93,3.29)   ;
\draw    (182.63,137.9) .. controls (155.66,151.59) and (145.41,198.47) .. (176.37,217.15) ;
\draw [shift={(177.81,217.98)}, rotate = 208.93] [color={rgb, 255:red, 0; green, 0; blue, 0 }  ][line width=0.75]    (10.93,-3.29) .. controls (6.95,-1.4) and (3.31,-0.3) .. (0,0) .. controls (3.31,0.3) and (6.95,1.4) .. (10.93,3.29)   ;
\draw    (225.23,221.66) .. controls (230.69,217.12) and (241.28,219.17) .. (249,185.09) ;
\draw [shift={(249.35,183.51)}, rotate = 462.23] [color={rgb, 255:red, 0; green, 0; blue, 0 }  ][line width=0.75]    (10.93,-3.29) .. controls (6.95,-1.4) and (3.31,-0.3) .. (0,0) .. controls (3.31,0.3) and (6.95,1.4) .. (10.93,3.29)   ;
\draw [color={rgb, 255:red, 208; green, 2; blue, 27 }  ,draw opacity=1 ]   (367.43,71.96) .. controls (364.85,67.1) and (364.3,53.84) .. (332.28,42.16) ;
\draw [shift={(330.8,41.63)}, rotate = 379.40999999999997] [color={rgb, 255:red, 208; green, 2; blue, 27 }  ,draw opacity=1 ][line width=0.75]    (10.93,-3.29) .. controls (6.95,-1.4) and (3.31,-0.3) .. (0,0) .. controls (3.31,0.3) and (6.95,1.4) .. (10.93,3.29)   ;
\draw    (317.34,134.63) .. controls (347.28,138.98) and (382.66,106.58) .. (368.13,73.47) ;
\draw [shift={(367.43,71.96)}, rotate = 424.12] [color={rgb, 255:red, 0; green, 0; blue, 0 }  ][line width=0.75]    (10.93,-3.29) .. controls (6.95,-1.4) and (3.31,-0.3) .. (0,0) .. controls (3.31,0.3) and (6.95,1.4) .. (10.93,3.29)   ;
\draw    (330.8,41.63) .. controls (323.72,42.19) and (316.24,34.42) .. (290.3,57.82) ;
\draw [shift={(289.1,58.91)}, rotate = 317.41999999999996] [color={rgb, 255:red, 0; green, 0; blue, 0 }  ][line width=0.75]    (10.93,-3.29) .. controls (6.95,-1.4) and (3.31,-0.3) .. (0,0) .. controls (3.31,0.3) and (6.95,1.4) .. (10.93,3.29)   ;
\draw [color={rgb, 255:red, 208; green, 2; blue, 27 }  ,draw opacity=1 ]   (330.07,215.08) .. controls (333.49,210.76) and (345.4,204.92) .. (343.19,170.9) ;
\draw [shift={(343.08,169.33)}, rotate = 445.65] [color={rgb, 255:red, 208; green, 2; blue, 27 }  ,draw opacity=1 ][line width=0.75]    (10.93,-3.29) .. controls (6.95,-1.4) and (3.31,-0.3) .. (0,0) .. controls (3.31,0.3) and (6.95,1.4) .. (10.93,3.29)   ;
\draw    (252.53,194.49) .. controls (260.61,223.64) and (304.53,242.97) .. (328.97,216.33) ;
\draw [shift={(330.07,215.08)}, rotate = 490.36] [color={rgb, 255:red, 0; green, 0; blue, 0 }  ][line width=0.75]    (10.93,-3.29) .. controls (6.95,-1.4) and (3.31,-0.3) .. (0,0) .. controls (3.31,0.3) and (6.95,1.4) .. (10.93,3.29)   ;
\draw    (343.08,169.33) .. controls (339.71,163.08) and (343.81,153.11) .. (311.94,138.79) ;
\draw [shift={(310.46,138.13)}, rotate = 383.65999999999997] [color={rgb, 255:red, 0; green, 0; blue, 0 }  ][line width=0.75]    (10.93,-3.29) .. controls (6.95,-1.4) and (3.31,-0.3) .. (0,0) .. controls (3.31,0.3) and (6.95,1.4) .. (10.93,3.29)   ;

\draw (269,138) node [anchor=north west][inner sep=0.75pt]   [align=left] {$\displaystyle a_{3}$};
\draw (198,87) node [anchor=north west][inner sep=0.75pt]   [align=left] {$\displaystyle a_{5}$};
\draw (214,141) node [anchor=north west][inner sep=0.75pt]   [align=left] {$\displaystyle a_{4}$};
\draw (286,87) node [anchor=north west][inner sep=0.75pt]   [align=left] {$\displaystyle a_{2}$};
\draw (240,55) node [anchor=north west][inner sep=0.75pt]   [align=left] {$\displaystyle a_{1}$};
\draw (239,105) node [anchor=north west][inner sep=0.75pt]   [align=left] {$\displaystyle w$};
\draw (154,78) node [anchor=north west][inner sep=0.75pt]   [align=left] {$\displaystyle w_{5}$};
\draw (184,172) node [anchor=north west][inner sep=0.75pt]   [align=left] {$\displaystyle w_{4}$};
\draw (285,175) node [anchor=north west][inner sep=0.75pt]   [align=left] {$\displaystyle w_{3}$};
\draw (319,74) node [anchor=north west][inner sep=0.75pt]   [align=left] {$\displaystyle w_{2}$};
\draw (236,23) node [anchor=north west][inner sep=0.75pt]   [align=left] {$\displaystyle w_{1}$};
\draw (187,-6) node [anchor=north west][inner sep=0.75pt]   [align=left] {$\displaystyle v_{1}$};
\draw (121,130) node [anchor=north west][inner sep=0.75pt]   [align=left] {$\displaystyle v_{5}$};
\draw (230,216) node [anchor=north west][inner sep=0.75pt]   [align=left] {$\displaystyle v_{4}$};
\draw (345,148) node [anchor=north west][inner sep=0.75pt]   [align=left] {$\displaystyle v_{3}$};
\draw (327,19) node [anchor=north west][inner sep=0.75pt]   [align=left] {$\displaystyle v_{2}$};
\end{tikzpicture}
\caption{If $Q$ admits an R-charge, then each maximal weak cut is a cut. \label{figRcut}} 
\end{figure}
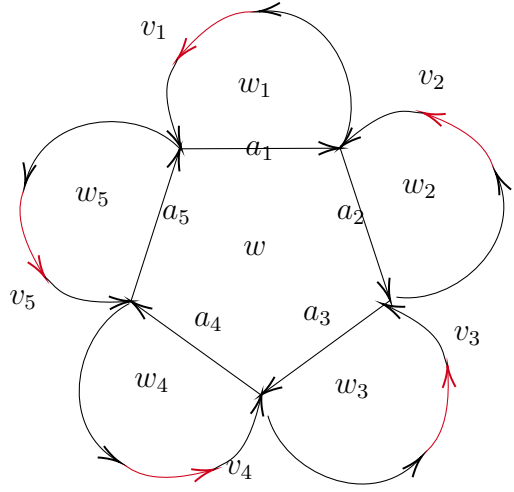

$ii)\Rightarrow iii)$. Consider a maximal weak cut $I$, and $R:\mathbb{R}^{Q_1}\to\mathbb{R}$ such that $R\circ\partial_2\geq R_I\circ\partial_2$, and $R_I\circ\partial_2(w)=2\Rightarrow R\circ\partial_2(w)=2$. We must show that $R\circ\partial_2= R_I\circ\partial_2$. Consider $(w=a_n...a_1)\in Q_2$ such that $R_I\circ\partial_2(w)=0$ \ie $a_k\not\in I$ for each $k$ (see Figure \ref{figRcut}): in particular one has $R\circ\partial_2(w)\geq0$. Because $I$ is maximal, each arrow $a_k\in w$ is contained in a simple cycle $w_k=v_ka_k\in Q_2$ such that $v_k$ contains exactly one arrow of $I$ (since otherwise $I\cup\{a_k\}$ would be a larger weak cut). The oriented cycle $\bar{w}=v_1...v_n$ is a product of simple oriented cycles, and satisfies: 
\begin{align}
    &R\circ\partial_2(\bar{w})\geq R_I\circ\partial_2(\bar{w})=2n\nn\\
    &\partial_2(w)=\sum_k\partial_2(w_k)-\partial_2(\bar{w})\nn\\
    \Rightarrow &R\circ\partial_2(w)=2n-R\circ\partial_2(\bar{w})\leq 0\nn\\
    \Rightarrow &R\circ\partial_2(w)=0=R_I\circ\partial_2(w)
\end{align}
where the third line holds because $R_I\circ\partial_2(w_k)=2$. By disjunction of cases, one has then $R\circ\partial_2=R_I\circ\partial_2$. When $R$ is an R-charge, one obtains then that $R_I$ is a R-charge, \ie $I$ is a cut. $\Box$ \medskip

\begin{lemma}\label{cutcycle}
    For a quiver with an R-charge and a cycle $w_0$ passing through all the nodes, the cuts are given by $I=\{a:i\to j|i>j\}$ for each labelling of the nodes such that $w_0:1\to 2\to ...\to n\to 1$.
\end{lemma}

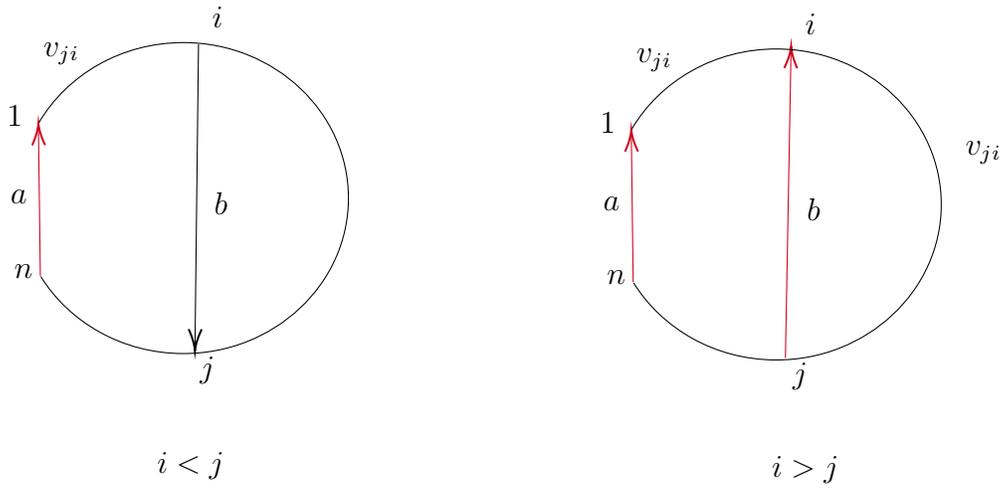
\begin{figure}
\centering
\begin{tikzpicture}[x=0.75pt,y=0.75pt,yscale=-1,xscale=1]

\draw [color={rgb, 255:red, 208; green, 2; blue, 27 }  ,draw opacity=1 ]   (129.34,122.21) -- (128.39,47.37) ;
\draw [shift={(128.37,45.37)}, rotate = 449.28] [color={rgb, 255:red, 208; green, 2; blue, 27 }  ,draw opacity=1 ][line width=0.75]    (10.93,-3.29) .. controls (6.95,-1.4) and (3.31,-0.3) .. (0,0) .. controls (3.31,0.3) and (6.95,1.4) .. (10.93,3.29)   ;
\draw    (209.12,5.4) -- (207.34,158.15) ;
\draw [shift={(207.32,160.15)}, rotate = 270.67] [color={rgb, 255:red, 0; green, 0; blue, 0 }  ][line width=0.75]    (10.93,-3.29) .. controls (6.95,-1.4) and (3.31,-0.3) .. (0,0) .. controls (3.31,0.3) and (6.95,1.4) .. (10.93,3.29)   ;
\draw  [draw opacity=0] (128.57,45.02) .. controls (142.81,20.78) and (170.26,4.41) .. (201.73,4.52) .. controls (247.72,4.67) and (284.88,39.93) .. (284.73,83.28) .. controls (284.59,126.62) and (247.19,161.64) .. (201.21,161.48) .. controls (170.64,161.38) and (143.96,145.76) .. (129.56,122.57) -- (201.47,83) -- cycle ; \draw   (128.57,45.02) .. controls (142.81,20.78) and (170.26,4.41) .. (201.73,4.52) .. controls (247.72,4.67) and (284.88,39.93) .. (284.73,83.28) .. controls (284.59,126.62) and (247.19,161.64) .. (201.21,161.48) .. controls (170.64,161.38) and (143.96,145.76) .. (129.56,122.57) ;
\draw [color={rgb, 255:red, 208; green, 2; blue, 27 }  ,draw opacity=1 ]   (428.34,125.44) -- (427.39,50.6) ;
\draw [shift={(427.37,48.6)}, rotate = 449.28] [color={rgb, 255:red, 208; green, 2; blue, 27 }  ,draw opacity=1 ][line width=0.75]    (10.93,-3.29) .. controls (6.95,-1.4) and (3.31,-0.3) .. (0,0) .. controls (3.31,0.3) and (6.95,1.4) .. (10.93,3.29)   ;
\draw [color={rgb, 255:red, 208; green, 2; blue, 27 }  ,draw opacity=1 ]   (505.12,163.63) -- (507.96,8.5) ;
\draw [shift={(508,6.5)}, rotate = 451.05] [color={rgb, 255:red, 208; green, 2; blue, 27 }  ,draw opacity=1 ][line width=0.75]    (10.93,-3.29) .. controls (6.95,-1.4) and (3.31,-0.3) .. (0,0) .. controls (3.31,0.3) and (6.95,1.4) .. (10.93,3.29)   ;
\draw  [draw opacity=0] (427.57,48.25) .. controls (441.81,24.01) and (469.26,7.64) .. (500.73,7.74) .. controls (546.72,7.9) and (583.88,43.16) .. (583.73,86.5) .. controls (583.59,129.85) and (546.19,164.87) .. (500.21,164.71) .. controls (469.64,164.61) and (442.96,148.99) .. (428.56,125.79) -- (500.47,86.23) -- cycle ; \draw   (427.57,48.25) .. controls (441.81,24.01) and (469.26,7.64) .. (500.73,7.74) .. controls (546.72,7.9) and (583.88,43.16) .. (583.73,86.5) .. controls (583.59,129.85) and (546.19,164.87) .. (500.21,164.71) .. controls (469.64,164.61) and (442.96,148.99) .. (428.56,125.79) ;

\draw (114.95,115.08) node [anchor=north west][inner sep=0.75pt]   [align=left] {$\displaystyle n$};
\draw (111.35,35.03) node [anchor=north west][inner sep=0.75pt]   [align=left] {$\displaystyle 1$};
\draw (214.87,-14.77) node [anchor=north west][inner sep=0.75pt]   [align=left] {$\displaystyle i$};
\draw (208.57,161.77) node [anchor=north west][inner sep=0.75pt]   [align=left] {$\displaystyle j$};
\draw (113.15,77.72) node [anchor=north west][inner sep=0.75pt]   [align=left] {$\displaystyle a$};
\draw (215.77,78.61) node [anchor=north west][inner sep=0.75pt]   [align=left] {$\displaystyle b$};
\draw (129.76,3.02) node [anchor=north west][inner sep=0.75pt]   [align=left] {$\displaystyle v_{j}{}_{i}$};
\draw (413.95,118.31) node [anchor=north west][inner sep=0.75pt]   [align=left] {$\displaystyle n$};
\draw (410.35,38.26) node [anchor=north west][inner sep=0.75pt]   [align=left] {$\displaystyle 1$};
\draw (513.87,-11.54) node [anchor=north west][inner sep=0.75pt]   [align=left] {$\displaystyle i$};
\draw (507.57,165) node [anchor=north west][inner sep=0.75pt]   [align=left] {$\displaystyle j$};
\draw (412.15,80.95) node [anchor=north west][inner sep=0.75pt]   [align=left] {$\displaystyle a$};
\draw (514.77,81.84) node [anchor=north west][inner sep=0.75pt]   [align=left] {$\displaystyle b$};
\draw (428.76,6.24) node [anchor=north west][inner sep=0.75pt]   [align=left] {$\displaystyle v_{j}{}_{i}$};
\draw (594.76,53) node [anchor=north west][inner sep=0.75pt]   [align=left] {$\displaystyle v_{j}{}_{i}$};
\draw (187,210) node [anchor=north west][inner sep=0.75pt]   [align=left] {$\displaystyle i< j$};
\draw (497,212) node [anchor=north west][inner sep=0.75pt]   [align=left] {$\displaystyle i >j$};
\end{tikzpicture}
\caption{Under assumptions of lemma \ref{cutcycle}, all cuts are of the form $I=\{a:i\to j| i>j\}$ \label{figcutcy}}
\end{figure}

Consider a cut $I$ of $Q$: it contains exactly one arrow $a\in w_0$: we choose the labelling of $Q_0$ such that $a:n\to 1$ (see Figure \ref{figcutcy}). Consider an arrow $(b:i\to j)\in Q_1$, and denote by $v_{ji}$ the minimal path going from $j$ to $i$ on the cycle $w_0$: $v_{ji}$ contains an arrow of $I$ if and only if $i<j$, and the simple oriented cycle $v_{ji}b\in Q_2$ contains exactly one arrow of the cut $I$, \ie $b\in I$ if and only if $j<i$. One concludes that each cut of $Q$ is of the form $I=\{b:i\in j|j<i$. for a specific cyclic ordering of $w_0$.\medskip 

\begin{figure}
    \centering
\begin{tikzpicture}[x=0.75pt,y=0.75pt,yscale=-1,xscale=1]

\draw    (233.32,89.59) -- (303.53,66.77) ;
\draw [shift={(305.43,66.16)}, rotate = 522] [color={rgb, 255:red, 0; green, 0; blue, 0 }  ][line width=0.75]    (10.93,-3.29) .. controls (6.95,-1.4) and (3.31,-0.3) .. (0,0) .. controls (3.31,0.3) and (6.95,1.4) .. (10.93,3.29)   ;
\draw    (305.43,66.16) -- (348.82,125.88) ;
\draw [shift={(350,127.5)}, rotate = 234] [color={rgb, 255:red, 0; green, 0; blue, 0 }  ][line width=0.75]    (10.93,-3.29) .. controls (6.95,-1.4) and (3.31,-0.3) .. (0,0) .. controls (3.31,0.3) and (6.95,1.4) .. (10.93,3.29)   ;
\draw    (350,127.5) -- (306.61,187.23) ;
\draw [shift={(305.43,188.84)}, rotate = 306] [color={rgb, 255:red, 0; green, 0; blue, 0 }  ][line width=0.75]    (10.93,-3.29) .. controls (6.95,-1.4) and (3.31,-0.3) .. (0,0) .. controls (3.31,0.3) and (6.95,1.4) .. (10.93,3.29)   ;
\draw    (305.43,188.84) -- (235.22,166.03) ;
\draw [shift={(233.32,165.41)}, rotate = 378] [color={rgb, 255:red, 0; green, 0; blue, 0 }  ][line width=0.75]    (10.93,-3.29) .. controls (6.95,-1.4) and (3.31,-0.3) .. (0,0) .. controls (3.31,0.3) and (6.95,1.4) .. (10.93,3.29)   ;
\draw    (233.32,165.41) -- (233.32,91.59) ;
\draw [shift={(233.32,89.59)}, rotate = 450] [color={rgb, 255:red, 0; green, 0; blue, 0 }  ][line width=0.75]    (10.93,-3.29) .. controls (6.95,-1.4) and (3.31,-0.3) .. (0,0) .. controls (3.31,0.3) and (6.95,1.4) .. (10.93,3.29)   ;
\draw    (305.43,66.16) .. controls (315.95,-16.09) and (168.49,18.15) .. (232.34,88.52) ;
\draw [shift={(233.32,89.59)}, rotate = 226.99] [color={rgb, 255:red, 0; green, 0; blue, 0 }  ][line width=0.75]    (10.93,-3.29) .. controls (6.95,-1.4) and (3.31,-0.3) .. (0,0) .. controls (3.31,0.3) and (6.95,1.4) .. (10.93,3.29)   ;
\draw    (352.14,126.8) .. controls (433.29,109.77) and (352.59,-18.31) .. (307.05,65.09) ;
\draw [shift={(306.36,66.36)}, rotate = 297.85] [color={rgb, 255:red, 0; green, 0; blue, 0 }  ][line width=0.75]    (10.93,-3.29) .. controls (6.95,-1.4) and (3.31,-0.3) .. (0,0) .. controls (3.31,0.3) and (6.95,1.4) .. (10.93,3.29)   ;
\draw    (306.95,189.44) .. controls (348.57,261.15) and (444.88,144.35) .. (351.41,127.26) ;
\draw [shift={(349.98,127.01)}, rotate = 369.57] [color={rgb, 255:red, 0; green, 0; blue, 0 }  ][line width=0.75]    (10.93,-3.29) .. controls (6.95,-1.4) and (3.31,-0.3) .. (0,0) .. controls (3.31,0.3) and (6.95,1.4) .. (10.93,3.29)   ;
\draw    (232.92,90.66) .. controls (158.13,54.87) and (144.37,205.63) .. (231.22,167.08) ;
\draw [shift={(232.54,166.49)}, rotate = 515.28] [color={rgb, 255:red, 0; green, 0; blue, 0 }  ][line width=0.75]    (10.93,-3.29) .. controls (6.95,-1.4) and (3.31,-0.3) .. (0,0) .. controls (3.31,0.3) and (6.95,1.4) .. (10.93,3.29)   ;
\draw    (232.54,166.54) .. controls (176.86,227.97) and (317.39,284.26) .. (305.28,190.01) ;
\draw [shift={(305.08,188.58)}, rotate = 441.89] [color={rgb, 255:red, 0; green, 0; blue, 0 }  ][line width=0.75]    (10.93,-3.29) .. controls (6.95,-1.4) and (3.31,-0.3) .. (0,0) .. controls (3.31,0.3) and (6.95,1.4) .. (10.93,3.29)   ;

\draw (257,57) node [anchor=north west][inner sep=0.75pt]   [align=left] {$\displaystyle a_{1}$};
\draw (238,87) node [anchor=north west][inner sep=0.75pt]   [align=left] {$\displaystyle i_{1}$};
\draw (223,-4) node [anchor=north west][inner sep=0.75pt]   [align=left] {$\displaystyle v_{i_{1}}{}_{i_{2}}$};
\draw (215,118) node [anchor=north west][inner sep=0.75pt]   [align=left] {$\displaystyle a_{5}$};
\draw (264,174) node [anchor=north west][inner sep=0.75pt]   [align=left] {$\displaystyle a_{4}$};
\draw (330,151) node [anchor=north west][inner sep=0.75pt]   [align=left] {$\displaystyle a_{3}$};
\draw (328,80) node [anchor=north west][inner sep=0.75pt]   [align=left] {$\displaystyle a_{2}$};
\draw (237,144) node [anchor=north west][inner sep=0.75pt]   [align=left] {$\displaystyle i_{5}$};
\draw (293,164) node [anchor=north west][inner sep=0.75pt]   [align=left] {$\displaystyle i_{4}$};
\draw (327,117) node [anchor=north west][inner sep=0.75pt]   [align=left] {$\displaystyle i_{3}$};
\draw (293,70) node [anchor=north west][inner sep=0.75pt]   [align=left] {$\displaystyle i_{2}$};
\draw (139,118) node [anchor=north west][inner sep=0.75pt]   [align=left] {$\displaystyle v_{i_{5}}{}_{i_{1}}$};
\draw (195,207) node [anchor=north west][inner sep=0.75pt]   [align=left] {$\displaystyle v_{i_{4}}{}_{i_{5}}$};
\draw (383,185) node [anchor=north west][inner sep=0.75pt]   [align=left] {$\displaystyle v_{i_{3}}{}_{i_{4}}$};
\draw (383,42) node [anchor=north west][inner sep=0.75pt]   [align=left] {$\displaystyle v_{i_{2}}{}_{i_{3}}$};
\end{tikzpicture}
\caption{Under assumptions of lemma \ref{cutcycle}, the set $I=\{a:i\to j| i>j\}$ is a cut. \label{figcutcy2}}
\end{figure}
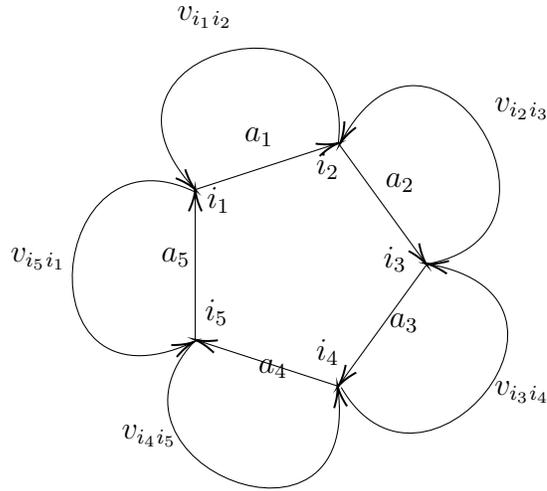

Conversely, consider the set $I=\{b:i\to j|j<i$\}, and a cycle $w: i_1\overset{a_1}{\to} i_2...i_r\overset{a_r}{\to} i_1\in Q_2$, such that $R\circ\partial_2(w)=2$ (see Figure \ref{figcutcy2}). We will show that $w$ contains exactly one arrow of $I$. The cycle $v_{i_1 i_r}...v_{i_2 i_1}$ is equal to the $m$-th iteration $w_0^m$, for some $m\in\mathbb{N}$. Since  $2=R\circ\partial_2(w)=\sum_{k=1}^rR\circ\partial_2( v_{i_{k+1}i_{k}}a_k)-mR\circ\partial_2(w_0)=2r-2m$, the number of iterations is $m=r-1$. But $w_0\in Q_2$ contains exactly one arrow of $I$, and each cycle $v_{i_{k+1}i_{k}}a_k\in Q_2$ contains exactly one arrow of $I$. Indeed, there are two options:  a) either $i_k>i_{k+1}$, and then $a_k\in I$ but $v_{i_{k+1}i_{k}}$ does not contain any arrow of $I$, or b)  $i_k<i_{k+1}$, and then $a_k\not\in I$ but $v_{i_{k+1}i_{k}}$ contains the arrow $n\to 1\in I$. Evaluating the R-charge, we get  $R_I\circ\partial_2(w)=\sum_{k=1}^rR_I\circ\partial_2( v_{i_{k+1}i_{k}}a_k)-(r-1)R_I\circ\partial_2(w_0)=2$. Therefore  $w\in Q_2$ contains exactly one arrow of $I$, \ie $I$ is a cut. $\Box$

\medskip

When $Q$ admits no R-charge (see Figures \ref{fighexa}, \ref{figpenta} for examples), the set $I=\{b:i\in j|j<i\}$ is a minimal strong cut of $Q$ which is not a cut: each simple oriented cycle contains at least one arrow of $I$, but there must be a cycle $w\in Q_2$ containing more that one arrow of $Q$. In particular, if $Q$ has an other cycle $w_1$ passing though all the nodes in a different order than $w_0$, then it has no R-charge. Indeed, for a labelling of the nodes such that $w_0:1\to 2\to ...\to n\to 1$, the cycle $w_1$ is of the form $1=\sigma(1)\to\sigma(2)\to ...\to\sigma(n)\to \sigma(1)$, with $\sigma$ a permutation of $\{1,...,n\}$ different from the identity. There must then be at least two arrows of $w_1$ such that $\sigma(i)>\sigma(i+1)$, and then $w_1$ contains at least two arrows of $I=\{a:i\to j|i>j\}$, \ie $I$ is not a cut, and then $Q$ has no R-charge. More generally, the criterion of the above lemma gives a simple algorithm to check if a quiver with a cycle passing through all the nodes has an R-charge. \medskip

\subsection{Current decomposition for Abelianized quivers}

\begin{lemma}\label{lemnonab}
    For $Q$ a biconnected quiver and $d$ a dimension vector with $d_i\geq 1$ for $i\in Q_0$, a strictly positive conserved current on $Q^d$ can be expressed as a sum of (not necessarily positive) currents circulating on the cycles in $p^{-1}(Q_2)$.
\end{lemma}

Proof: Consider a strictly positive conserved current $\lambda$ on $Q^d$: in particular $Q^d$, and therefore $Q$, is strongly connected from $i)$ of lemma \ref{lemgraphcoho}, and $\lambda$ can be expressed as a sum of positive currents circulating on the cycles in $Q^d_2$. One must then show that for $Q$ a biconnected strongly connected quiver each cycle in $Q^d_2$ can be expressed as a linear combination of cycles in $p^{-1}(Q_2)$, \ie $\partial_2(\mathbb{R}^{Q^d_2})\subset\partial_2(\mathbb{R}^{p^{-1}(Q_2)})$. One can construct the Abelianized quiver by using a finite sequence of elementary steps, where one node is split into two nodes at each step. Each of these steps preserves the fact that the quiver is biconnected and strongly connected, thus it suffices to prove the statement for an elementary step.\medskip

\begin{figure}
    \centering
    \begin{tikzpicture}[x=0.75pt,y=0.75pt,yscale=-1,xscale=1]

\draw    (143.93,120.26) -- (236.85,-40.21) ;
\draw [shift={(237.85,-41.94)}, rotate = 480.07] [color={rgb, 255:red, 0; green, 0; blue, 0 }  ][line width=0.75]    (10.93,-3.29) .. controls (6.95,-1.4) and (3.31,-0.3) .. (0,0) .. controls (3.31,0.3) and (6.95,1.4) .. (10.93,3.29)   ;
\draw    (237.85,-41.94) -- (350.98,-37.98) ;
\draw [shift={(352.98,-37.91)}, rotate = 182] [color={rgb, 255:red, 0; green, 0; blue, 0 }  ][line width=0.75]    (10.93,-3.29) .. controls (6.95,-1.4) and (3.31,-0.3) .. (0,0) .. controls (3.31,0.3) and (6.95,1.4) .. (10.93,3.29)   ;
\draw    (352.98,-37.91) -- (557.42,120.37) ;
\draw [shift={(559,121.6)}, rotate = 217.75] [color={rgb, 255:red, 0; green, 0; blue, 0 }  ][line width=0.75]    (10.93,-3.29) .. controls (6.95,-1.4) and (3.31,-0.3) .. (0,0) .. controls (3.31,0.3) and (6.95,1.4) .. (10.93,3.29)   ;
\draw    (559,121.6) -- (354.55,282.56) ;
\draw [shift={(352.98,283.79)}, rotate = 321.78999999999996] [color={rgb, 255:red, 0; green, 0; blue, 0 }  ][line width=0.75]    (10.93,-3.29) .. controls (6.95,-1.4) and (3.31,-0.3) .. (0,0) .. controls (3.31,0.3) and (6.95,1.4) .. (10.93,3.29)   ;
\draw    (352.98,283.79) -- (224.7,285.11) ;
\draw [shift={(222.7,285.13)}, rotate = 359.40999999999997] [color={rgb, 255:red, 0; green, 0; blue, 0 }  ][line width=0.75]    (10.93,-3.29) .. controls (6.95,-1.4) and (3.31,-0.3) .. (0,0) .. controls (3.31,0.3) and (6.95,1.4) .. (10.93,3.29)   ;
\draw    (222.7,285.13) -- (144.79,122.06) ;
\draw [shift={(143.93,120.26)}, rotate = 424.46000000000004] [color={rgb, 255:red, 0; green, 0; blue, 0 }  ][line width=0.75]    (10.93,-3.29) .. controls (6.95,-1.4) and (3.31,-0.3) .. (0,0) .. controls (3.31,0.3) and (6.95,1.4) .. (10.93,3.29)   ;
\draw    (143.93,120.26) -- (351.4,282.56) ;
\draw [shift={(352.98,283.79)}, rotate = 218.04] [color={rgb, 255:red, 0; green, 0; blue, 0 }  ][line width=0.75]    (10.93,-3.29) .. controls (6.95,-1.4) and (3.31,-0.3) .. (0,0) .. controls (3.31,0.3) and (6.95,1.4) .. (10.93,3.29)   ;
\draw    (143.93,120.26) -- (351.04,69.79) ;
\draw [shift={(352.98,69.32)}, rotate = 526.31] [color={rgb, 255:red, 0; green, 0; blue, 0 }  ][line width=0.75]    (10.93,-3.29) .. controls (6.95,-1.4) and (3.31,-0.3) .. (0,0) .. controls (3.31,0.3) and (6.95,1.4) .. (10.93,3.29)   ;
\draw    (352.98,-37.91) -- (145.52,119.05) ;
\draw [shift={(143.93,120.26)}, rotate = 322.89] [color={rgb, 255:red, 0; green, 0; blue, 0 }  ][line width=0.75]    (10.93,-3.29) .. controls (6.95,-1.4) and (3.31,-0.3) .. (0,0) .. controls (3.31,0.3) and (6.95,1.4) .. (10.93,3.29)   ;
\draw    (352.98,176.56) -- (145.86,120.78) ;
\draw [shift={(143.93,120.26)}, rotate = 375.07] [color={rgb, 255:red, 0; green, 0; blue, 0 }  ][line width=0.75]    (10.93,-3.29) .. controls (6.95,-1.4) and (3.31,-0.3) .. (0,0) .. controls (3.31,0.3) and (6.95,1.4) .. (10.93,3.29)   ;
\draw    (352.98,176.56) -- (557.07,122.11) ;
\draw [shift={(559,121.6)}, rotate = 525.06] [color={rgb, 255:red, 0; green, 0; blue, 0 }  ][line width=0.75]    (10.93,-3.29) .. controls (6.95,-1.4) and (3.31,-0.3) .. (0,0) .. controls (3.31,0.3) and (6.95,1.4) .. (10.93,3.29)   ;
\draw    (559,121.6) -- (354.92,69.81) ;
\draw [shift={(352.98,69.32)}, rotate = 374.24] [color={rgb, 255:red, 0; green, 0; blue, 0 }  ][line width=0.75]    (10.93,-3.29) .. controls (6.95,-1.4) and (3.31,-0.3) .. (0,0) .. controls (3.31,0.3) and (6.95,1.4) .. (10.93,3.29)   ;
\draw    (352.98,69.32) -- (352.98,-35.91) ;
\draw [shift={(352.98,-37.91)}, rotate = 450] [color={rgb, 255:red, 0; green, 0; blue, 0 }  ][line width=0.75]    (10.93,-3.29) .. controls (6.95,-1.4) and (3.31,-0.3) .. (0,0) .. controls (3.31,0.3) and (6.95,1.4) .. (10.93,3.29)   ;
\draw    (352.98,69.32) -- (352.98,174.56) ;
\draw [shift={(352.98,176.56)}, rotate = 270] [color={rgb, 255:red, 0; green, 0; blue, 0 }  ][line width=0.75]    (10.93,-3.29) .. controls (6.95,-1.4) and (3.31,-0.3) .. (0,0) .. controls (3.31,0.3) and (6.95,1.4) .. (10.93,3.29)   ;
\draw    (352.98,283.79) -- (352.98,178.56) ;
\draw [shift={(352.98,176.56)}, rotate = 450] [color={rgb, 255:red, 0; green, 0; blue, 0 }  ][line width=0.75]    (10.93,-3.29) .. controls (6.95,-1.4) and (3.31,-0.3) .. (0,0) .. controls (3.31,0.3) and (6.95,1.4) .. (10.93,3.29)   ;
\draw (129.03,173.14) node [anchor=north west][inner sep=0.75pt]   [align=left] {$\displaystyle ( d,1)$};
\draw (147.02,22.27) node [anchor=north west][inner sep=0.75pt]   [align=left] {$\displaystyle ( a,1)$};
\draw (265.42,-65.26) node [anchor=north west][inner sep=0.75pt]   [align=left] {$\displaystyle u$};
\draw (449.31,16.6) node [anchor=north west][inner sep=0.75pt]   [align=left] {$\displaystyle ( b_{0} ,2)$};
\draw (283.12,291.76) node [anchor=north west][inner sep=0.75pt]   [align=left] {$\displaystyle v$};
\draw (449.18,209.14) node [anchor=north west][inner sep=0.75pt]   [align=left] {$\displaystyle ( c_{1} ,2)$};
\draw (410.74,66.03) node [anchor=north west][inner sep=0.75pt]   [align=left] {$\displaystyle ( c_{0} ,2)$};
\draw (236.29,69.58) node [anchor=north west][inner sep=0.75pt]   [align=left] {$\displaystyle ( c_{0} ,1)$};
\draw (227.87,218.05) node [anchor=north west][inner sep=0.75pt]   [align=left] {$\displaystyle ( c_{1} ,1)$};
\draw (247.97,-8.13) node [anchor=north west][inner sep=0.75pt]   [align=left] {$\displaystyle ( b_{0} ,1)$};
\draw (238.26,128.07) node [anchor=north west][inner sep=0.75pt]   [align=left] {$\displaystyle ( b_{1} ,1)$};
\draw (409.51,161.77) node [anchor=north west][inner sep=0.75pt]   [align=left] {$\displaystyle ( b_{1} ,2)$};
\draw (326.73,14.21) node [anchor=north west][inner sep=0.75pt]   [align=left] {$\displaystyle u_{0}$};
\draw (325.27,201.93) node [anchor=north west][inner sep=0.75pt]   [align=left] {$\displaystyle u_{1}$};
\draw (327.95,108.64) node [anchor=north west][inner sep=0.75pt]   [align=left] {$\displaystyle v_{0}$};
\draw (223,3) node [anchor=north west][inner sep=0.75pt]   [align=left] {\textcolor[rgb]{0.95,0.05,0.05}{{\LARGE +}}};
\draw (207,190) node [anchor=north west][inner sep=0.75pt]   [align=left] {\textcolor[rgb]{0.95,0.05,0.05}{{\LARGE +}}};
\draw (293,105) node [anchor=north west][inner sep=0.75pt]   [align=left] {\textcolor[rgb]{0.95,0.05,0.05}{{\LARGE +}}};
\draw (397,24) node [anchor=north west][inner sep=0.75pt]   [align=left] {\textcolor[rgb]{0.95,0.05,0.05}{{\LARGE +}}};
\draw (401,191) node [anchor=north west][inner sep=0.75pt]   [align=left] {\textcolor[rgb]{0.95,0.05,0.05}{{\LARGE +}}};
\draw (300,21) node [anchor=north west][inner sep=0.75pt]   [align=left] {\textcolor[rgb]{0.29,0.56,0.89}{{\LARGE $-$}}};
\draw (300,200) node [anchor=north west][inner sep=0.75pt]   [align=left] {\textcolor[rgb]{0.29,0.56,0.89}{{\LARGE $-$}}};
\draw (397,105) node [anchor=north west][inner sep=0.75pt]   [align=left] {\textcolor[rgb]{0.29,0.56,0.89}{{\LARGE $-$}}};
\draw (106,111) node [anchor=north west][inner sep=0.75pt]   [align=left] {$\displaystyle ( i,1)$};
\draw (565,110) node [anchor=north west][inner sep=0.75pt]   [align=left] {$\displaystyle ( i,2)$};

\end{tikzpicture}

    \caption{Proof of the fact that a simple cycle of $Q^d$ can be expressed as a signed sum of cycles in $p^{-1}(Q_2)$.
        \label{fignonab}}
\end{figure}
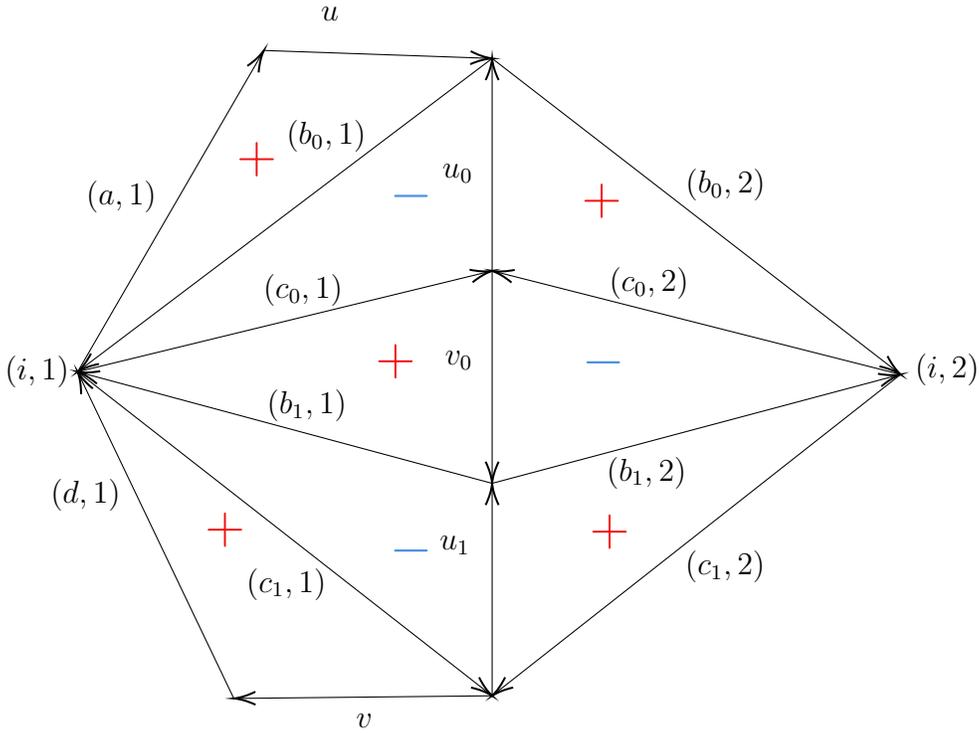

Consider a biconnected strongly connected quiver $Q$, a node $i\in Q_0$ and the dimension vector $d$ such that $d_i=2$ and $d_j=1$ for $j\neq i$. Consider a cycle $w\in Q^d_2-p^{-1}(Q_2)$: it  necessarily passes
through the two nodes $(i,1),(i,2)$, \ie it is of the form\footnote{Since $i$ is the only non-Abelian node, we abuse notation and denote by $(a,k)$ and $(b,k)$ the lift of the arrows $a:i\to j$ 
and $b:j\to i$ on $Q$ to the $k$-th copy of the node $i$ on $Q^d$.} $(d,1)v(c,2)(b,2)u(a,1)$, where $a,c$ are arrows of $Q$ with source $i$, $b,d$ arrows of $Q$ with target $i$, and $u$ and $v$ paths of $Q$ avoiding $i$, such that $dvc,bua\in Q_2$ (see Figure \ref{fignonab}). We claim that there is a sequence of arrows $b_0=b,b_1,...,b_n$ with target $i$, a sequence of arrows $c_0,...,c_n=c$ with source $i$, and sequences of paths $u_1,...,u_n$ and $v_1,...,v_{n-1}$ in $Q$, such that $b_ku_kc_k$ and $b_{k+1}v_kc_k$ are simple oriented cycles in $Q_2$ for $1\leq k\leq n-1$. Then:
\begin{align}
    \partial_2\left(
    (d,1)v(c,2)(b,2)u(a,1)\right)
    =&\partial_2\left( (d,1)v(c,1)+(b,1)u(a,1) \right. \nn\\ 
 & \hspace*{-4cm}
 +\sum_{k=0}^{n-1}
    \left((b_k,2)u_k(c_k,2)-(b_k,1)u_k(c_k,1)+(b_{k+1},1)
    v_k(c_k,1)-(b_{k+1},2)v_k(c_k,2) \right)
\end{align}
where each of the cycles appearing on the right belongs to $p^{-1}(Q_2)$, i.e. project to a simple cycle of $Q$. Thus $\partial_2((d,1)v(c,2)(b,2)u(a,1))\in \partial_2(\mathbb{R}^{p^{-1}(Q_2)})$, and then $\partial_2(\mathbb{R}^{Q^d_2})\subset\partial_2(\mathbb{R}^{p^{-1}(Q_2)})$ which concludes the proof.\medskip

Proof of the claim: Consider a strongly connected biconnected quiver $Q$, and a node $i\in Q_0$. Consider the equivalence $\sim$ on the arrows of $Q$ with source or target $i$, generated by $a\sim b$, with $a$ (resp. $b$) with source (resp. target) $i$ if there is a path $v$ in $Q$ such that $bva$ is a simple oriented cycle in $Q_2$. We need to show 
that there is a single equivalence class under $\sim$. Let $S$
be the set of equivalence classes. To each $s\in S$ of $\sim$ one can associate the subquiver $Q^s\subset Q-\{i\}$ whose nodes are the nodes $j$ such that there is a path $v:j\to i$ passing only once through $i$ and ending by an arrow in the given equivalence class.\medskip

Since $Q$ is strongly connected, for each $j\in Q_0-\{i\}$ there is a path from $j$ to $i$, which up to truncating the path, can be considered as passing only once through $i$, \ie $j\in Q_0^s$ for at least one $s\in S$. Consider $j\in Q_0^s$, $j'\in Q_0^{s'}$ and $u:j\to j'$ an oriented path of $Q$ avoiding $i$. Consider paths $v:j\to i$, $v':j'\to i$ passing only once through $i$ and ending respectively by $b\in s$, $b'\in s'$. Because $Q$ is strongly connected, there is a path $u':i\to j$ beginning by an arrow $a$ with source $i$: the two cycles $vu'$ (resp. $v'uu'$) contain a simple cycle containing $ba$ (resp. $b'a$), and then $b\sim a\sim b'$, \ie $s=s'$. For $j=j'$ and $u$ the trivial path at $j$, one obtains that the $Q_0^s$ form a partition of $Q_0^s-\{i\}$, and for $u=a:j\to j'$ an arrow of $Q$ one obtains that $Q^s$ and $Q^{s'}$ are disconnected in $Q-\{i\}$. If there were different equivalence classes under $\sim$, $Q-\{i\}$ would be disconnected, contradicting the assumption that $Q$ is biconnected. Thus  there is a single equivalence class under $\sim$, which proves the claim. $\Box$\medskip

\subsection{Higgs branch}

\begin{lemma}\label{lemgenvan}
    Let $Q$ be a quiver, $C\subset Q_2$ a set of cycles of $Q$ such that there exists a subset $I\subset Q_1$ of arrows such that each cycle of $C$ contains exactly one arrow of $I$.  For any dimension vector $d\in \mathbb{N}^Q_0$, there exists a dense open subset $U_{C,d}\subset\mathbb{C}^C$ such that for 
    any $(\nu_w)_{w\in C}\in U_{c,d}$ and any $d$-dimensional representation $\phi$ of the quiver $Q$ with potential
    $W=\sum_{w\in C}\nu_w w$, the trace $\Tr(w)$ vanishes on the representation $\phi$ for all cycles $w\in C$ appearing in the potential.
\end{lemma}

Proof: We follow an argument of Kontsevich quoted in the proof  of \cite[Prop. 3.1]{Efimov2011QuantumCV}. For $d\in\mathbb{N}^{Q_0}$, denote by $M_d$ the smooth quasi-projective connected space of $d$-dimensional representations of $Q$. Now, consider a potential cut $I$ of $W$, such that $\Tr(W)=\sum_{a\in I}\Tr(a\partial_a W)$: by homogeneity,
the critical points of $\Tr(W)$ on $M_d$ necessarily have $\Tr(W)=0$. Let $f:
M_d \rightarrow \mathbb{C}^C$ be the map which associates to any $d$-dimensional representation of $Q$ the vector 
$(\Tr(w)_{w\in C}$. Away from the locus $f^{-1}(0)$, the image of this map
descends to the projective space $\mathbb{P}^{|C|-1}$. The trace of the potential $W=\sum_{w\in C}\nu_w w$
is obtained by composing $f$ with a linear form, corresponding to 
a hyperplane section of $\mathbb{P}^{|C|-1}$. By
applying Bertini's theorem to the complete linear system $\Tr(W)^{-1}(0)$, one finds  that, for $(\nu_w)_{w\in C}$ in a dense open subset $U_{C,d}\subset\mathbb{C}^C$, $\Tr(W)^{-1}(0)$ is a smooth connected strict sub-variety of $M_d$ away from the zero locus $f^{-1}(0)$.
In particular, its tangent space at any point $x\in\Tr(W)^{-1}(0)-f^{-1}(0))$ is strictly included in the tangent
space of $M_d$,
\begin{align}
    T_x(\Tr(W)^{-1}(0))\subsetneq T_x(M_d)
\end{align}
hence $\delta(\Tr(W))|_x\neq 0$.
It follows that the critical points of  $(\Tr(W)$ lie in $f^{-1}(0)$, hence $\Tr(w)=0$ for $w\in C$ on a $d$-dimensional representation of $(Q,W)$. Specializing to the dimension vector $d=(1,...,1)$, one obtains that for $(\nu_w)_{w\in C}\in U_C$ a dense open subset of $\mathbb{C}^C$, and for any Abelian representation of $(Q,W)$, the cycles $w$ vanish for all $w\in C$. $\Box$.\medskip

Now, for any weak cut $I\subset Q_1$, we denote by $Q_2^I$ the set of cycles containing exactly one arrow of $I$, and for $W=\sum_{w\in Q_2}\nu_w w$, we define $W_I:=\sum_{w\in Q_2^I}\nu_w w$. Let $p_I:\mathbb{C}^{Q_2}\to\mathbb{C}^{Q_2^I}$ be the natural projection.

\begin{definition}\label{defgenpot}
    A potential $W=\sum_{w\in Q_2}\nu_w w$ is said to be generic if $(\nu_w)_{w\in Q_2}$ is in the dense open subset $\bigcap_{I}p_I^{-1}(U_{Q_2^I})$. 
\end{definition}

\begin{proposition}\label{prophiggsweak}
    Consider a quiver with generic potential $(Q,W)$. If there exists a self-stable Abelian representation of $(Q,W)$, then for each weak cut $I$:
    \begin{align}
        |I|\leq |Q_1-I|-|Q_0|+1
    \end{align}
\end{proposition}

Proof: Let $\phi$ be a $\zeta$-stable representation of $(Q,W)$, with $\zeta,W$ generic, and $I$ a weak cut of $Q$. Denote by $J$ the set of arrows of $Q$ vanishing in $\phi$. The representation $\phi$ gives a $\zeta$-stable representation of $Q_J$, the quiver where the arrows of $J$ have been removed: in particular, the stability of $\phi$ implies that $Q_J$ is connected (otherwise, $\phi$ would be a direct sum of representations supported on the connected components). Consider the set $K\subset I-J$ of arrows $(a:i\to j)\in I-J$ such that $i$ and $j$ are connected in $Q_{J\cup I}$. The quiver $Q_{J\cup K}$ obtained by removing all arrows in $K$ is then still connected. Sending the arrows of $K$ to $0$ in $\phi$, one obtains a representations $\psi$ of $Q_{J\cup K}$ without vanishing arrows. Because $Q_{J\cup K}$ is connected, the gauge group $(\mathbb{C}^\ast)^{Q_0}$ scaling the nodes acts on the space $(\mathbb{C}^\ast)^{Q_1-J-K}$ of representations of $Q_{J\cup K}$ without vanishing arrows with stabilizer $\mathbb{C}^\star$, giving a smooth moduli space of dimension $|Q_1-J-K|-|Q_0|+1$.\medskip

Consider the potential $W_I$ obtained from the generic potential $W$ by keeping only the cycles $w\in Q_2$ which contain an arrow of $I$. Consider $L\subset J\cup K$ the set of arrows contained in a cycle $w\in Q_2$ such that $R_{J\cup K}\circ\partial_2(w)= R_I\circ\partial_2(w)=2$. Consider $a\in L$:
\begin{itemize}
    \item If $a\in I$, because $I$ is a weak cut, any cycle $w\in Q_2$ which contains $a$ is a cycle of $W_I$, and contains no other arrow of $I$, and in particular no other arrows of $K$, \ie:
    \begin{align}
        \partial_a W_I|_{\psi}=\partial_a W|_{\psi}=\partial_a W|_\phi=0
    \end{align}
    \item Suppose $a\in J-I$. There is a cycle $w\in Q_2$ containing $a$ such that $R_{J\cup K}\circ\partial_2(w)= R_I\circ\partial_2(w)=2$. Hence $w$ contains an arrow $b\in I$ different from $a$, which cannot be in $K$, hence $b\in I-K$. Denote by $i\sim j$ the equivalence relation identifying vertices $i$ and $j$ which are connected in $Q_{J\cup I}$: one has $t(b)\sim s(a)$ and $t(a)\sim s(b)$ because $w$ contains no other arrows of $J\cup I$. Consider an other cycle $w'$ containing $a$. 
    \begin{itemize}
        \item If $w'$ contains an other arrow of $J$, then $0=\partial_a w'|_\psi=\partial_a w'|_\phi$.
        \item If $w'$ contains no other arrow of $J$ and an arrow $c\in K\subset I-J$, by definition, $s(c)\sim t(c)$: using $t(a)\sim s(c)$ and $t(c)\sim s(a)$ because $w'$ contains no other arrows of $J\cup I$, one obtains $t(b)\sim s(b)$, a contradiction because $b\in I-K$.
        \item If $w'$ contains no other arrow of $J$ and no arrow of $I$, then $t(a)\sim s(a)$, and then $t(b)\sim s(b)$, a contradiction because $b\in I-K$.
        \item In the remaining case $w'$ contains one arrow of $J$ and one arrow of $I-K$, then $w'$ is a cycle of $W_I$ and  $\partial_a w'|_\psi=\partial_a w'|_\phi$
    \end{itemize}
    By disjunction of cases one has $\partial_a W|_\psi=\partial_a W|_\phi=0$, and because the only cycles contributing to $\partial_a W|_\psi$ are cycles of $W_I$, one has $\partial_a W_I|_\psi=\partial_a W|_\psi=0$.
\end{itemize}
By disjunction on cases, $\partial_a W_I|_\psi=0$ for $a\in L$, hence $\psi$ is a representation of the quiver with relations $(Q_{J\cup K},\partial_L W_I)$ without vanishing arrows.
The tangent space of the moduli space of representations of $(Q_{J\cup K},\partial_{L}W_I)$ at $\psi$ is given by the intersection of the kernel of the $|L|$ differential forms $\delta(\partial_a W_I)$ for $a\in L$.\medskip

As in \eqref{relrel}, a linear relation $\sum_{a\in L}\tilde{\psi}_a \delta(\partial_aW_I)=0$ between these differential forms yields a representation $\bar{\psi}=(\psi_a,\epsilon \tilde{\psi}_b)_{a\in Q_1-L,b\in L}$ over $\mathbb{C}\epsilon/\epsilon^2$ such that $\delta_b W|_{\bar{\psi}}=0$ for $b\in Q_1-J\cup K$. For an arrow $b\in J\cup K$, because $\partial_b|_\psi=0$, $\partial_b|_{\bar{\psi}}$ is at most of order $\epsilon$, and $\bar{\psi}_b$ is at most of order $\epsilon$, hence $\bar{\psi}_b\partial_b|_{\bar{\psi}}=0$. Then $\bar{\psi}_b\partial_b|_{\bar{\psi}}=0$ for $b\in Q_1$. Since the each cycle of the potential $W_I$ contains exactly one arrow of $I$, Lemma \ref{lemgenvan} implies that the cycles of $W_I$ vanish in $\bar{\psi}$. By definition, each arrow $a\in L\subset J\cup K$ is contained in a cycle $w$ of $W_I$ in which it is the only arrow of $J\cup K$. In particular  all the other arrows are in $Q_1-J-K$ and are non-vanishing: then $\epsilon \tilde{\psi}_a\prod_{a\neq b\in w}\psi_b=0$, hence $\tilde{\psi}_a=0$. The $|L|$ differential forms $\delta(\partial_a W_I)$ for $a\in L$ are then independent, hence:
\begin{align}
    |Q_1-J-K|-|Q_0|+1-|L|\geq 0
\end{align}
Defining
\begin{align}
    R(a)=\left\{\begin{array}{ll}
        0 & \mbox{if } a\in Q_1-J \\
        1 & \mbox{if } a\in J-L\\
        2 & \mbox{if } a\in L\cap J
    \end{array}\right.\quad\quad 
    R'(a)=\left\{\begin{array}{ll}
        0 & \mbox{if } a\in Q_1-K \\
        1 & \mbox{if } a\in K-L\\
        2 & \mbox{if } a\in K\cap L
    \end{array}
\right. 
\end{align}
one finds
\begin{align}\label{ineqweak}
    \sum_{a\in Q_1}(R+R')(a)\leq|Q_1|-|Q_0|+1
\end{align}

We will now show that the following inequality
holds for all $w\in Q_2$:
\begin{align}\label{ineqRI}
    R_I\circ\partial_2(w)\leq (R+R')\circ\partial_2(w)
\end{align}
By disjunction of cases,
\begin{itemize}
    \item If $w$ contains no arrow of $I$, one has directly $R_I\circ\partial_2(w)=0\leq (R+R')\circ\partial_2(w)$.
    \item Suppose that $w$ contains an arrow $a\in I$. If $w$ contains no arrow of $J$, then $t(a)$ and $s(a)$ are connected in $Q_{J\cup I}$, hence $a\in K$. Then in any cases, $w$ contains an arrow of $J\cup K$. If it contains a single arrow of $J\cup K$, then by definition it is an arrow of $L$, contributing to $R+R'$ by $2$. If it contains at least two arrows of $J\cup K$, then each of them contributes to $R+R'$ by $1$. By disjunction of cases $R_I\circ\partial_2(w)=2\leq (R+R')\circ\partial_2(w)$.
\end{itemize}
The inequality \eqref{ineqRI} follows, and can be rewritten:
\begin{align}
    (R_I-R')\circ\partial_2(w)\leq     R\circ\partial_2(w)\;\quad\forall\;w\in Q_2
\end{align}

We now assume that $\zeta$ is a self-stability condition. Because the positive conserved current $\lambda=(1+\delta_a+|\psi_a|^2)_{a\in Q_1}$ (with $\delta_a\ll1$) corresponding with the self-stable representation $\psi$ is a sum of positive currents supported of cycles in $Q_2$ from Lemma \ref{lemgraphcoho}, it follows that:
\begin{align}
    (R_I-R')(\lambda)\leq R(\lambda)
\end{align}
We deduce then:
\begin{align}
    2|I|+\sum_{a\in Q_1}(R_I-R')(a)\delta_a&\leq\sum_{a\in Q_1}(R_I-R')(a)(1+\delta_a+|\phi_a|^2)+\sum_{a\in Q_1}R'(a)\nn\\&=(R_I-R')(\lambda)+\sum_{a\in Q_1}R'(a)\nn\\&\leq R(\lambda)+\sum_{a\in Q_1}R'(a)=\sum_{a\in Q_1}R(a)(1+\delta_a+|\phi_a|^2)+\sum_{a\in Q_1}R'(a)\nn\\&=\sum_{a\in Q_1}(R+R')(a)+\sum_{a\in Q_1}\delta_aR(a)\nn\\&\leq|Q_1|-|Q_0|+1+\sum_{a\in Q_1}\delta_aR(a)\nn\\
    \implies 2|I|&\leq|Q_1|-|Q_0|+1
\end{align}
Here we have used the fact that $(R_I-R')(a)\geq 0$ in the first line, the fact that $\phi_a=0$ when $R(a)\neq 0$ in the fourth line, \eqref{ineqweak} in the fifth line and the fact that $\delta\ll 1$. in the last line. $\Box$

\bibliography{refscaling.bib}
\bibliographystyle{utphys.bst}
\end{document}